\documentclass[11pt]{article}
\pdfoutput=1
\usepackage{fullpage}
\usepackage{tensor}
\usepackage{xcolor}
\usepackage{cite}
\usepackage{graphicx}
\usepackage{amsmath}
\usepackage{amssymb}
\usepackage{caption}
\usepackage{subfig}
\usepackage{url}
\usepackage[toc,page]{appendix}
\usepackage[utf8x]{inputenc}

\title{}
\date{}
\renewcommand{\vec}[1]{\mbox{\boldmath$ #1 $}}

\def\beq{\begin{equation}}
\def\eeq{\end{equation}}
\allowdisplaybreaks
\newcommand \slsh [1] {\not\!{#1}}

\begin{document}
\bibliographystyle{utphys}

\newcommand\n[1]{\textcolor{red}{(#1)}} 
\newcommand{\diff}{\mathop{}\!\mathrm{d}}
\newcommand{\lb}{\left}
\newcommand{\rb}{\right}
\newcommand{\f}{\frac}
\newcommand{\pd}{\partial}
\newcommand{\tr}{\text{tr}}
\newcommand{\fdiff}{\mathcal{D}}
\newcommand{\im}{\text{im}}
\let\caron\v
\renewcommand{\v}{\mathbf}
\newcommand{\T}{\tensor}
\newcommand{\R}{\mathbb{R}}
\newcommand{\C}{\mathbb{C}}
\newcommand{\Z}{\mathbb{Z}}
\newcommand{\msbar}{\ensuremath{\overline{\text{MS}}}}
\newcommand{\DIS}{\ensuremath{\text{DIS}}}
\newcommand{\abar}{\ensuremath{\bar{\alpha}_S}}
\newcommand{\bb}{\ensuremath{\bar{\beta}_0}}
\newcommand{\rc}{\ensuremath{r_{\text{cut}}}}
\newcommand{\Nd}{\ensuremath{N_{\text{d.o.f.}}}}
\newcommand{\red}[1]{{\color{red} #1}}
\setlength{\parindent}{0pt}

\titlepage
\begin{flushright}
QMUL-PH-21-40\\
CERN-TH-2021-136 \\
\end{flushright}

\vspace*{0.5cm}

\begin{center}
{\bf \Large Threshold resummation of new partonic channels at
  next-to-leading power}

\vspace*{1cm} \textsc{Melissa van Beekveld\footnote{
melissa.vanbeekveld@physics.ox.ac.uk}$^a$,
  Leonardo Vernazza\footnote{leonardo.vernazza@to.infn.it}$^{b,c}$ 
  and Chris D. White\footnote{christopher.white@qmul.ac.uk}$^d$}  \\

\vspace*{0.5cm} $^a$Rudolf Peierls Centre for Theoretical Physics, 20 Parks Road, Oxford OX1 3PU, UK\\ 

\vspace*{0.5cm} $^b$INFN, Sezione di Torino, Via P. Giuria 1, I-10125 Torino, Italy\\

\vspace*{0.5cm} $^c$Theoretical Physics Department, CERN, CH-1211 Geneva 23, Switzerland\\

\vspace*{0.5cm} $^d$Centre for Research in String Theory, School of
Physics and Astronomy, \\
Queen Mary University of London, 327 Mile End
Road, London E1 4NS, UK\\

\end{center}

\vspace*{0.5cm}

\begin{abstract}
Collider observables involving heavy particles are subject to large logarithmic terms near threshold, which must be summed to all orders in perturbation theory to obtain sensible results. Relatively recently, this resummation has been extended to next-to-leading power in the threshold variable, using a variety of approaches. In this paper, we consider partonic channels that turn on only at next-to-leading power, and show that it is possible to resum leading logarithms using well-established diagrammatic techniques in Quantum Chromodynamics. We first consider deep inelastic scattering, where we reproduce the results of a recent study using an effective theory approach. Next, we consider the quark-gluon channel in both Drell-Yan and Higgs boson production, showing that an explicit all-order form for the leading logarithmic partonic cross section can be obtained. Our results agree with previous conjectures based on fixed-order results.
\end{abstract}

\vspace*{0.5cm}

\section{Introduction}
\label{sec:intro}

The ever-increasing precision of experimental data from the Large Hadron Collider, together with the clear lack of any signatures of new physics, necessitate continual improvements of our understanding of the Standard Model. The dominance of the strong force in collider environments makes it especially important that the theory of Quantum Chromodynamics (QCD) is better understood. For a given observable computed in perturbative QCD, progress can be made either by including subleading fixed orders in the strong coupling constant, or by including certain kinematically enhanced contributions at all orders in perturbation theory, a process known as
{\it resummation}. In this paper, we will be concerned with color-neutral scattering processes containing a heavy (or off-shell) finale-state particle produced near threshold (i.e.~$q\bar{q} \rightarrow \gamma^*/Z$ or $gg\rightarrow h$). In all such processes, one may define a {\it threshold
  variable} $\xi$ which vanishes at the threshold itself. As is
well-known, the differential partonic cross section with respect to this variable has the following generic form:
\begin{equation}
\frac{d \sigma}{d \xi} \, \sim \, 
\sigma_0 \sum_{n = 0}^{\infty} \left( \frac{\alpha_s}{\pi}
\right)^n \left[\sum_{m = 0}^{2 n - 1} \left[ \, c_{n m}^{(-1)} \left(
  \frac{\ln^m \xi}{\xi} \right)_+ + \, c_{nm}^{(0)} \, \ln^m \xi + \ldots \, \right]+ \, c_{n}^{(\delta)} \,
  \delta(\xi)\right]. 
\label{thresholddef}
\end{equation}
Here, an overall normalisation constant $\sigma_0$ originating from the leading-order (LO) cross section is dressed by higher-order contributions, involving a series of terms in the threshold variable $\xi$ that diverge as $\xi\rightarrow 0$, albeit integrably so. The first set of terms in the second sum (with variable $m$) and the $\delta(\xi)$ contribution are associated with leading power (LP) in a systematic expansion in the threshold variable. The logarithmic counting is such that those logarithms with $m=2n-1$ at ${\cal O}(\alpha_s^n)$ are called the leading-logarithmic (LL) terms, with $m=2n-2$ the next-to-leading logarithmic (NLL) terms, and so on.
The LP terms are well known to originate from the emission of soft and collinear radiation. Following the pioneering work of refs.~\cite{Parisi:1980xd,Curci:1979am,Sterman:1987aj,Catani:1989ne,Catani:1990rp,Gatheral:1983cz,Frenkel:1984pz, Sterman:1981jc} based on diagrammatic arguments, several approaches have been developed for LP resummation, including using Wilson lines~\cite{Korchemsky:1993xv,Korchemsky:1993uz}, renormalisation group arguments~\cite{Forte:2002ni}, and Soft
Collinear Effective Theory (SCET)~\cite{Becher:2006nr,Schwartz:2007ib,Bauer:2008dt, Chiu:2009mg}. Recent pedagogical reviews of different approaches may be found in e.g.~refs.~\cite{Luisoni:2015xha,Becher:2014oda,Campbell:2017hsr}.\\

Until relatively recently, much less has been known about the second set of logarithmic terms in eq.~(\ref{thresholddef}), which constitute
next-to-leading power (NLP) in the threshold variable $\xi$. Their precise origin remains unknown to this date, but it has already been shown that their numerical contribution cannot be neglected~\cite{Kramer:1996iq, Ball:2013bra, Bonvini:2014qga, Anastasiou:2015ema, Anastasiou:2016cez, vanBeekveld:2019cks,  vanBeekveld:2021hhv, Ajjath:2021lvg}, therefore constituting a necessity to understand and resum them. Following previous work in Quantum Electrodynamics (QEC)~\cite{Low:1958sn,Burnett:1967km,DelDuca:1990gz,Grunberg:2009yi}, refs.~\cite{Laenen:2008gt,Laenen:2010uz} used a mixture of diagrammatic and path-integral methods to argue that certain NLP terms should indeed be resummable. References~\cite{Soar:2009yh, Moch:2009hr,Moch:2009mu,deFlorian:2014vta,Presti:2014lqa} reached a similar (and indeed more general) conclusion, using results in fixed-order perturbation theory to conjecture some all-order forms for NLP terms in processes including deep-inelastic scattering (DIS), Drell-Yan (DY) production of an off-shell vector boson, and Higgs boson production. Following more formal work showing that next-to-soft physics can be related to asymptotic properties of scattering amplitudes at null infinity in both gauge theories and gravity~\cite{Cachazo:2014fwa,Casali:2014xpa} (itself
related~\cite{White:2014qia} to the earlier work of
refs.~\cite{Gross:1968in,White:2011yy}), there has been more
widespread interest in exploring the properties of NLP terms, which could not be more timely given the numerical motivation mentioned above. Examples using direct QCD arguments include developing factorisation theorems for NLP contributions that extend their LP counterparts~\cite{Bonocore:2015esa,Bonocore:2016awd,Bonocore:2020xuj,Gervais:2017yxv, Gervais:2017zky,Gervais:2017zdb,Laenen:2020nrt}; carrying out
fixed-order studies that aim to motivate such formulae~\cite{DelDuca:2017twk,vanBeekveld:2019prq,Bonocore:2014wua,Bahjat-Abbas:2018hpv,Ebert:2018lzn,Boughezal:2018mvf,Boughezal:2019ggi}; resumming NLP contributions by combining factorisation and renormalisation group arguments~\cite{Ajjath:2020ulr,Ajjath:2020sjk,Ajjath:2020lwb,Ahmed:2020caw,Ahmed:2020nci,Ajjath:2021lvg}; and resumming specific contributions~\cite{Bahjat-Abbas:2019fqa}. There is also an ever-growing body of work examining NLP effects in SCET, including identifying relevant operators contributing at NLP order and/or their mixing under renormalisation~\cite{Kolodrubetz:2016uim,Moult:2016fqy,Feige:2017zci,Beneke:2017ztn,Beneke:2018rbh,Bhattacharya:2018vph,Beneke:2019kgv,Bodwin:2021epw}; development of factorisation formulae~\cite{Moult:2019mog,Beneke:2019oqx,Liu:2019oav,Liu:2020tzd}; and explicit studies for particular observables, either at fixed-order or resummed~\cite{Boughezal:2016zws,Moult:2017rpl,Chang:2017atu,Moult:2018jjd,Beneke:2018gvs,Ebert:2018gsn,Beneke:2019mua,Moult:2019uhz,Liu:2020ydl,Liu:2020eqe,Wang:2019mym,Beneke:2020ibj}. This has proceeded in tandem with direct QCD approaches, with a highly useful exchange of ideas and results between what are often cast as opposing formalisms. Indeed, it is always useful to have complementary viewpoints on the same underlying physics, such that a more varied toolbox can be employed in extending the frontier of QCD perturbation theory yet further.\\

With this spirit in mind, we turn our attention in this paper to a particular class of NLP contributions present DIS and the production of a colourless off-shell or heavy boson, such as DY and Higgs production. Much of the above-mentioned work has focused on the study of NLP corrections to partonic cross sections which have the same initial state as the Born contribution. Indeed, these partonic channels are the only ones relevant at LP, and it has been explicitly demonstrated that one can resum their LL NLP contributions in both DY and Higgs production, using SCET~\cite{Beneke:2018gvs,Beneke:2019mua} or direct QCD~\cite{Bahjat-Abbas:2019fqa}, and in agreement with previous conjectures~\cite{Moch:2009hr}. However, it is also the case that new partonic channels can open up at next-to-leading order (NLO) and beyond that contribute LL NLP logarithms, and must be counted alongside their counterparts in the kinematically leading partonic channel. Na\"{i}vely, one expects that one should indeed be able to resum such contributions: once a subleading partonic channel has been turned on, the cross section is already at next-to-leading power, and thus any further emissions must be maximally soft and collinear. The known resummation properties of the latter should then guarantee NLP resummation for these terms, an expectation that turns out to be ultimately correct. Nevertheless, turning this observation into a practical resummation formula is not as easy as it might seem.  A perusal of the QCD literature shows that the NLP coefficients $c_{nm}^{(0)}$ appearing in eq.~(\ref{thresholddef}) for various processes of
interest do not have an obvious exponential form, even for the highest power of the logarithm at each order of $\alpha_s$. This is in stark contrast to the LL terms at LP, and also the LL NLP terms in the leading partonic channel.\\

For DIS, important progress was made in ref.~\cite{Vogt:2010cv}, which considered the kinematically subleading gluon initial state, and presented an all-order conjecture for the LL NLP terms in the off-diagonal DGLAP splitting function $P_{qg}(x)$, as well as the coefficients appearing in the partonic structure function. Similar conjectures could be made for the related splitting function $P_{gq}$, using a Higgs-induced DIS process, where an incoming gluon fuses with a Higgs boson at Born level. Recently, ref.~\cite{Beneke:2020ibj} confirmed these results within the framework of SCET, using the assumption that the one-loop virtual corrections to the new partonic channel at NLO exponentiate, and showing how such exponentiation can be obtained through re-factorisation (see also \cite{Liu:2019oav,Liu:2020tzd}). Reference \cite{Presti:2014lqa}
provided further conjectures for the LL NLP terms arising from subleading partonic channels in DY and Higgs production, which to date remain unproven.\\

In this paper, we will demonstrate the validity of the conjectures
made in refs.~\cite{Vogt:2010cv,Presti:2014lqa} using well-established direct QCD arguments (see
e.g.~refs.~\cite{Gribov:1972ri,Gribov:1972rt,Dokshitzer:1977sg}, and refs.~\cite{Dokshitzer:1978hw,Dokshitzer:1991wu} for pedagogical reviews). We will calculate all-order forms for subleading partonic cross sections in the LL approximation at LP and NLP, using diagrammatic arguments that enable us to straightforwardly obtain the form of the fully real radiative contribution at each order in the coupling, before fixing the form of the virtual corrections using known constraints. We will then be able to obtain closed forms for the various splitting and coefficient functions presented in refs.~\cite{Vogt:2010cv,Presti:2014lqa}, finding full agreement. In DIS, our results overlap with the SCET approach of ref.~\cite{Beneke:2020ibj} (which connects to the work done in
ref.~\cite{Moult:2019uhz} on the NLP thrust distribution), but our approach is entirely complementary: where that paper examines purely virtual corrections to subleading partonic channels at each order, we consider the opposite extreme of the fully real contribution. Further, we derive the all-order structure of the $qg$ partonic cross sections by direct computation, rather than via the renormalization-group 
evolution arguments as discussed in section 4 of ref.~\cite{Beneke:2020ibj}. Aside from this, our results go further than ref.~\cite{Beneke:2020ibj}, which did not discuss DY or Higgs production.\\

The structure of our paper is as follows. In section~\ref{sec:DIS}, we derive the all-order LL NLP terms in the off-diagonal splitting functions in (Higgs-induced) DIS, together with the appropriate coefficient functions. In sections~\ref{sec:DY} and~\ref{sec:Higgs}, we extend our arguments to both DY and Higgs boson production, showing how one can obtain all-order LL forms for the coefficient functions for quark-gluon initial states. We discuss our results and conclude in section~\ref{sec:discuss}. Certain technical details are contained in appendices~\ref{app:integrals} and \ref{app:transition}.

\section{Splitting functions and coefficient functions in DIS}
\label{sec:DIS}

\begin{figure}
\centering
\mbox{\captionsetup[subfigure]{oneside,margin={0.7cm,0cm},skip=0.5cm}
\subfloat[]{
	\includegraphics[width=0.35\textwidth]{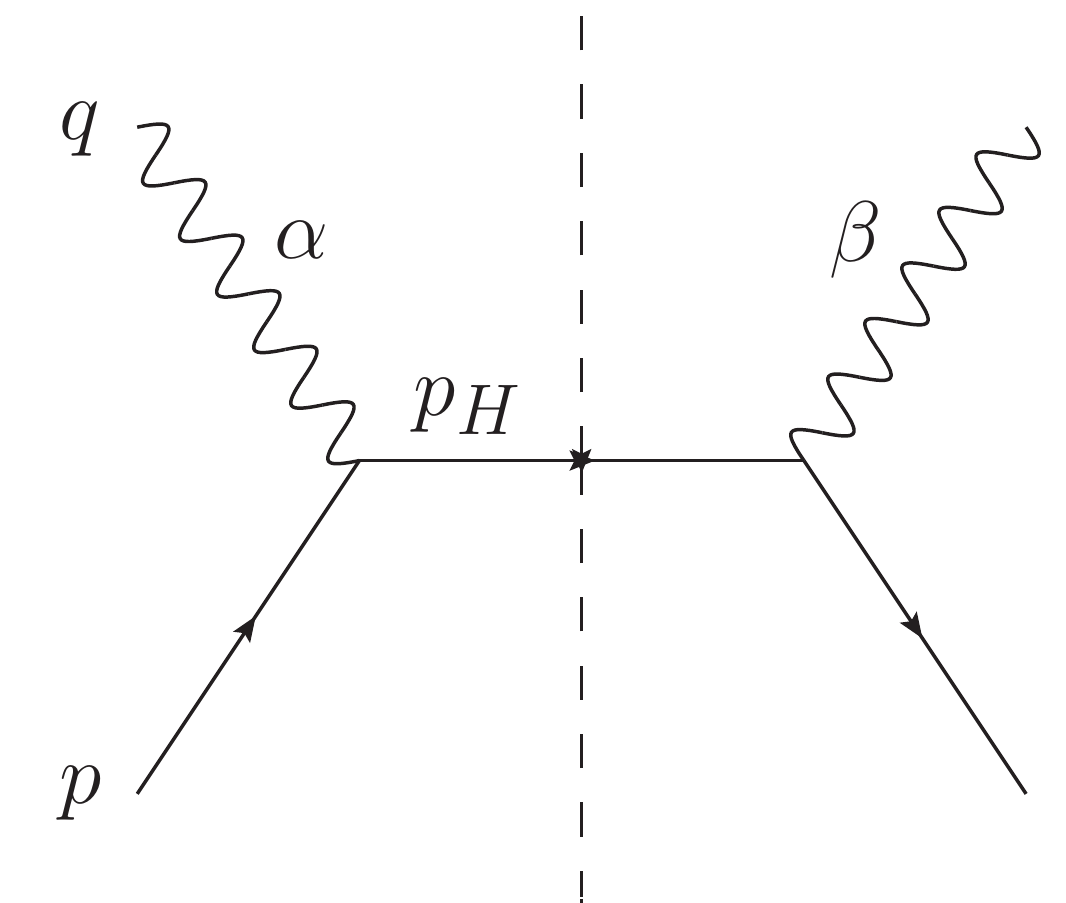}
	\label{fig:DISLOa}
}
\hspace{3cm}
\captionsetup[subfigure]{oneside,margin={0.7cm,0cm},skip=0.5cm}
\subfloat[]{
	\includegraphics[width=0.35\textwidth]{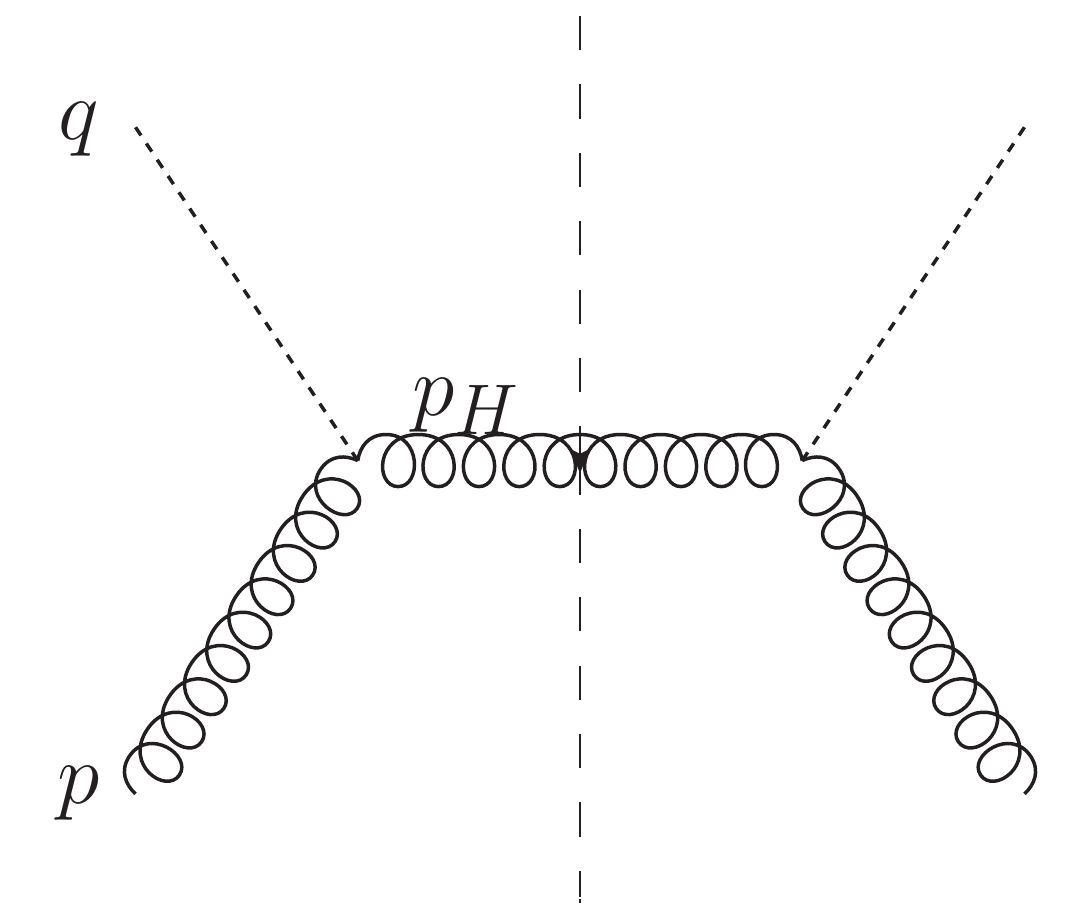}
	\label{fig:DISLOb}
}}
\caption{LO squared amplitudes for: (a) DIS; (b) Higgs-induced DIS, where the Higgs boson couples to the gluon via an effective coupling. }
\label{fig:DISLO}
\end{figure}

In this section, we consider the deep-inelastic scattering of a
virtual photon with a proton. At LO, the photon couples to a valence quark, leading to the process
\begin{equation}
q(p)+\gamma^*(q)\rightarrow q(p_H)\,,
\label{DISproc}
\end{equation}
whose squared matrix element is depicted in figure~\ref{fig:DISLOa}.
We will also have reason to consider the alternative process of {\it Higgs-induced DIS}, in which the virtual photon is replaced by a Higgs boson, and the valence quark by a gluon:
\begin{equation}
g(p)+h(q)\rightarrow g(p_H)\,.
\label{ghproc}
\end{equation}
It is assumed that the Higgs couples to gluons via a suitable
effective coupling (e.g.~a top-quark loop with the top mass taken to infinity), but the precise details need not concern us here. The squared LO Feynman diagram for eq.~(\ref{ghproc}) is shown in figure~\ref{fig:DISLOb}. \\

In both of the above processes, we can define the conventional Bj\"orken variable
\begin{equation}
x=\frac{Q^2}{2p\cdot q},\quad Q^2=-q^2.
\label{xdef}
\end{equation}
In the DIS case, we can then define the structure function
\begin{equation}
F_2(x,Q^2)=\int {\rm d} \Phi\,
T_2^{\alpha\beta} \,\overline{|{\cal M}_{{\rm DIS}}|^2}_{\alpha\beta}\,,
\label{F2def}
\end{equation}
where the integral denotes a sum over all possible radiative final states, $\overline{|{\cal M}_{{\rm DIS}}|^2}_{\alpha\beta}$ is the squared amplitude averaged (summed) over initial (final) state partonic colours and spins, and we have labelled photon Lorentz indices as in figure~\ref{fig:DISLOa}. Here the open Lorentz indices
$\alpha$ and $\beta$ belong to the initial-state photon on either side of the final-state cut, and we have introduced the projector
\begin{equation}
T_2^{\alpha\beta}=-\frac{1}{4\pi}\frac{1}{1-\epsilon}\left(\eta^{\alpha\beta}
+(3-2\epsilon)\frac{q^2}{(p\cdot q)^2}p^\alpha p^\beta\right),
\label{T2def}
\end{equation}
where we work in $d=4-2\epsilon$ spacetime dimensions. For the Higgs process of figure~\ref{fig:DISLOb}, we may similarly define the structure function
\begin{equation}
  F_{\phi}(x,Q^2)=\int {\rm d}\Phi \, \overline{|{\cal M}_{\rm DIS}|^2}\,,
  \label{Fphig}
\end{equation}
where $\overline{|{\cal M}_{\rm DIS}|^2}$ is the summed/squared amplitude in this case, and no projection is needed to get a scalar quantity, due to the scalar nature of the virtual Higgs boson. \\

\begin{figure}
	\centering
  \mbox{	\captionsetup[subfigure]{oneside,margin={0.9cm,0cm},skip=0.5cm}
  	\subfloat[ ]{
  		\includegraphics[width=0.35\textwidth]{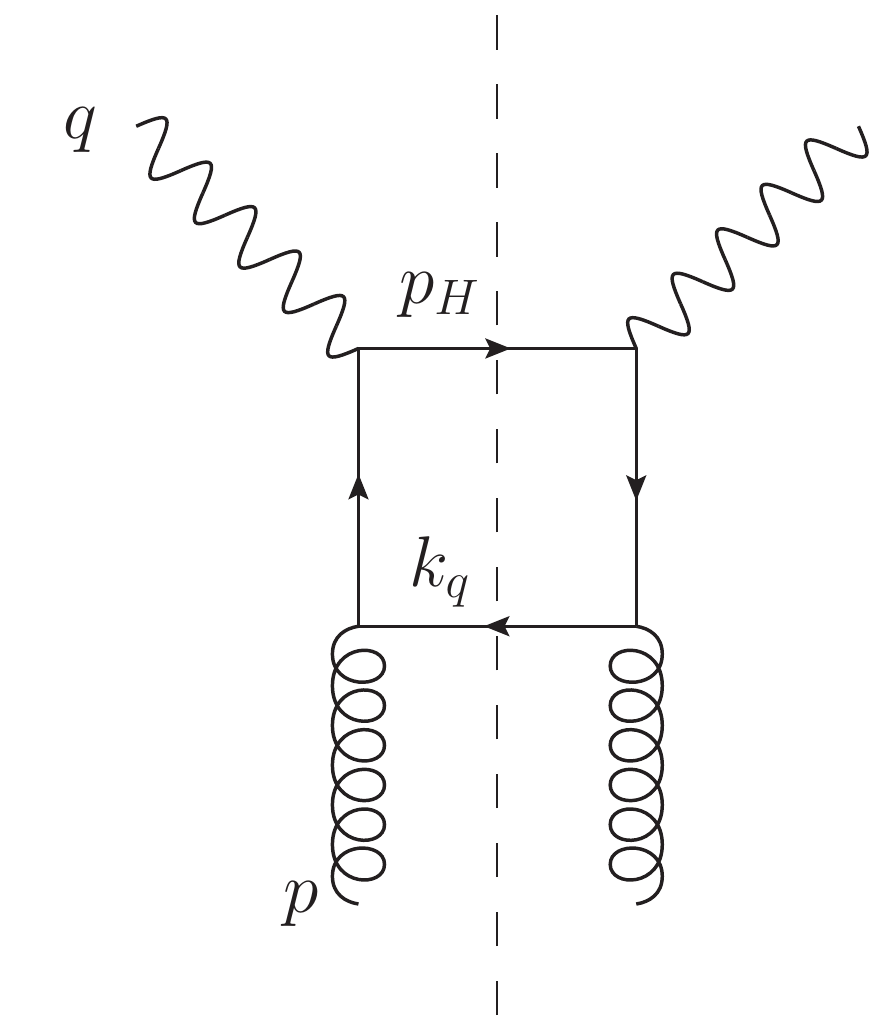}
  		\label{fig:DISNLOa}
  	}
  \hspace{3cm}
  \captionsetup[subfigure]{oneside,margin={1.0cm,0cm},skip=0.5cm}
  \subfloat[]{
  	\includegraphics[width=0.35\textwidth]{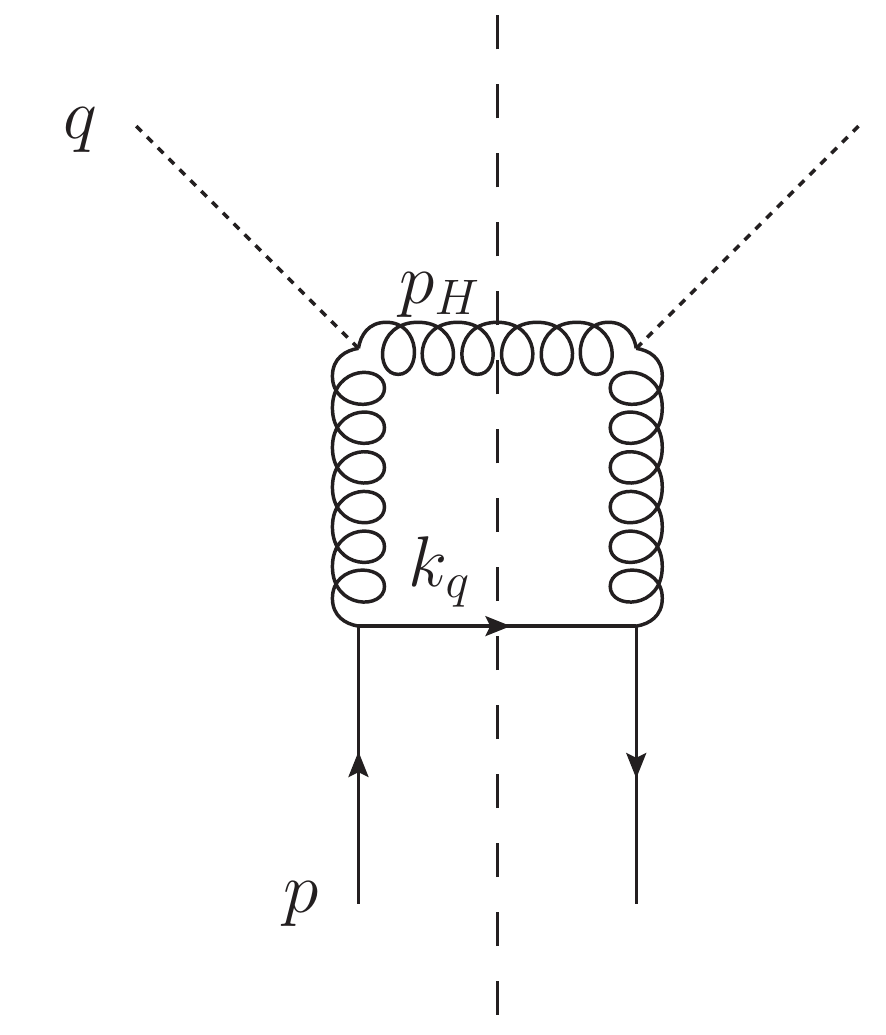}
  	\label{fig:DISNLOb}
  }}
    \caption{(a) NLO correction to DIS, where a gluon couples
      indirectly to the photon; (b) the analogous channel in
      Higgs-induced DIS.}
    \label{fig:DISNLO}
\end{figure}

Dressing the LO processes of figure~\ref{fig:DISLO} with additional radiation will open up different partonic initial states at NLO and beyond. For normal DIS, there is the process of figure~\ref{fig:DISNLOa}, in which a gluon rather than a quark is present in the initial state. There is also a crossed box diagram for this channel, where the connection of the quark legs is interchanged in the complex conjugate amplitude (not shown in figure~\ref{fig:DISNLO}). We will argue that, due to various choices to be described later, this diagram will not contribute at LL NLP. 
The equivalent process for Higgs-induced DIS is that of
figure~\ref{fig:DISNLOb} in which the initial state contains a quark. Once the new partonic channels have opened up at NLO, one may emit further radiation, which becomes complicated very quickly as the perturbative order increases due to the multitude of possible Feynman diagrams. However, we are after the LL behaviour at threshold, which necessarily corresponds to all additional radiation being maximally soft and/or collinear. We will see that it is then possible to arrange things so that only a small set of Feynman diagrams contribute. \\

A first restriction that allows us to eliminate contributing diagrams stems from the fact that the processes of figure~\ref{fig:DISNLO} are already suppressed by a power of the threshold variable. They involve an emission of a quark with the soft momentum $k_q$ which, after summing over spins, leads to a factor
\begin{displaymath}
\sum_{\rm spins} u(k_q)\bar{u}(k_q)=\slsh{k_q}
\end{displaymath}
in the squared matrix element. For a soft gluon emission with momentum $k$, the sum over polarisation states is ${\cal O}(k^0)$, which indeed has one less power of soft momentum than the quark case shown above. The emission of the fermion puts the diagram at NLP, and after this first emission one only needs to consider further (maximally soft and collinear) gluon radiation (i.e.~the same radiation as one would need to consider at LP for diagonal channels). Let us write the momentum of each additional parton (including that present at NLO) using a Sudakov decomposition
\begin{equation}
  k_\mu=\alpha p_\mu+\beta q'_\mu+k_{\perp\,\mu},\quad k_\perp\cdot p=
  k_\perp\cdot q'=0\,,
  \label{Sudakov}
\end{equation}
where we have introduced the vector
\begin{equation}
  q'=q+x p\,,
  \label{q'def}
\end{equation}
which is null from eq.~(\ref{xdef}). For future use, we also note the
relations
\begin{equation}
  p+q=(1-x)p+q',\quad p\cdot q'=p\cdot q\neq 0\,.
  \label{pqrels}
\end{equation}
In eq.~(\ref{Sudakov}), the $d$-dimensional vector
\begin{equation}
  k_\perp=(0,\vec{k}_\perp,0)\,,
  \label{kperpdef}
\end{equation}
containing the $(d-2)$-vector $\vec{k}_\perp$, constitutes the momentum transverse to the incoming beams. \\

\begin{figure}
  \begin{center}
  	
  	\centering
  	\mbox{	\captionsetup[subfigure]{oneside,margin={0.3cm,0cm},skip=0.5cm}
  		\subfloat[]{
  			\includegraphics[width=0.3\textwidth]{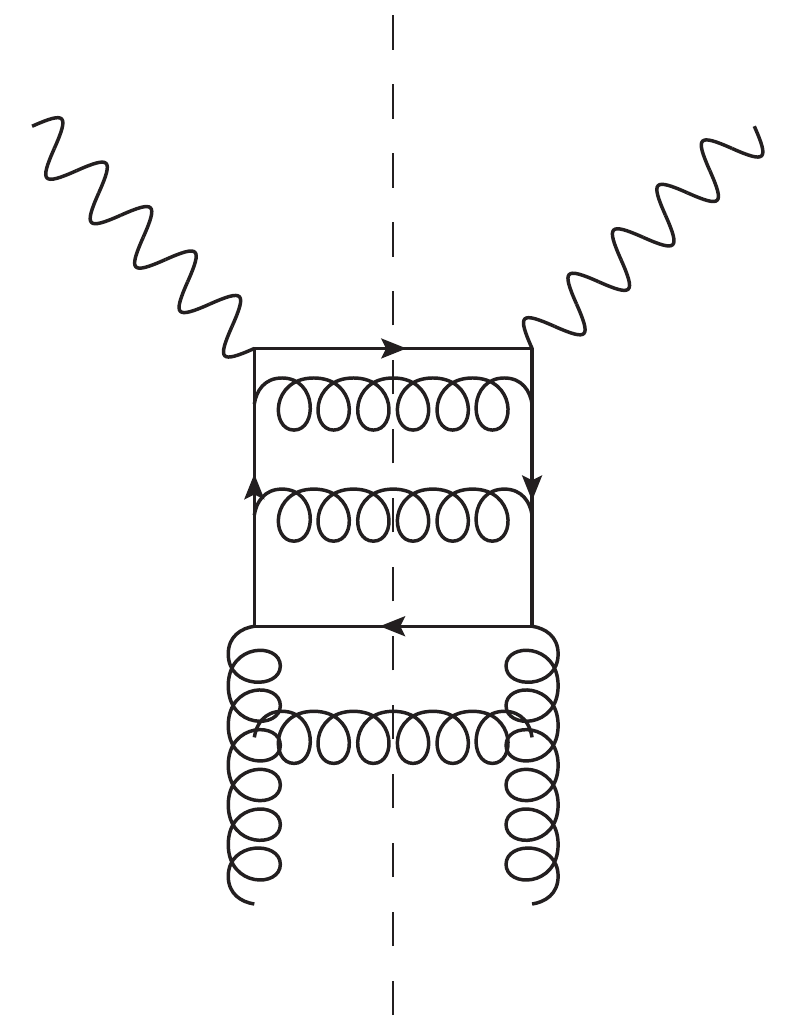}%
  			\label{fig:laddera}}
  	
  		\hspace{2.5cm}
  		\captionsetup[subfigure]{oneside,margin={0.2cm,0cm},skip=0.5cm}
  		\subfloat[]{
  			\includegraphics[width=0.3\textwidth]{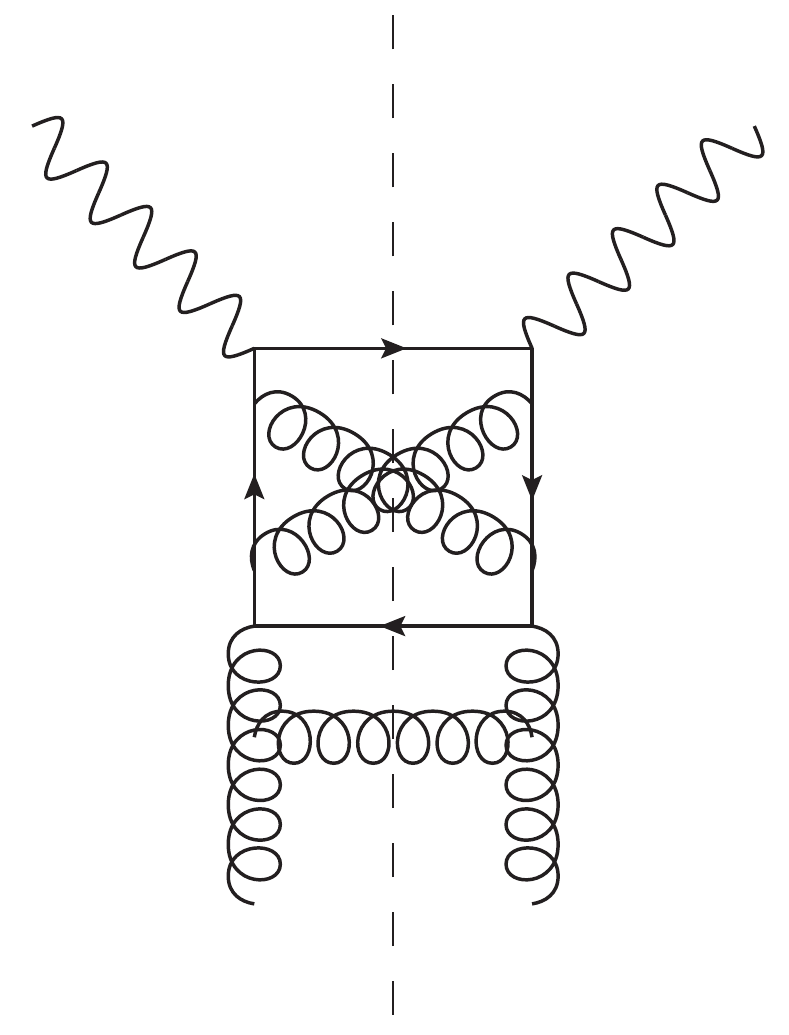}	\label{fig:ladderb}
  		}
  	}
    \caption{(a) A ladder graph contributing to the gluon channel in DIS at NLP LL; (b) a crossed-ladder graph.}
    \label{fig:ladder}
  \end{center}
\end{figure}

A second restriction on the number of Feynman diagrams that contribute arises as follows. As argued in detail in refs.~\cite{Gribov:1972ri,Gribov:1972rt,Dokshitzer:1977sg}, leading logarithms only arise from the kinematic region in which the transverse momenta of the emitted partons are strongly ordered. Furthermore, one may reduce the set of relevant Feynman diagrams for the squared matrix element to those having a pure ladder form, as shown in figure~\ref{fig:laddera}. Crossed ladders, such as the graph in figure~\ref{fig:ladderb}, do not contribute at LL. In non-abelian theories such as QCD, this property is not
guaranteed in general gauges, but can be made manifest by choosing to define the polarisation states of the emitted gluons in a particular way. Upon choosing a reference vector $c_\mu$, one may define physical gluon polarisation vectors $\epsilon_\mu(k)$ via the simultaneous requirements
\begin{equation}
  k\cdot \epsilon(k)=c\cdot\epsilon(k)=0\,.
  \label{physpols}
\end{equation}
If in addition $c$ is a null vector ($c^2=0$), the sum over physical gluon polarisation states has the form
\begin{equation}
  \sum_{{\rm pols.}}\epsilon_\mu^\dag (k)\epsilon_\nu(k)=
  -\eta_{\mu\nu}+\frac{k_\mu c_\nu + k_\nu c_\mu}{c\cdot k}\,.
  \label{polsum}
\end{equation}
As explained in detail in refs.~\cite{Dokshitzer:1978hw,Dokshitzer:1991wu}, the kinematic dominance of uncrossed gluon ladders occurs for the explicit choice
\begin{equation}
  c=q'\,,
  \label{c=q'}
\end{equation}
i.e.~the same vector that occurs in the Sudakov decomposition of eq.~(\ref{Sudakov}). Although we will explicitly check whether the same holds at NLP, the choice of eq.~\eqref{c=q'} will allow us to straightforwardly obtain the purely real corrections to the processes of figure~\ref{fig:DISNLO}, at arbitrary order in the coupling, in the LL approximation. \\

Summarising, we have established that to compute the NLP LL contribution of normal or Higgs-induced DIS, we need to consider the NLO diagrams of figure~\ref{fig:DISNLO} dressed with $n$ ordered soft-gluon emissions, after choosing a particularly beneficial form of the reference vector used in the sum over physical polarisations of the $n$ gluons. If $n$ additional gluons are emitted, we will have to integrate over an $(n+2)$-body phase space, whose treatment will now be discussed. 

\subsection{Phase space for multiparton final states}
\label{sec:phasespace}

Focusing on the case of conventional DIS process for concreteness, we will need to consider integrating the squared matrix element for the following process:
\begin{equation}
		g(p)+\gamma^*(q)\rightarrow q(p_H)+q(k_q)+\sum_{i=1}^n g(k_i)\,,
		\label{DISprocn}
\end{equation}
where we label the soft-quark momentum present in the NLO matrix element by $k_q$, and any additional gluon momenta by $\{k_i\}$. The phase space for this process is
\begin{align}
  \int {\rm d}\Phi^{(n+2)}&=(2\pi)^d\int \frac{{\rm d}^d p_H}{(2\pi)^{d-1}}
  \, \delta_+(p_H^2)\int
  \frac{{\rm d}^d k_q}{(2\pi)^{d-1}}\, \delta_+(k_q^2)\left[\prod_{i=1}^n\int\frac{{\rm d}^d k_i}
    {(2\pi)^{d-1}}\delta_+(k_i^2)\right]\notag\\
  &\hspace{7cm}\times
  \delta^{(d)}\left(p+q-p_H-k_q
  -\sum_{i=1}^n k_i\right)\notag\\
  &=(2\pi)^{(n+2)(1-d)+d}\int {\rm d}^d k_q\,  \delta_+(k_q^2)\left[
    \prod_{i=1}^n \int{\rm d}^d k_i\delta_+(k_i^2)\right]
  \delta_+\left[\left(p+q-k_q-\sum_{i=1}^n k_i\right)^2\right],
  \label{phasespace1}
\end{align}
where we use the conventional notation
\begin{equation}
  \delta_+(k^2)\equiv \theta(k^0)\delta(k^2)\,,
\label{delta+}
\end{equation}
and we have used the $d$-dimension delta function to carry out the integral over $p_H$ in the second line of
eq.~(\ref{phasespace1}). Each emitted gluon momentum may be expanded using the Sudakov decomposition of eq.~(\ref{Sudakov}), giving
\begin{equation}
k_i=\bar{\alpha}_i p+\bar{\beta}_i q'+k_{i\perp}.
	\label{suds}
\end{equation}
One may find the variables $\bar{\alpha}_i$ and $\bar{\beta}_i$ by contracting on both sides with $p$ and $q'$:
\begin{equation}
  \bar{\alpha}_i=\frac{q'\cdot k}{p\cdot q},\quad \bar{\beta}_i=\frac{p\cdot k}{p\cdot q}\,,
  \label{alphabetadef}
\end{equation}
where we use that $p\cdot q' = p\cdot q$.  Furthermore, one may rewrite the measure in an integral over $k$ as follows:
\begin{equation}
  {\rm d}^d k_i=p\cdot q\, {\rm d}\bar{\alpha}_i\,{\rm d}\bar{\beta}_i\,d^{d-2}k_{i,\perp}
  =\frac{p\cdot q}{2}{\rm d}\bar{\alpha}_i\,{\rm d}\bar{\beta}_i\,{\rm d} \vec{k}_{i,\perp}^2
  (\vec{k}_{i,\perp}^2)^{\frac{d-4}{2}}{\rm d}\Omega_{d-2}^{(i)}\,,
\label{intk}
\end{equation}
where we have used the notation of eq.~(\ref{kperpdef}), and also introduced the differential solid angle in the transverse directions of the gluon with momentum $k_i$, ${\rm d}\Omega^{(i)}_{d-2}$. To derive eq.~(\ref{intk}), one may first pick the following parametrisation for $p$ and $q$ (see e.g. ref.~\cite{vanBeekveld:2019prq}):
\begin{align}
  p&=\frac{s+Q^2}{2\sqrt{s}}(1,0,\ldots,0,1),\quad
  q=\left(\frac{s-Q^2}{2\sqrt{s}},0,\ldots,0,-\frac{(s+Q^2)}{2\sqrt{s}}
  \right),
\label{pqparam}
\end{align}
where we have introduced the partonic squared centre-of-mass energy
\begin{equation}
  s=(p+q)^2= 2p\cdot q (1-x)\,.
\label{sdef}
\end{equation}
Then eq.~(\ref{pqparam}) together with eq.~(\ref{xdef}) implies
\begin{equation}
	p=\frac{Q^2}{2\sqrt{s}}\frac{1}{x}
	\left(1,0,\ldots,0,1\right),\quad
	q'=\frac{Q^2}{2\sqrt{s}}\frac{1-x}{x}\left(1,0,\ldots,0,-1
	\right),
	\label{pq'param}
\end{equation}
so that eqs.~(\ref{Sudakov}, \ref{alphabetadef}) yield the
explicit variable transformation
\begin{equation}
  \bar{\alpha}_i=\frac{(1-x)}{\sqrt{s}}(k_i^0+k_i^z),\quad
  \bar{\beta}_i=\frac{1}{\sqrt{s}}(k_i^0-k_i^z)\,,
\label{alphatrans}
\end{equation}
from which eq.~(\ref{intk}) follows as required. \\

Using eqs.~(\ref{pqrels}, \ref{alphabetadef}) and writing a similar decomposition for quark momentum $k_q$ with variables $\alpha_q$ ($\beta_q$) along the direction of $p$ ($q'$), the argument of the delta function in eq.~(\ref{phasespace1}) can be written as
\begin{equation}
  \left((1-x)p+q'-k_q-\sum_{i=1}^n k_i\right)^2
  =2p\cdot q\left[(1-x)\left(1-\beta_q-\sum_{i=1}^n \bar{\beta}_i\right)
    -\alpha_q-\sum_{i=1}^n\bar{\alpha}_i\right]+\ldots,
  \label{deltaarg}
\end{equation}
where we have neglected terms which are quadratic in $k_q$ and/or $\{k_i\}$, which correspond to correlations between different emitted partons. As discussed above, the emitted parton momenta are required to be soft in the LL limit. Therefore, these quadratic terms will be power-suppressed with respect to the terms included in eq.~(\ref{deltaarg}). Given that our matrix-element will already be at NLP, we do not need to keep the phase-space correlations~\footnote{Note that such terms were considered for kinematically leading channels in DY and Higgs production in ref.~\cite{Bahjat-Abbas:2019fqa}, where they could be neglected at NLP LL, but not at NLP NLL order. Here, however, the phase space correlations will only contribute at NNLP, due to there being no LP contribution to the squared matrix element.}. The phase space of eq.~(\ref{phasespace1}) then becomes
\begin{align}
  \int {\rm d}\Phi^{(n+2)}&=(2\pi)^{-(n+1)d}\,\pi^{n+2}\,(p\cdot q)^n\,
  \int {\rm d}\alpha_q {\rm d}\beta_q {\rm d}\vec{p}_{q,\perp}^2 {\rm d}\Omega_{d-2}^{(1)}
  (\vec{p}_{q,\perp}^2)^{\frac{d-4}{2}}\, \delta_+\left(2p\cdot q\,\alpha_q\,
  \beta_q-\vec{p}_{q,\perp}^2\right)\notag\\
  &\times \left[\prod_{i=1}^n \int{\rm d}\bar{\alpha}_i {\rm d}\bar{\beta}_i
    {\rm d}\vec{k}_{i,\perp}^2 (\vec{k}_{i,\perp}^2)^{\frac{d-4}{2}}
    {\rm d}\Omega_{d-2}^{(i)}\,\delta_+\left(2p\cdot q\,\bar{\alpha_i}\,\bar{\beta}_i
    -\vec{k}_{i,\perp}^2\right)\right]\notag\\
  &\times \delta\left[(1-x)\left(1-\beta_q-\sum_{i=1}^n\bar{\beta}_i\right)
    -\alpha_q-\sum_{i=1}^n\bar{\alpha}_i\right],
  \label{phasespace2}
\end{align}
where we have used eq.~(\ref{intk}) in squaring the emitted parton momenta inside the on-shell delta functions for $k_q$ and $\{k_i\}$. We may use these delta functions to carry out the integrals over $\vec{p}_{q,\perp}^2$ and $\{\vec{k}_{i,\perp}^2\}$, and can simplify the result further by assuming that the matrix elements we are going to integrate will not depend on any of the transverse solid angles, which will indeed turn out to be the case. We get
\begin{align}
\int {\rm d} \Phi^{(n+2)}&=2^{-\frac12(n+1)(d+4)}\, \pi^{n+2-(n+1)d}
\, (p\cdot q)^{n+(n+1)\frac{(d-4)}{2}}\, \Omega_{d-2}^{n+1}\notag\\
&\hspace{3cm}\times \,\int {\rm d} \alpha_q\,{\rm d} \beta_q(\alpha_q\,\beta_q)^{\frac{d-4}{2}}
\left[\prod_{i=1}^n \int {\rm d} \bar{\alpha}_i\,{\rm d} \bar{\beta}_i
  (\bar{\alpha}_i\,\bar{\beta}_i)^{\frac{d-4}{2}}\right]\notag\\
&\hspace{4cm} \times \, \delta\left[(1-x)\left(1-\beta_q-\sum_{i=1}^n\bar{\beta}_i\right)
    -\alpha_q-\sum_{i=1}^n\bar{\alpha}_i\right].
\label{phasespace3}
\end{align}
Next, we may use the standard result
\begin{equation}
  \Omega_{d-2}=\frac{2\pi^{\frac{d-2}{2}}}{\Gamma\left(\frac{d-2}{2}\right)}\,,
  \label{omegad-2}
\end{equation}
as well as rescaling
\begin{equation}
  \beta_q\rightarrow \frac{\beta_q}{(1-x)},\quad
  \bar{\beta}_i\rightarrow \frac{\bar{\beta}_i}{(1-x)},
  \label{betarescale}
\end{equation}
to obtain
\begin{align}
  \int {\rm d}\Phi^{(n+2)}&=\frac{2\pi}{(4\pi)^{\frac{(n+1)d}{2}}}
 \, (Q^2)^{n+(n+1)\frac{(d-4)}{2}} \, 
\frac{x^{-(n+1)\frac{(d-4)}{2}-n}\, 
	(1-x)^{-(n+1)\frac{d-2}{2}}}{\Gamma\left(\frac{d-2}{2}\right)^{n+1}}\notag\\
  &\times \int {\rm d}\alpha_q\,{\rm d}\beta_q\, (\alpha_q\,\beta_q)^{\frac{d-4}{2}}
\left[\prod_{i=1}^n \int {\rm d}\bar{\alpha}_i\,{\rm d}\bar{\beta}_i
  (\bar{\alpha}_i\,\bar{\beta}_i)^{\frac{d-4}{2}}\right]\, \delta\left[1-x-\alpha_q-\beta_q
    -\sum_{i=1}^n(\bar{\alpha}_i+\bar{\beta}_i)\right].
\label{phasespace4}
\end{align}
Notice that the final delta function links all of the Sudakov
variables together. We can decouple this dependence by using the
identity
\begin{equation}
  \delta(u)=\int_{-i\infty}^{i\infty}\frac{{\rm d}T}{2\pi i} \, e^{Tu},
\label{deltaid}
\end{equation}
such that eq.~(\ref{phasespace4}) becomes
\begin{align}
  \int {\rm d}\Phi^{(n+2)}&=\frac{2\pi}{(4\pi)^{\frac{(n+1)d}{2}}}
 \, (Q^2)^{n+(n+1)\frac{(d-4)}{2}}
\frac{x^{-(n+1)\frac{(d-4)}{2}-n}
	(1-x)^{-(n+1)\frac{d-2}{2}}}{\Gamma\left(\frac{d-2}{2}\right)^{n+1}}\, \int_{-i\infty}^{i\infty}\frac{{\rm d}T}{2\pi i}\, e^{T(1-x)}\notag\\
  &
 \times   \int {\rm d}\alpha_q\,{\rm d}\beta_q\, (\alpha_q\,\beta_q)^{\frac{d-4}{2}}\, 
  e^{-T(\alpha_q+\beta_q)}
\left[\prod_{i=1}^n \int {\rm d}\bar{\alpha}_i\,{\rm d}\bar{\beta}_i
  \,(\bar{\alpha}_i\,\bar{\beta}_i)^{\frac{d-4}{2}}
  \, e^{-T(\bar{\alpha}_i+\bar{\beta}_i)}\right].
\label{phasespace5}
\end{align}
Provided all of the $\alpha$ and $\beta$ integrals can be carried out, the final integral over $T$ has the form of an inverse Laplace transform. \\

In this section, we have derived a convenient form for the $(n+2)$-body phase space in the LL approximation. Although we considered the process of eq.~(\ref{DISprocn}), we can also apply this result to Higgs-induced DIS, given that the phase space is insensitive to the identity of the emitted partons in the final state, and that the definition of the Bj\"orken $x$ variable is the same. Before moving on, we note that is more convenient at NLO (the case $n=0$ above) to use an alternative form of eq.~(\ref{phasespace5}), in which the dependence on $(1-x)$ has not been scaled out of $\beta_1$, and the delta function is left intact. From eqs.~(\ref{phasespace3}, \ref{omegad-2}), one finds
\begin{equation}
  {\rm d}\Phi^{(2)}=\frac{2\pi}{(4\pi)^{\frac{d}{2}}}\frac{(Q^2)^{\frac{(d-4)}{2}}}
  {\Gamma(\frac{d-2}{2})}x^{-\frac{d-4}{2}}\,{\rm d}\alpha_q\, {\rm d}\beta_q
 \,  (\alpha_q \beta_q)^{\frac{d-4}{2}}\, \delta[(1-x)(1-\beta_q)-\alpha_q].
  \label{dPhi2}
\end{equation}
Let us now proceed to calculate all-order forms for the kinematically subleading partonic structure functions in (Higgs-induced) DIS, where we will first examine the relevant structure functions at NLO. 

\subsection{The quark structure function in Higgs-induced DIS}
\label{sec:HiggsDISNLO}

We start our discussion by computing the NLO structure of the Higgs-induced DIS process. One reason to consider this process is that it allows for a straightforward calculation of the off-diagonal splitting function $P_{gq}$, as argued in refs.~\cite{Vogt:2010cv,Beneke:2020ibj}. As shown in
figure~\ref{fig:DISNLOb}, the quark initial state turns on at NLO. To normalise our results, we first need the result for the LO process of figure~\ref{fig:DISLOb}, whose squared matrix element (summed/averaged over colours and spins) is
\begin{equation}
	\overline{|{\cal M}_{gh\rightarrow g}|^2}
	=\frac{|\lambda|^2}{d-2}\left(-\eta_{\alpha\beta}+\frac{q'_\alpha p_\beta
		+q'_\beta p_\alpha}{q'\cdot p}\right)
	\left(-\eta^{\alpha\beta}+\frac{{q'}^\alpha p_2^\beta
		+{q'}^\beta p_2^\alpha}{q'\cdot p_2}\right),
	\label{DISLOcalc1}
\end{equation}
where we have used the gluon polarisation choice of eq.~(\ref{c=q'})
for the incoming and outgoing gluons, and denoted the Higgs effective
coupling to gluons by 
\begin{displaymath}
\lambda \eta^{\mu\nu},
\end{displaymath}
where $\mu$, $\nu$ are the Lorentz indices of the gluon entering the effective vertex. Contracting indices and combining with the phase space, one finds the following LO contribution to the structure function of eq.~(\ref{Fphig}):
\begin{equation}
  {\cal F}_{\phi}(x,Q^2)\Big|_{\rm LO}=\frac{2\pi|\lambda|^2}{2p\cdot q}
  \, \delta(1-x)\,.
  \label{FphigLO}
\end{equation}
For this and subsequent structure functions, we will divide all higher-order contributions by the prefactor $\sigma_0$ of the delta-function appearing at LO, thus introducing the normalised quantity
\begin{equation}
  W_{i}(x,Q^2)\equiv\frac{1}{\sigma_0}{\cal F}_i(x,Q^2).
  \label{Wdef}
\end{equation}
Now let us calculate the LO contribution to the quark structure function, given by the diagram of figure~\ref{fig:DISNLOb}. One finds a squared matrix
element (summed/averaged over spins and colours) 
\begin{align}
  \overline{|{\cal M}_{qh\rightarrow qg}|^2}&=\frac{C_F}{2}g_s^2\mu^{4-d}|\lambda|^2
  \frac{{\rm Tr}[\slsh{p}\gamma^\beta\slsh{k_q}\gamma^\alpha]}
       {(2p\cdot k_q)^2}\left(-\eta_{\alpha\beta}+
       \frac{q'_\alpha p_{H,\beta}+q'_\beta p_{H,\alpha}}{q'\cdot p_H}
       \right)\notag\\
       &=\frac{C_F}{8}\frac{g_s^2\mu^{4-d}|\lambda|^2}{(p\cdot k_q)^2}
       \left[(d-2){\rm Tr}[\slsh{p}\slsh{k}_q]
         +\frac{1}{q'\cdot p_2}\Big(
         {\rm Tr}[\slsh{p}\slsh{p}_H\slsh{k}_q\slsh{q}']
         +{\rm Tr}[\slsh{p}\slsh{q}'\slsh{k}_q\slsh{p}_H]
         \Big)\right],
       \label{HiggsNLO1}
\end{align}
where $C_F = (N_c^2-1)/(2N_c) = 4/3$ is the Casimir of the fundamental representation, $N_c$ is the number of colours, $g_s^2 = 4\pi \alpha_s$ denotes the (dimensionless) coupling of QCD, and $\mu$ the dimensional regularization scale. Evaluating the trace terms and using the momentum conservation condition
\begin{equation}
p_H=(1-x)p+q'-k_q\,,
  \label{momcon}
\end{equation}
the trace contribution appearing in eq.~(\ref{HiggsNLO1}) simplifies as follows:
\begin{equation} 
{\rm Tr}[\slsh{p}\slsh{p}_H\slsh{k}_q\slsh{q}']
+{\rm Tr}[\slsh{p}\slsh{q}'\slsh{k}_q\slsh{p}_H]
=16\, p\cdot q\,k_q\cdot q'.
\label{trace}
\end{equation}
Converting to Sudakov variables, eq.~(\ref{HiggsNLO1}) then becomes
\begin{equation}
\overline{|{\cal M}_{qh\rightarrow qg}|^2}=\frac{C_F \,g_s^2 \mu^{4-d}|\lambda|^2}{2p\cdot q}\left[\frac{d-2}{\beta_q}+\frac{4\alpha_q}{\beta_q^2 [1-x-\alpha_q]}\right],
  \label{HiggsNLO2}
\end{equation}
where the first (second) term come from the first (second) term in the gluon polarisation sum of eq.~(\ref{polsum}). Combining with the phase space of eq.~(\ref{dPhi2}), one
finds
\begin{align}
	\int {\rm d}\Phi^{(2)}\overline{|{\cal M}_{qh\rightarrow qg}|^2}
	&\,\,=\,\, \frac{2\pi}{(4\pi)^{\frac{d}{2}}}\frac{1}{\Gamma\left(\frac{d-2}{2}\right)}
	(Q^2)^{\frac{d-4}{2}}\left(\frac{1-x}{x}\right)^{\frac{d-4}{2}}
	\frac{C_F \,g_s^2\mu^{4-d}|\lambda|^2}{2p\cdot q}\notag\\
	&\hspace{2cm} \times\int_0^1
	{\rm d}\beta_q[\beta_q(1-\beta_q)]^{\frac{d-4}{2}}
	\left[\frac{d-2}{\beta_q}+\frac{4(1-\beta_q)}{\beta_q^3}\right].
	\label{HiggsNLO3}
\end{align}
The integral on the second line is given by 
\begin{align}
  \int_0^1
  {\rm d}\beta_q[\beta_q(1-\beta_q)]^{\frac{d-4}{2}}\, 
  \left[\frac{d-2}{\beta_q}+\frac{4(1-\beta_q)}{\beta_q^3}\right]
  &\,=\, \frac{\Gamma\left(\frac{d-4}{2}\right)\Gamma\left(\frac{d-2}{2}\right)}
       {\Gamma(d-3)}+\frac{2\Gamma\left(\frac{d-8}{2}\right)
         \Gamma\left(\frac{d}{2}\right)}{\Gamma(d-8)}\notag\\
       &\,=\,-\frac{2}{\epsilon}+\ldots\,.
\label{HiggsNLO4}
\end{align}
Note that the second term on the first line is ${\cal O}(\epsilon)$, and therefore does not contribute to LL behaviour. This suggests that we do not need to include the second term in the gluon polarisation tensor for $p_H$, and we will return to this point when discussing the higher-order corrections. Combining eq.~(\ref{HiggsNLO4}) with the remaining matrix element and phase space factors, one finds
\begin{equation}
	\int {\rm d}\Phi^{(2)}\overline{|{\cal M}_{qh\rightarrow qg}|^2} \,=\,
	\frac{\alpha_s}{4\pi}\left(\frac{4\pi\mu^2}{Q^2}\right)^{\epsilon} \frac{x^{\epsilon}}{\Gamma(1-\epsilon)}\left(\frac{2\pi |\lambda|^2}
	{2\, p\cdot q}\right)\left[-\frac{2C_F(1-x)^{-\epsilon}}
	{\epsilon}+ \mathcal{O}(\varepsilon)\right].
	\label{HiggsNLO5}
\end{equation}
Throughout, we will follow refs.~\cite{Vogt:2010cv,Presti:2014lqa} by defining perturbative coefficients of all quantities $X$ by
\begin{equation}
	X=\sum_n a_s^n X^{(n)}\,,\quad a_s=\frac{\alpha_s}{4\pi}\,.
	\label{Xdef}
\end{equation}
Collecting only the single pole in the limit that $x\rightarrow 1$,  setting the renormalisation scale equal to the hard scale of the process ($\mu = Q$)~\footnote{This is allowed at LL, since scale-dependence only contributes at NLL.}, and recognising the LO normalisation factor from eq.~(\ref{FphigLO}), we
find that
\begin{equation}
  W_{\phi,q}^{(1)}(x)=-\frac{2C_F}{\epsilon}(1-x)^{-\epsilon}\,.
\label{Tphiq1res}
\end{equation}
That this is indeed NLP in the threshold expansion can be seen by comparing with eq.~(\ref{thresholddef}), where the threshold variable in this case is $\xi=(1-x)$. For later use, it is convenient to express this result in Mellin space, where the Mellin transform of a function $f(x)$ is defined by
\begin{equation}
	f(N)=\int_0^1{\rm d}x \, x^{N-1} \,f(x)\,,
	\label{Mellin}
\end{equation}
where it is clear from the arguments of the function whether we are in $N$-space or $x$-space. In Mellin space eq.~(\ref{Tphiq1res}) becomes
\begin{equation}
  W_{\phi,q}^{(1)}(N)=-\frac{2C_F}{\epsilon}\frac{N^\epsilon}{N}\,,
\label{Tphiq1resN}
\end{equation}
where we have taken the $N\rightarrow \infty$ limit, and we keep only the dependence on LL terms~\footnote{Note that LL is defined to mean the maximum power of $\ln(N)$ at each power of $\alpha_s$ and $\epsilon$. Thus, the factor $N^\epsilon$ in eq.~(\ref{T2g1N}) generates LL terms at each power of $\epsilon$ when expanded.}. Having calculated the first non-zero contribution to the subleading partonic channel in Higgs-induced DIS, we now proceed to calculate its all-order structure function.

\subsubsection{All-order structure function in Higgs-induced DIS}
\label{sec:Tphiqcalc}

\begin{figure}
  \begin{center}
  	\centering
  	\mbox{	\captionsetup[subfigure]{oneside,margin={0.8cm,0cm},skip=0.5cm}
  		\subfloat[]{
  			\includegraphics[width=0.35\textwidth]{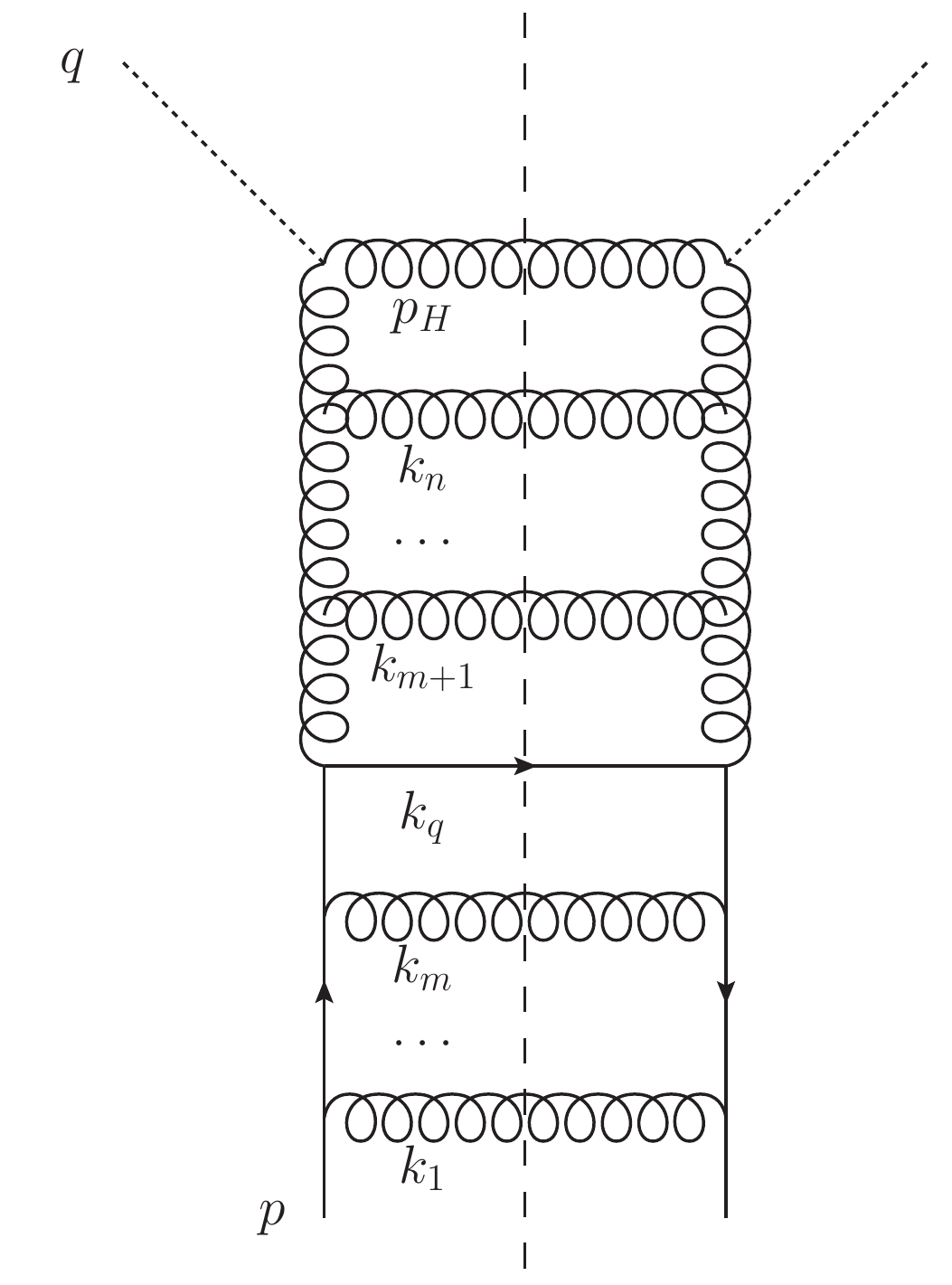}\label{fig:gluonladdera}
  		}
  		
  		\hspace{2.5cm}
  		\captionsetup[subfigure]{oneside,margin={0.8cm,0cm},skip=0.5cm}
  		\subfloat[]{
  			\includegraphics[width=0.35\textwidth]{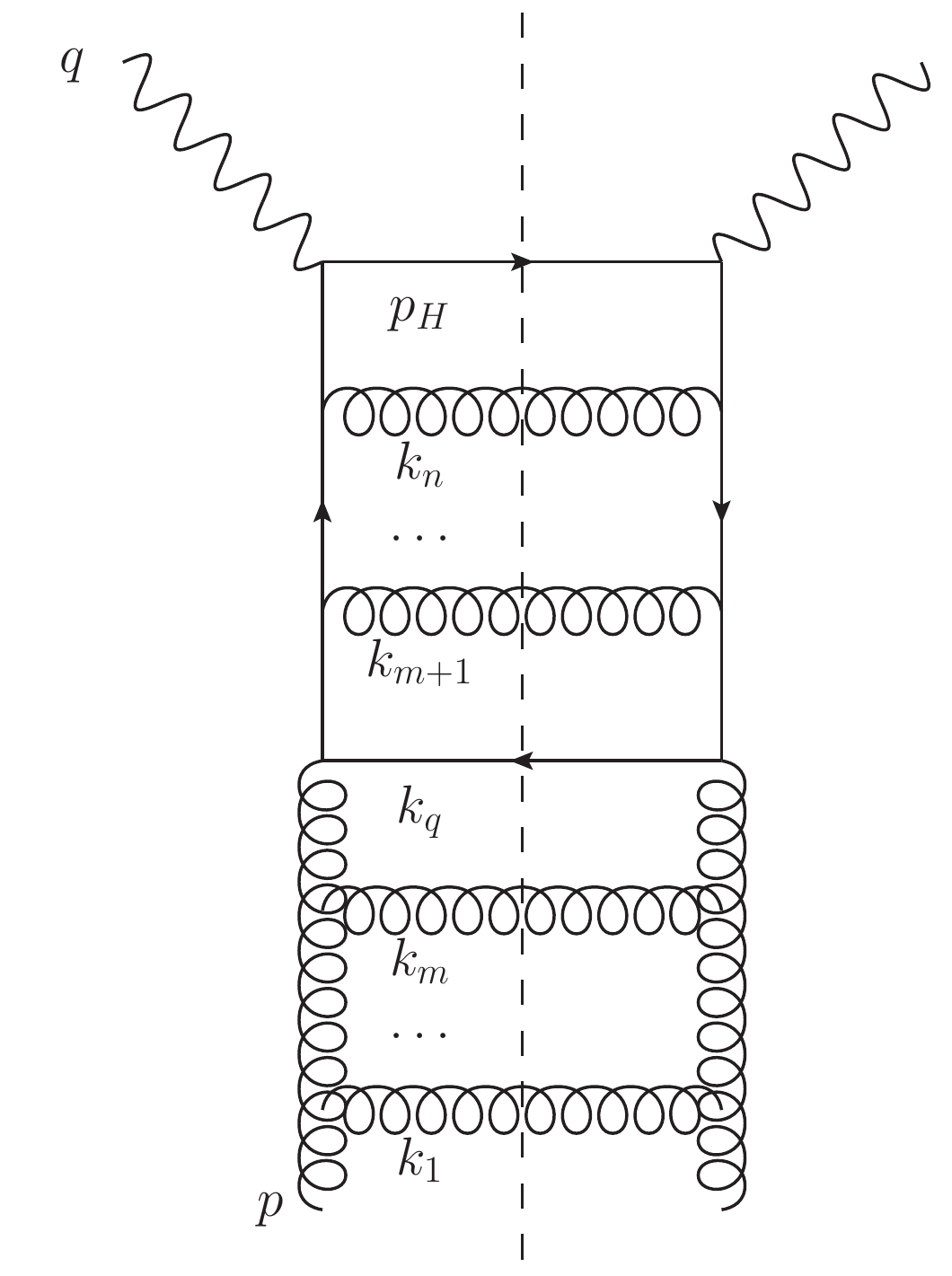}	\label{fig:gluonladderb}
  		}
  	}
    \caption{(a) Ladder graph contributing to $T^{(n)}_{\phi,q}$, where all possible values of $m$ must be considered; (b) similar but for $T^{(n)}_{2,g}$.}
    \label{fig:gluonladder}
  \end{center}
\end{figure}

To derive all-order forms for the relevant splitting and coefficient functions, we now need to dress these processes with arbitrary numbers of additional soft gluons. As discussed at the beginning of section~\ref{sec:DIS}, we can calculate $W_{\phi,q}^{(n)}$ by considering general uncrossed `ladder graphs' such as that of figure~\ref{fig:gluonladdera}, which has $m$ gluon `rungs' connecting the quark legs in the lower part, and $n-m$ gluons connecting the gluons in the upper part of the diagram. Furthermore, as we have argued that only maximally soft gluons will contribute at NLP, one may apply the well-known {\it eikonal Feynman rule}
\begin{equation}
  V^\mu=\pm g_s\mu^{\frac{4-d}{2}}{\bf T}_i  \frac{p_i^\mu}{p_i\cdot k}\,,
  \label{eikrule}
\end{equation}
for emission of a soft gluon of momentum $k$ from a hard leg of momentum $p_i$, where ${\bf T}_i$ is a colour generator in the appropriate representation, and the sign $+$ for an outgoing or $-$ for an incoming hard particle. 
The summed and averaged squared matrix element for the diagram of figure~\ref{fig:gluonladdera} is then found to be
\begin{align}
\overline{|{\cal M}_{qh\rightarrow qg_1\dots g_n}|^2} \notag &  \\
&\hspace{-2.0cm}
=\,\frac{|\lambda|^2 C_F^{m+1}C_A^{n-m}
  	g_s^{2(n+1)}}{8\mu^{(d-4)(n+1)}}\left(\prod_{i=1}^n\frac{2q \cdot p\,p\cdot k_i}
  {q'\cdot k_i}\right){\rm Tr}[\slsh{p}\gamma^\beta\slsh{k}_q\gamma^\alpha]
  \left(-\eta_{\alpha\beta}+\frac{q'_\alpha p_{H,\beta}+q'_{\beta}p_{H,\alpha}}
       {q'\cdot p_H}\right)\notag\\
&\hspace{-1.8cm}
\times \frac{1}{(p\cdot k_1)^2[p\cdot(k_1+k_2)]^2\ldots
         [p\cdot(k_1+\ldots+k_m+k_q)]^2\ldots
         [p\cdot(k_1+\ldots +k_n+k_q)]^2}.
  \label{gluoncalc1}
\end{align}
At NLO, we found that the second term in the remaining gluon polarisation tensor did not contribute at LL order. Let us assume the same thing will happen here (we will explicitly check this later). Then eq.~(\ref{gluoncalc1}) simplifies to
\begin{align}
\overline{|{\cal M}_{qh\rightarrow qg_1\dots g_n}|^2}&=\frac{d-2}{2}\, |\lambda|^2C_F^{m+1}C_A^{n-m}
  \frac{2^ng_s^{2(n+1)}}{(p\cdot q)^{n+1}\mu^{(d-4)(n+1)}}\, \beta_q\left(\prod_{i=1}^n
\frac{\bar{\beta}_i}{\bar{\alpha}_i}\right)\notag\\
&\times\frac{1}{\bar{\beta}_1^2(\bar{\beta}_1+\bar{\beta_2})^2
\ldots(\bar{\beta_1}+\ldots+\bar{\beta}_m+\beta_q)^2\ldots
(\bar{\beta}_1+\ldots+\bar{\beta}_n+\beta_q)^2},
\label{gluoncalc2}
\end{align}
where we have introduced the Sudakov variables from
eq.~(\ref{suds}). To make things look a little more symmetric, let us
introduce the variables
\begin{equation}
(1-x)b_i=\begin{cases}\bar{\beta}_i,\quad 1\leq i\leq m;\\
\beta_q,\quad i=m+1;\\
\bar{\beta}_{i-1},\quad m+2\leq i\leq n+1.
\end{cases}
\label{bidef}
\end{equation}
Upon combining with the phase space of eq.~(\ref{phasespace5})
(suitably relabelled), we will end up with the integral
\begin{align}
	&\int_{-i\infty}^{i\infty}\frac{{\rm d}T}{2\pi i}e^{T(1-x)}
	\frac{1}{T^{(n+1)d-4n-3}}\int {\rm d}\alpha_q\,\alpha_q^{\frac{d-4}{2}}e^{-\alpha_q}
	\left(\prod_{i=1}^n \int {\rm d}\bar{\alpha}_i\,\bar{\alpha}_i^{\frac{d-6}{2}}
	e^{-\bar{\alpha}_i}\right)\notag\\
	&\hspace{4cm}\times \left(\prod_{i=1}^{n+1}\int {\rm d} b_i\,b_i^{\frac{d-2}{2}}\right)
	\frac{e^{-\sum_{i=1}^{n+1}b_i}}{b_1^2(b_1+b_2)^2\ldots 
		(b_1+b_2+\ldots+b_{n+1})^2}\notag\\
	&=\Gamma^n\left(\frac{d-4}{2}\right)\Gamma\left(\frac{d-2}{2}\right)
	\frac{(1-x)^{(n+1)(d-4)}}{\Gamma[(n+1)d-4n-3]}\notag\\
	&\hspace{4cm}\times\left(\prod_{i=1}^{n+1}\int {\rm d}b_i\,b_i^{\frac{d-2}{2}}\right)
	\frac{e^{-\sum_{i=1}^{n+1}b_i}}{b_1^2(b_1+b_2)^2\ldots 
		(b_1+b_2+\ldots+b_{n+1})^2}\,,
	\label{gluoncalc3}
\end{align}
where we have already scaled out the $T$ dependence from the Sudakov variables in the first line, and carried out the $(\alpha_q,\bar{\alpha}_i)$ integrals in the second line, as well as the inverse Laplace transform in $T$ using
\begin{eqnarray}
\int_{-i\infty}^{i\infty}\frac{{\rm d}T}{2\pi i} {\rm e}^{T (1-x)}T^{-\alpha} = \frac{(1-x)^{\alpha-1}}{\Gamma(\alpha)}\,. \label{eq:laplace}
\end{eqnarray}
To carry out the remaining integrals, we may recall the above remarks that crossed-ladder contributions are kinematically subleading. This means that we may add those contributions, ignoring the colour factor, such that one may replace the factor
\begin{displaymath}
\frac{1}{b_1(b_1+b_2)\ldots (b_1+b_2+\ldots+b_{n+1})}
\end{displaymath}
with a sum over all permutations $\pi$ (see Appendix~\ref{app:integrals} for the justification of this replacement). This means we can write
\begin{align}
&\left[\frac{1}{b_1(b_1+b_2)\ldots (b_1+b_2+\ldots+b_{n+1})}\right]^2
\rightarrow \notag\\
& \hspace{4cm}\frac{1}{(n+1)!}
\left[\sum_\pi \frac{1}{b_{\pi_1}(b_{\pi_1}+b_{\pi_2})\ldots
	(b_{\pi_1}+b_{\pi_2}+\ldots+b_{\pi_{n+1}})}\right]^2.
\label{bireplace}
\end{align}
To explain the combinatorial factor on the right-hand side, note that expanding the brackets gives $[(n+1)!]^2$ terms in total (including identical contributions). The diagonal terms are simply relabellings of the original term on the left-hand
side of eq.~(\ref{bireplace}), in which each denominator is explicitly squared. There are $(n+1)!$ such terms, and we must correct for this overcounting. We do not have to worry about the cross-terms: these correspond to the kinematic parts of crossed-ladder graphs, and thus are kinematically subleading. The right-hand side of eq.~(\ref{bireplace}) is now written in a form that allows us to apply the eikonal identity, i.e.
\begin{equation}
\sum_\pi \frac{1}{b_{\pi_1}(b_{\pi_1}+b_{\pi_2})\ldots
(b_{\pi_1}+b_{\pi_2}+\ldots+b_{\pi_{n+1}})}=
\prod_{i=1}^{n+1}\frac{1}{b_i}\,,
\label{eikidbi}
\end{equation}
in each bracket, so that the $\{b_i\}$ integrals simply become
\begin{align}
& \left(\prod_{i=1}^{n+1}\int {\rm d}b_i\,b_i^{\frac{d-2}{2}}\right)
	\frac{e^{-\sum_{i=1}^{n+1}b_i}}{b_1^2(b_1+b_2)^2\ldots 
		(b_1+b_2+\ldots+b_{n+1})^2}\, \notag \\
&\hspace{3cm} \rightarrow \frac{1}{(n+1)!}\prod_{i=1}^{n+1}\int {\rm d} b_i\, b_i^{\frac{d-6}{2}}e^{-b_i}
=\frac{\Gamma^{n+1}\left(\frac{d-4}{2}\right)}{(n+1)!}
=\frac{1}{(n+1)!}\left(-\frac{1}{\epsilon}\right)^{n+1}+\ldots\, .\, 
\label{biints2}
\end{align}
For the reader who is not convinced by this argument, we provide a direct computation of the integrals in eq.~(\ref{gluoncalc3}) in appendix~\ref{app:integrals}, finding the same result. Indeed, this also justifies the statement that crossed ladders are kinematically subleading and would only contribute beyond NLP LL, i.e.~at NLP NLL. \\

We may now substitute our result for the $\{b_i\}$ integrals from eq.~(\ref{biints2}) into eq.~(\ref{gluoncalc3}), and then combine with the remaining factors from the phase space and matrix element. Dividing by the LO normalisation and keeping only the leading $\epsilon$ behaviour, one finds
\begin{equation}
		W_{\phi,q}^{(n+1)}(x)
		=\left(\sum_{m=0}^{n} C_F^{m+1}C_A^{n-m}\right)
		\frac{2^{2n+1}}{(n+1)!}
		\bigg(\frac{\mu^2}{Q^2 (1-x)}\bigg)^{(n+1)\epsilon}
		\left(-\frac{1}{\epsilon}\right)^{2n+1}+\ldots,
		\label{Tphiq}
\end{equation}
where we have used the fact that the kinematic part of each ladder diagram is the same to immediately sum over all possible colour structures. In $N$ space one then finds up to LL
\begin{equation}
	W_{\phi,q}^{(n+1)}=-\left(\sum_{m=0}^{n}
	C_F^{m+1}C_A^{n-m}\right)\frac{2}{\epsilon}\frac{N^\epsilon}{N}
	\left(\frac{4 N^\epsilon}{\epsilon^2}\right)^n \frac{1}{(n+1)!}\,.
	\label{Tphiqresult}
\end{equation}
We can resum the tower of higher-order contributions into a closed form:
\begin{align}
	W_{\phi,q}\Big|_{\rm LL}&=\sum_{n=1}^\infty
	a_s^n W_{\phi,q}^{(n)}\notag\\
	&=-\frac{2a_sC_F}{\epsilon}\frac{N^\epsilon}{N}
	\frac{1}{C_F-C_A}\left(\frac{4a_s N^\epsilon}{\epsilon^2}\right)^{-1}
	\left\{\exp\left[\frac{4a_s C_F N^\epsilon}
	{\epsilon^2}\right]
	-\exp\left[\frac{4a_s C_A N^\epsilon}{\epsilon^2}\right]
	\right\}.
	\label{Tphiqresult2}
\end{align}
In understanding this result, it is useful to take note of the identity
\begin{equation}
		\frac{C_F^{n+1}-C_A^{n+1}}{C_F-C_A}=\sum_{m=0}^n C_F^m C_A^{n-m},
		\label{colid}
\end{equation}
which demonstrates that the inverse factor of
$(C_F-C_A)$ in eq.~(\ref{Tphiqresult2}) simply combines with the colour factors arising from expanding the exponentials, to reproduce the democratic sum of factors of the form $C_F^p C_A^q$ in eq.~(\ref{Tphiqresult}). \\

We now come back to verify an assumption we made above, namely that the second term in the gluon polarisation tensor of eq.~(\ref{gluoncalc1}) gives a subleading contribution and can therefore be neglected. The relevant contribution to the squared matrix element is
\begin{align}
&  \frac{|\lambda|^2}{8}C_F^{m+1}C_A^{n-m}g_s^{2(n+1)}\mu^{(4-d)(n+1)}
  \left(\prod_{i=1}^n \frac{2q\cdot p\,p\cdot k_i}
       {q'\cdot k_i}\right){\rm Tr}[\slsh{p}\gamma^\beta\slsh{k}_q\gamma^\alpha]
       \left(\frac{q'_\alpha p_{H,\beta}+q'_\beta p_{H,\alpha}}
            {q'\cdot p_{H}}\right)\notag\\
            &\hspace{1cm} \times
            \frac{1}{(p\cdot k_1)^2[p\cdot(k_1+k_2)]^2\ldots
         [p\cdot(k_1+\ldots+k_m+k_q)]^2\ldots
         [p\cdot(k_1+\ldots +k_n+k_q)]^2}.            
\label{term2calc1}
\end{align}
The trace combination appearing on the first line is
\begin{align}
  (q'_\alpha p_{H,\beta}+q'_\beta p_{H,\alpha})
  {\rm Tr}[\slsh{p}\gamma^\beta\slsh{k}_q\gamma^\alpha]
  &=8(p\cdot p_H \, k_q\cdot q'-p\cdot k_q \, p_H\cdot q'+p\cdot q' \, k_q\cdot p_H)
  \notag\\
  &=16 \, k_q\cdot q' \, p\cdot q +\ldots,
\label{traceterm}
\end{align}
where we have used eq.~(\ref{momcon}) to eliminate $p_H$, and neglected terms which are quadratic in soft momenta $\{p_q,k_i\}$, which is consistent with neglecting such quadratic terms in the phase space. Substituting this back into eq.~(\ref{term2calc1}) and transforming to Sudakov variables gives
\begin{align}
	&2|\lambda|^2 C_F^{m+1}C_A^{n-m}g_s^{2(n+1)}\mu^{(4-d)(n+1)}
	2^n(p\cdot q)^{-n-1}\frac{\alpha_q}{1-x-\alpha_q-\sum_{i=1}^n \bar{\alpha}_i}
	\left[\prod_{i=1}^n\frac{\bar{\beta}_i}{\bar{\alpha}_i}\right]\notag\\
	& \hspace{3cm} \times \frac{(1-x)^{n+2}}{\bar{\beta}_1^2(\bar{\beta}_1+\bar{\beta_2})^2
		\ldots(\bar{\beta_1}+\ldots+\bar{\beta}_m+\beta_q)^2\ldots
		(\bar{\beta}_1+\ldots+\bar{\beta}_n+\beta_q)^2}\, ,
	\label{term2calc2}
\end{align}
where we have rescaled the $\bar{\beta}_i$ and $\beta_q$ variables by $1/(1-x)$.  We now notice that the delta function in the phase space of eq.~\eqref{phasespace4} allows one to make  the replacement
\begin{equation}
  1-x-\alpha_q-\sum_{i=1}^n \bar{\alpha}_i\rightarrow \beta_q
  +\sum_{i=1}^n\bar{\beta}_i\,,
  \label{alphas}
\end{equation}
so that eq.~(\ref{term2calc3}) becomes
\begin{align}
		&2|\lambda|^2 C_F^{m+1}C_A^{n-m}g_s^{2(n+1)}\mu^{(4-d)(n+1)}
		2^n (p\cdot q)^{-n-1}(1-x)^{n+2}\,\alpha_q
		\left[\prod_{i=1}^n\frac{\bar{\beta}_i}{\bar{\alpha}_i}\right]\notag\\
		& \hspace{0.5 cm}\times \frac{1}{\bar{\beta}_1^2(\bar{\beta}_1+\bar{\beta_2})^2
			\ldots(\bar{\beta_1}+\ldots+\bar{\beta}_m+\beta_q)^2\ldots
			(\bar{\beta}_1+\ldots+\bar{\beta}_{n-1}+\beta_q)^2
			(\bar{\beta}_1+\ldots+\bar{\beta}_n+\beta_q)^3}.
		\label{term2calc3}
\end{align}
Note the difference with eq.~(\ref{gluoncalc2}); here we have an additional factor of $\alpha_q$ in the numerator and $\left[\sum_i \bar{\beta}_i + \beta_q\right]^{-1}$ in the denominator, whereas we are missing the factor of $\beta_q$ with respect to eq.~(\ref{gluoncalc2}). 
Upon combining with the phase space of eq.~(\ref{phasespace5}) and scaling the $T$ dependence out of the Sudakov variables, the afore-mentioned difference will result in an additional factor of $T$ such that the overall
$T$ integral becomes
\begin{align}
  \int_{-i\infty}^{i\infty}\frac{{\rm d}T}{2\pi i}
 \, e^{T(1-x)}\, \frac{1}{T^{(n+1)(d-4)}}
    &\, =\, \frac{1}{1-x}\, \frac{(1-x)^{(n+1)(d-4)}}{\Gamma[(n+1)(d-4)]} \notag\\
    & \, = \,- \frac{2(n+1)\epsilon}{1-x} \left(1-x\right)^{-2(n+1)\epsilon} + \ldots\, .
  \label{Tint}
\end{align}
There are $2(n+1)$ remaining integrals over the rescaled Sudakov parameters $\bar{\alpha}_i$, $\bar{\beta}_i$, $\beta_q$ and $\alpha_q$, each of which may potentially contribute at most one singularity in $\epsilon$. However, contrary to eq.~\eqref{gluoncalc3}, the $\alpha_q$ integral is now not singular due to the additional factor of $\alpha_q$ in eq.~\eqref{term2calc3}. Together, the total contribution to the structure function from the second term in the gluon polarisation tensor must be ${\cal  O}(\epsilon^{-2n})$ or higher. This is indeed subleading compared to the contributions we have already considered, and thus we were justified in neglecting it~\footnote{Note also that the factor of $1/(1-x)$ in eq.~(\ref{Tint}) gets cancelled against a factor of $(1-x)^2$ in the matrix element squared contribution, eq.~\eqref{term2calc3}.}. \\
\subsubsection{Adding the virtual corrections}
\label{sec:virtual}
By construction, eq.~(\ref{Tphiqresult2}) only includes the pure real emission contributions at each order in perturbation theory. There are also virtual corrections, and in a direct calculation of the structure function one must include all possible channels, with different numbers of real and virtual gluons, combined with appropriate phase spaces. As is well-known, however, it is not necessary to do this at LL order: all of our additional gluon emissions are associated with the emission of soft gluons. These lead to infrared singularities, which must largely cancel between real and virtual graphs (after phase space integration)~\cite{Bloch:1937pw,Kinoshita:1962ur,Lee:1964is}, leaving only collinear poles that can be absorbed into the parton distribution functions. To be more specific, the purely real result of eq.~(\ref{Tphiqresult2}) is ${\cal O}(\epsilon^{-2n+1})$ at ${\cal  O}(\alpha_s^n)$, which must reduce to ${\cal O}(\epsilon^{-n})$ upon combination with the virtual corrections. We may then fix the latter by the argument made below, which is a variant of the {\it soft gluon unitarity} requirement that has previously been adopted at LP~\cite{Catani:1989ne}.\\

The effect of the virtual corrections is to modify the real emission contributions at each order, removing singularities which are simultaneously soft and collinear. Thus, they will modify the double poles in eq.~(\ref{Tphiqresult2}) as follows:
\begin{equation}
  \frac{N^\epsilon}{\epsilon^2}\rightarrow
  \frac{N^\epsilon+\lambda}{\epsilon^2}\,,
  \label{Nepsmod}
\end{equation}
for some constant $\lambda$, where the second term on the right-hand side is down a power of $N^\epsilon$ due to having swapped a phase space integral for a real gluon with an integral over a virtual momentum. Making this modification everywhere in eq.~(\ref{Tphiqresult2}), one obtains the ansatz~\footnote{Using this form we have implicitly assumed that the virtual corrections exponentiate. However, we do not have to make this assumption, and we come back to this point in section~\ref{sec:qgallorder}. }
\begin{align}
  W_{\phi,q}\Big|_{\rm LL}&=-\frac{2a_s C_F}{\epsilon}
  \frac{N^\epsilon}{N}\frac{1}{C_F-C_A}
 \left(\frac{4a_s(N^\epsilon+\lambda_1)}
       {\epsilon^2}\right)^{-1}
  \exp\left[\frac{4a_s(\lambda_2 C_F
      +\lambda_3 C_A)}{\epsilon^2}\right]
 \notag\\
       &\quad\times\left\{\exp\left[\frac{4a_s C_F
           (N^\epsilon+\lambda_4)}
    {\epsilon^2}\right]
  -\exp\left[\frac{4a_s C_A (N^\epsilon+\lambda_5)}{\epsilon^2}\right]
  \right\}.
  \label{Tphiqansatz}
\end{align}
We have been extra general in allowing for the possibility of an overall prefactor, consisting of exponentiated virtual corrections that modify both terms in the bracket equally. As a result, there is some redundancy in this parametrisation, as we will see. At $\mathcal{O}(a_s)$ no poles will get removed by the virtual corrections (as indeed, no virtual correction exists at that order for the $qg$ channel), and $W_{\phi,q}^{(1)}\Big|_{\rm LL}$ obtained from eq.~\eqref{Tphiqansatz} should be equal to the purely real correction obtained by expanding eq.~\eqref{Tphiqresult2} up to $\mathcal{O}(a_s)$. Upon doing so we find the condition
\begin{align}
W_{\phi,q}^{(1)}\Big|_{\rm LL} = -\frac{2 C_F}{N(C_F - C_A)} \frac{C_F(\lambda_4 + 1) - C_A (\lambda_5 + 1)}{\epsilon (\lambda_1 + 1)} \equiv -\frac{2 C_F}{N}\frac{1}{\epsilon}\,,
\end{align}
which we use as a constraint on $\lambda_1$. Starting from ${\cal O}(a_s^2)$, we may impose that all poles $\epsilon^{-m}$ with $n+1\leq m\leq 2n-1$ vanish due to the virtual corrections, leading to the solution
\begin{eqnarray}
	\lambda_5 = -\frac{C_A \lambda_3 + C_F \lambda_2 + C_A}{C_A}\,,\quad \lambda_4 = -\frac{C_A \lambda_3 + C_F \lambda_2 + C_F}{C_F}\,,\quad  \lambda_1 = -1\,.
\end{eqnarray}
Upon using this in eq.~\eqref{Tphiqresult2}, we see that the $\lambda_2$ and $\lambda_3$ coefficients cancel, and we find the solution
\begin{align}
  W_{\phi,q}\Big|_{\rm LL}&\,=\,-\frac{2a_s C_F}{\epsilon}\frac{N^\epsilon}{N}
  \frac{1}{C_F-C_A}\left(\frac{4a_s(N^\epsilon-1)}{\epsilon^2}\right)^{-1}
  \notag\\
  &\quad\quad\quad\times \left\{\exp\left[\frac{4a_s C_F(N^\epsilon-1)}
    {\epsilon^2}\right]-\exp\left[\frac{4a_s C_A(N^\epsilon-1)}{\epsilon^2}
    \right]
  \right\},
\label{Tphiqresult3}
\end{align}
which is our final result for the all-order structure function in Higgs induced DIS. Understandably, the effect of the virtual corrections has been to simply remove the double pole wherever it appears, and one may question the pedantic nature of our above procedure in this case. Later on when discussing the DY and Higgs production channels, however, we will see a case that is not so simple {\it  a priori}, and thus we have trodden carefully here. 

\subsubsection{Discussion of the result}
Eq.~(\ref{Tphiqresult3}) is the all-order LL form for the quark component of the structure function of eq.~(\ref{Fphig}). As such, we may check it against ref.~\cite{Vogt:2010cv}, which conjectures an all-order form for the $C_F$ terms, again at LL order only:
\begin{equation}
  \left.W_{\phi,q}^{(n)}\right|_{C_F^n}=\frac{1}{n!}W_{\phi,q}^{(1)}
  \left(W_{2,q}^{(1)}\right)^{n-1}.
  \label{Tiqn}
\end{equation}
Here $W_{\phi,q}^{(1)}$ has been given in $N$-space in eq.~(\ref{Tphiq1resN}), and we have also introduced the LP NLO contribution to the conventional DIS structure function of eq.~(\ref{F2def}):
\begin{equation}
  W_{2,q}^{(1)}=4C_F\frac{(N^\epsilon-1)}{\epsilon^2}+\ldots.
  \label{T2q1}
\end{equation}
It is straightforward to check that upon setting $C_A\rightarrow 0$ in eq.~(\ref{Tphiqresult3}) and expanding in $a_s$, one verifies eq.~(\ref{Tiqn}). Reference \cite{Vogt:2010cv} further conjectured that all remaining colour structures would be obtained by replacing
\begin{displaymath}
  C_F^{n+1}\rightarrow \sum_{m=0}^{n} C_F^{m+1}C_A^{n-m}
\end{displaymath}
at ${\cal O}(a_s^n)$. Again, this agrees precisely with our result in eq.~(\ref{Tphiqresult3}). \\

Equation \eqref{Tphiqresult3} also reproduces the result found in eq.~(3.50) of ref.~\cite{Beneke:2020ibj}. Therefore, it may be useful to explicitly mention the differences between the methods used here and in \cite{Beneke:2020ibj}, which, taken together, gives us a more comprehensive understanding of the $qg$ DIS cross section. In eq.~\eqref{Tphiqresult2} we obtain by direct computation the all-order contribution to the total cross section, due to the real emission diagrams in figure~\ref{fig:gluonladder}, calculated in the limit in which the radiated particles are soft. Instead, in \cite{Beneke:2020ibj} one obtains the all-order contribution after calculating the virtual hard diagrams (see eq.~(2.23) there). There, the result is obtained first by assuming the exponentiation of the one-loop virtual contribution; subsequently, the exponentiation is justified within a re-factorisation approach in SCET, which allows one to write down a two-step renormalisation group evolution for the short-distance coefficient responsible for the virtual-diagram contributions to the total cross section. The two towers of contributions (all-real soft-emission and all-virtual hard diagrams) contain equivalent information, which is sufficient to reconstruct the full DIS cross section. This is possible because of the existence of consistency conditions, related to pole cancellations. In \cite{Beneke:2020ibj} such conditions have been obtained by requiring the finiteness of the hadronic cross section, while here we require that the partonic cross section has at most poles $\epsilon^{-n}$ at order $n$, after real and virtual contributions to the partonic cross section have been summed. The application of these conditions leads to the characteristic pattern $N^{\epsilon}-1$ appearing in the exponents of eq.~\eqref{Tphiqresult3}, also pointed out in eq.~(3.53) of  \cite{Beneke:2020ibj}.
    
\subsection{The gluon structure function in DIS at NLO}
\label{sec:DISNLO}
After having discussed the quark structure function in Higgs-induced DIS, we now consider the case of conventional DIS to extract the gluon structure function. Following ref.~\cite{Vogt:2010cv}, we may define $W_{2,g}$ to be the gluon contribution to the structure function of eq.~(\ref{F2def}), with the LO normalisation divided out. To be more precise, the LO structure function is straightforwardly found to be
\begin{equation}
	F_2(x,Q^2)\Big|_{\rm LO}=\sigma_0\, \delta(1-x)+{\cal O}(\epsilon),
	\quad \sigma_0=e_q^2,
	\label{F2LO}
\end{equation}
such that one defines
\begin{equation}
	W_{2,g} = \frac{1}{\sigma_0} \int {\rm d}\Phi\,  
	T_2^{\alpha\beta} \, \overline{|{\cal M}_{g\gamma^* \rightarrow q\bar{q}}|^2}_{\alpha\beta}\,.
\end{equation}
The gluon channel first occurs at NLO, and the squared amplitude is given by the diagram of figure~\ref{fig:DISNLOa}, where in line with the comments above, and refs.~\cite{Dokshitzer:1978hw,Dokshitzer:1991wu}, we can ignore the crossed box diagram as it is kinematically subleading. Averaging (summing) over initial (final) state colours and spins, one finds a squared matrix element
\begin{align}
	\overline{|{\cal M}_{g\gamma^* \rightarrow q\bar{q}}|^2}_{\alpha\beta}
	=T_R\frac{e_q^2 g_s^2\mu^{4-d}}{(d-2)}
	\frac{{\rm Tr}[\slsh{k}_q\gamma^\nu(\slsh{p}-\slsh{k}_q)
		\gamma_\beta\slsh{p}_H
		\gamma_\alpha(\slsh{p}-\slsh{k}_q)\gamma^\mu]}{(p-k_q)^2}
	\left(-\eta_{\mu\nu}+\frac{q'_\mu p_\nu+q'_\nu p_\mu}{p\cdot q'}
	\right),
	\label{DISNLOcalc1}
\end{align}
where we have used the gluon polarisation choice of eq.~(\ref{c=q'}), and introduced the normalisation factor of the QCD colour generators in the fundamental representation, $T_R=1/2$.  At LL level, all propagators must be maximally soft and/or collinear, so that we may take the 4-momentum of the emitted quark $k_q\rightarrow 0$ in the numerator. Introducing the Sudakov variables from eq.~(\ref{suds}) and projecting onto $T_2^{\alpha\beta}$ defined in eq.~(\ref{T2def}), one finds
\begin{equation}
	T_2^{\alpha\beta} \, \overline{|{\cal M}_{g\gamma^* \rightarrow q\bar{q}}|^2}_{\alpha\beta}  =T_R\frac{(d-2)e_q^2 g_s^2\mu^{4-d}}{2\pi}
	\frac{(1-\beta_q)}{\beta_q}\,.
	\label{DISNLOcalc2}
\end{equation}
Combining this with the phase space of eq.~(\ref{dPhi2}), one may carry out the $\alpha_q$ integral using the delta function, yielding
\begin{align}
	\int {\rm d}\Phi^{(2)} T_2^{\alpha\beta} \overline{|{\cal M}_{g\gamma^* \rightarrow q\bar{q}}|^2}_{\alpha\beta} &=
	\left(\frac{\alpha_s}{4\pi}\right)\, 2T_R\, (1-x)^{-\epsilon} \left(\frac{\mu^2}{Q^2}\right)^{\epsilon}\, \int {\rm d}\beta_q
	\,  \beta_q^{\frac{d-6}{2}}\, (1-\beta_q)^{\frac{d-2}{2}}+\ldots\notag\\
	&=\left(\frac{\alpha_s}{4\pi}\right)\, 2T_R\, (1-x)^{-\epsilon}\,\left(\frac{\mu^2}{Q^2}\right)^{\epsilon}\, 
	\frac{\Gamma(\frac{d-4}{2})\Gamma(\frac{d}{2})}{\Gamma(d-2)}+\ldots\notag\\
	&=\left(\frac{\alpha_s}{4\pi}\right)\left[-\frac{2T_R(1-x)^{-\epsilon}}
	{\epsilon}\right]
	+\ldots,
	\label{DISNLOcalc3}
\end{align}
where the ellipsis denotes terms that are suppressed by powers of $\epsilon$, and thus do not contribute at LL order. The contents of the square brackets constitute the first non-zero contribution to the gluon channel for DIS. Note, however, that we have only included a single quark in the diagram of figure~\ref{fig:DISNLOa}. We must instead include all possible massless (anti)-quark flavours, which amounts to simply multiplying eq.~(\ref{DISNLOcalc3}) by a factor of $2n_f$. From eq.~(\ref{DISNLOcalc3}), we find the ${\cal O}(\alpha_s)$ gluon structure function
\begin{equation}
	W_{2,g}^{(1)}(x)=-\frac{2n_f}{\epsilon}(1-x)^{-\epsilon}\, ,
	\label{T2g1}
\end{equation}
which in Mellin space becomes. 
\begin{equation}
	W_{2,g}^{(1)}(N)
	=-\frac{2n_f}{\epsilon}\frac{N^\epsilon}{N}+\ldots\,,
	\label{T2g1N}
\end{equation}
in the $N\rightarrow \infty$ limit. Note that one may obtain eq.~(\ref{T2g1N}) directly from the Higgs-induced DIS NLO result of eq.~(\ref{Tphiq1resN}) by the simple colour replacement
\begin{equation}
	C_F\rightarrow n_f,
	\label{colreplace}
\end{equation}
as remarked in ref.~\cite{Vogt:2010cv}.

\subsubsection{All order structure function in DIS}
\label{sec:T2gcalc}

Similarly to the Higgs-induced DIS case, we can also calculate an all-order LL form for the structure function $W_{2,g}$ in conventional DIS. This will be given by the ladder diagram of figure~\ref{fig:gluonladderb}, where again all emitted partons are soft. Dressing eq.~(\ref{DISNLOcalc1}) with the requisite eikonal Feynman rules and taking $k_q\rightarrow 0$ where possible, one obtains the (summed and averaged) squared matrix element
\begin{align}
  \overline{|{\cal M}_{g\gamma^*\rightarrow q\bar{q}g_1\dots g_n}|^2}&=\frac{T_R\, C_A^m\, C_F^{n-m} e_q^2 g_s^{2(n+1)}\mu^{(4-d)(n+1)}}
           {4(d-2)}\left(\prod_{i=1}^n\frac{2p\cdot q\, p\cdot k_i}
           {q'\cdot k_i}\right)\notag\\
           &\times
           {\rm Tr}[\slsh{k}_q\gamma^\mu\slsh{p}\gamma_\alpha\slsh{p}_H
             \gamma_\beta\slsh{p}\gamma^\nu]\left(
           -\eta_{\mu\nu}+\frac{q'_\mu p_\nu + q'_\nu p_\mu}{p\cdot q'}\right)
           \notag\\
           &\times \frac{T_{2}^{\alpha\beta}}
               {(p\cdot k_1)^2[p\cdot(k_1+k_2)]^2\ldots
         [p\cdot(k_1+\ldots+k_m+k_q)]^2\ldots
         [p\cdot(k_1+\ldots +k_n+k_q)]^2}.
  \label{DISallorder1}
\end{align}
Substituting the projector of eq.~(\ref{T2def}), contracting indices and carrying out the trace gives
\begin{align}
 \overline{|{\cal M}_{g\gamma^*\rightarrow q\bar{q}g_1\dots g_n}|^2}&=\frac{T_R\, C_A^m\, C_F^{n-m} e_q^2 g_s^{2(n+1)}}
           {2\pi\, \mu^{(d-4)(n+1)}}(p\cdot q)^{-n}\left(\prod_{i=1}^n\frac{\bar{\beta}_i}
           {\bar{\alpha}_i}\right)\frac{\beta_q(1-\beta_q-\sum_i\bar{\beta}_i)}
           {\bar{\beta}_1^2(\bar{\beta}_1+\bar{\beta}_2)^2\ldots
             (\bar{\beta}_1+\ldots \bar{\beta}_n+\beta_q)^2}\,,
      \label{DISallorder2}
\end{align}
where we have used momentum conservation to replace $p_H$, and introduced the usual Sudakov variables. To make the $\beta_q$ and $\bar{\beta}_i$ integrals easier, we can borrow the trick from the previous section of symmetrising over all crossed ladders. One may also use the delta function that appears in eq.~(\ref{phasespace2}) to
replace
\begin{equation}
  1-\beta_q-\sum_{i}\bar{\beta}_i\rightarrow
\frac{1}{1-x}\left(
  \alpha_q+\sum_i \bar{\alpha}_i\right).
\label{betareplace}  
\end{equation}
We then get
\begin{align}
 \overline{|{\cal M}_{g\gamma^*\rightarrow q\bar{q}g_1\dots g_n}|^2}&=\frac{T_R\, C_A^m\, C_F^{n-m} e_q^2 g_s^{2(n+1)}\mu^{(4-d)(n+1)}}
           {2\pi}\frac{(p\cdot q)^{-n}}{(n+1)!}\frac{2^n(d-2)}{1-x}
           \left(\prod_{i=1}^n\frac{1}{\bar{\alpha}_i\bar{\beta}_i}\right)
           \frac{\alpha_q+\sum_i\bar{\alpha}_i}{\beta_q}.
           \label{DISallorder3}
\end{align}
After rescaling the $\bar{\beta}_i$ and $\beta_q$ variables by a factor of $1/(1-x)$ and combining with the phase space, all integrals may be carried out similarly to the previous case, and one ultimately finds
\begin{equation}
\int {\rm d}\Phi^{(n+2)} \, \overline{|{\cal M}_{g\gamma^*\rightarrow q\bar{q}g_1\dots g_n}|^2}\,=\,
-\frac{2T_R a_s N^\epsilon}{\epsilon}
\left(\frac{4a_s N^\epsilon}{\epsilon^2}\right)^n \frac{C_A^m C_F^{n-m}}
{(n+1)!}\,,
\label{DISallorder4}
\end{equation}
which is valid at NLP LL order. 
As in the Higgs-induced case, we must sum over all values of $m$ to include all ladder diagrams, where $m$ gluons couple to the gluon leg (lower part of the diagram), and $n-m$ to the quark leg (upper part of the diagram). We must also multiply by $2n_f$ to take account of all (anti-)quark species that could be coupling to the gluon. The result is that the pure real emission contribution to $W_{2,g}^{(n+1)}$ can be easily obtained from eq.~(\ref{Tphiqresult}) by replacing a single power of $C_F$ with $n_f$, and interchanging $C_F\leftrightarrow C_A$ elsewhere, as in fact was already noted as part of the conjectures in ref.~\cite{Vogt:2010cv}. \\

Given the above replacements, it is not necessary to repeat the soft gluon unitarity argument in order to furnish eq.~(\ref{DISallorder4}) with virtual corrections. We can simply take the final result of eq.~(\ref{Tphiqresult3}) for the Higgs case, and make the necessary colour factor replacements to obtain
\begin{align}
  W_{2,g}\Big|_{\rm LL}&=-\frac{2a_s n_f}{\epsilon}\frac{N^\epsilon}{N}
  \frac{1}{C_A-C_F}\left(\frac{4a_s(N^\epsilon-1)}{\epsilon^2}\right)^{-1}
  \notag\\
  &\quad\times \left\{\exp\left[\frac{4a_s C_A(N^\epsilon-1)}
    {\epsilon^2}\right]-\exp\left[\frac{4a_s C_F(N^\epsilon-1)}{\epsilon^2}
    \right]
  \right\}.
\label{T2gresult}
\end{align}

\subsection{Resummed form for splitting and coefficient functions}

We have now derived all-order forms for the partonic structure functions for subleading channels in (Higgs-induced) DIS. As a consequence, we can derive resummed forms for the off-diagonal splitting functions $P_{qg}$ and $P_{gq}$, and also for the coefficient functions which control the finite parts of the structure functions~\footnote{This can be done following the steps described in section 3.2.3 of ref.~\cite{Beneke:2020ibj}. Nevertheless, we find it pedagogical to provide an independent derivation in what follows.}.
To define things more precisely, let us recall the mass factorisation procedure, by which those infrared poles remaining in
the structure functions after combining real and virtual contributions can be factorised as follows:
\begin{equation}
W_{a,k}=\widetilde{C}_{a,i}Z_{ik}\,,
\label{Takfac}
\end{equation}
or, in matrix form, 
\begin{equation}
\left(\begin{array}{c}W_{a,q}\\ W_{a,g}\end{array}\right)
=\left(\begin{array}{cc}\widetilde{C}_{a,q} & 
\widetilde{C}_{a,g}\end{array}\right)\left(
\begin{array}{cc} Z_{qq} & Z_{qg} \\
Z_{gq} & Z_{gg}
\end{array}\right).
\end{equation}
Here, $\widetilde{C}_{a,i}$ is an infrared finite coefficient function, and the {\it transition function} matrix ${\bf Z}\equiv \{Z_{ik}\}$ collects all the infrared divergences. The latter is related to the DGLAP
splitting functions, contained in the matrix
\begin{equation}
{\bf P}=\left(\begin{array}{cc} P_{qq} & P_{qg} \\
P_{gq} & P_{gg}\end{array}\right).
\label{Pdef}
\end{equation}
By definition, this satisfies
\begin{equation}
{\bf P}=\frac{d{\bf Z}}{d\ln Q^2}{\bf Z}^{-1},
\label{Peq}
\end{equation}
where $Q^2$ is the hard scale of the process, which has been taken to be equal to both the factorisation and renormalisation scales. Note that the ordering of the matrix and its inverse in eq.~(\ref{Peq}) are important, given the matrix-valued nature of ${\bf Z}$. In Mellin space, it is also conventional to define the anomalous dimension matrix
\begin{equation}
{\bf \gamma}\equiv \left(\begin{array}{cc} \gamma_{qq} & \gamma_{qg}\\
\gamma_{gq} & \gamma_{gg}\end{array}\right)=-{\bf P}(N)\,.
\label{gamdef}
\end{equation}
Let us focus explicitly on the function $P_{gq}$, which can be obtained from the Higgs-induced DIS process, as discussed in refs.~\cite{Vogt:2010cv,Beneke:2020ibj}. Once we have extracted $P_{gq}$ we may use the colour replacements to straightforwardly get $P_{qg}$. One can obtain $P_{gq}$ from the transition function $Z_{gq}$ using the relation~\footnote{As usual, we normalise perturbative coefficients in terms of $a_s=\alpha_s/(4\pi)$. However, following convention, $\gamma_{ij}^{(n)}$ is defined to be the coefficient of $a_s^{n+1}$.}
\begin{equation}
Z_{gq}^{(n)}=\frac{1}{n!}\sum_{m=0}^{n-1}
\frac{\gamma_{gq}^{(m)}}{\epsilon^{n-m}}\sum_{k=0}^{n-m-1}
\frac{(m+k)!}{k!}
\left(\gamma_{qq}^{(0)}\right)^{k}\left(\gamma_{gg}^{(0)}\right)^{n-m-1-k}.
\label{Zgqfinal}
\end{equation} 
This equation was stated without proof in ref.~\cite{Vogt:2010cv}, but we prove it in appendix~\ref{app:transition}. We are after obtaining $\gamma_{gq}^{(n-1)}$, which can be found directly from $Z_{gq}^{(n)}$ by extracting the ${\cal O}(\epsilon^{-1})$  term of eq.~\eqref{Zgqfinal}. This can be seen by requiring that $k=0$ and $n-m-1-k = n-m-1 = 0$ in eq.~\eqref{Zgqfinal}, resulting in
\begin{equation}
Z_{gq}^{(n)}\Big|_{k = 0, \, m = n-1}=
\frac{1}{\epsilon} \frac{\gamma_{gq}^{(n-1)}}{n}\,,
\label{Zgqeps-1}
\end{equation}
from which we see we indeed only need the ${\cal O}(\epsilon^{-1})$ part of $Z_{gq}^{(n)}$ to obtain $\gamma^{(n-1)}_{qg}$. The explicit form of the mass factorisation formula for $W_{\phi,q}$
is
\begin{equation}
W_{\phi,q}=\widetilde{C}_{\phi,q}Z_{qq}+\widetilde{C}_{\phi,g} Z_{gq}\,,
\label{Tphiqsol}
\end{equation}
where our aim is now to find both the quantities $\widetilde{C}_{\phi,q}$ and $Z_{gq}$ on the right-hand side. That we only have a single equation for two unknowns corresponds to the fact that the splitting and coefficient functions are not unique but defined only up to a choice of factorisation scheme. However, imposing the $\overline{\rm MS}$ scheme such that the transition functions contain only poles (and some particular numerical constants that need not explicitly concern us) is sufficient for us to find the above quantities, as we will see. Given that $Z_{gq}$ and $\widetilde{C}_{\phi,q}$ already start at NLP, we will need the remaining elements on the right-hand side of eq.~(\ref{Tphiqsol}) at LP, where they are fixed by standard resummation arguments. Quoting from ref.~\cite{Vogt:2010cv}, we have
\begin{equation}
  Z_{qq}=\exp\left[\frac{4a_s C_F\ln N}{\epsilon}\right],\quad
  \widetilde{C}_{\phi,g}=\exp\left[\frac{4a_s C_A(N^\epsilon-1-\epsilon\ln N)}
    {\epsilon^2}\right].
  \label{Zqq}
\end{equation}
That is, the LL transition function $Z_{qq}$ can be obtained by simply exponentiating the pole in the NLO result. Likewise, for the gluon coefficient function in Higgs-induced DIS, one removes the pole terms from the NLP result and exponentiates what is left. \\

As the coefficient function $\widetilde{C}_{\phi,q}$ is necessarily finite, one may rearrange eq.~(\ref{Tphiqsol}) to obtain the constraint
\begin{equation}
  \widetilde{C}_{\phi,q}
  =\frac{W_{\phi,q}}{Z_{qq}}-\frac{Z_{gq}\widetilde{C}_{\phi,g}}{Z_{qq}}
  \sim{\cal O}(\epsilon^0)\,.
  \label{Cconstraint}
\end{equation}
This equation implies that all $\epsilon$ poles must cancel between the first and second terms on the right-hand side. In particular, this must be true for the ${\cal O}(\epsilon^{-1})$ contribution, which we will
need to consider to get the single $\epsilon$ pole term of $Z_{gq}$ en route to the anomalous dimension $\gamma_{gq}$, as dictated by eq.~(\ref{Zgqeps-1}). For the first term in eq.~(\ref{Cconstraint}), we may substitute the results of eqs.~(\ref{Tphiqresult3}, \ref{Zqq}), and rearrange to get
\begin{align}
  \frac{W_{\phi,q}}{Z_{qq}}&=-\frac{C_F}{C_F-C_A}\frac{1}{2N\ln N}
  f(-\epsilon \ln N)
  \left\{\exp\left[\frac{4a_s C_F(N^\epsilon-1-\epsilon\ln N)}
    {\epsilon^2}\right]\right.\notag\\
  &\left.\quad
  -\exp\left[\frac{4a_s C_A(N^\epsilon-1-\epsilon\ln N)}
    {\epsilon^2}-\frac{4a_s (C_F-C_A)\ln N}{\epsilon}\right]
  \right\},
  \label{TZ1}
\end{align}
where we have isolated the function
\begin{equation}
  f(x)=\frac{x}{e^x-1}=\sum_{m=0}^\infty \frac{B_m}{m!}x^m,
  \label{fdef}
\end{equation}
which acts as the exponential generating function for the Bernoulli numbers $\{B_m\}$. We want to find the ${\cal O}(\epsilon^{-1})$ part of eq.~(\ref{TZ1}), for which we can note that the entire first line is finite as $\epsilon\rightarrow 0$. In the second line, we can recognise the coefficient function $\widetilde{C}_{\phi,g}$ from eq.~(\ref{Zqq}), such that the second term of eq.~(\ref{TZ1}) together with the prefactor can be written as
\begin{align}
\left.  \frac{W_{\phi,q}}{Z_{qq}}\right|_{\rm poles}\sim
  \frac{\widetilde{C}_{\phi,g}}{2N\ln N}\frac{C_F}{C_F-C_A}\sum_{n=1}^\infty a_s^n
  \frac{[4(C_A-C_F)]^n\ln^n N}{n!\epsilon^n}
  \sum_{m=0}^\infty
  \frac{B_m(-\epsilon\ln N)^m}{m!}.
  \label{TZ2}
\end{align}
By defining 
\begin{equation}
  Z_{gq}=\sum_{n=1}^\infty \sum_{m=-n}^{-1}a_s^n \epsilon^m Z_{gq}^{(n,m)},
  \label{Zgqexpand}
\end{equation}
where $m$ labels the power of $\epsilon$, 
we can write eq.~(\ref{Cconstraint}) as
\begin{align}
\widetilde{C}_{\phi,q} &\equiv {\cal O}(\epsilon^0)\notag \\
& =  \widetilde{C}_{\phi,g}
  \sum_{n=1}^\infty a_s^n
  \left[\frac{1}{2N\ln N}\frac{C_F}{C_F-C_A}
    \frac{[4(C_A-C_F)]^n\ln^n N}{n!\epsilon^n}
  \sum_{m=0}^\infty
  \frac{B_m(-\epsilon\ln N)^m}{m!}\right.\notag\\
 &\hspace{2cm}\left.\quad - \frac{1}{Z_{qq}}
\sum_{m=-n}^{-1} \epsilon^m Z_{gq}^{(n,m)}
  \right],
  \label{TZ3}
\end{align}
where we have extracted a common factor of $\widetilde{C}_{\phi,g}$. The above equation implies that the pole cancellation must apply within the square brackets. We remind the reader that we are in particular interested in the cancellation that happens for the $\mathcal{O}(\epsilon^{-1})$ part of this equation. For this reason, we can neglect the inverse factor of $Z_{qq}$, as it will only contribute further poles in $\epsilon$. In the first term, the ${\cal O}(\epsilon^{-1})$ contribution can only arise from the term in the second sum with $m=n-1$. In the second term, we need the term with $m=-1$. Demanding that these contributions cancel each other leads to the following result:
\begin{align}
  Z_{gq}^{(n,-1)}=-\frac{2a_s C_F}{N}[4 a_s(C_F-C_A)\ln^2 N]^{n-1}
  \frac{B_{n-1}}{n!(n-1)!}\,.
  \label{Zgq-1}
\end{align}
Combining this with eqs.~(\ref{Zgqeps-1}, \ref{gamdef}) yields
\begin{equation}
  P_{gq}(N)\Big|_{\rm LL}=\frac{2a_s C_F}{N}{\cal B}_0[4a_s(C_F-C_A)\ln^2 N]\,,
  \label{Pgqres}
\end{equation}
where
\begin{equation}
  {\cal B}_0(x)=\sum_{n=0}^\infty \frac{B_n}{(n!)^2}x^n\,.
  \label{B0def}
\end{equation}
Equation~(\ref{Pgqres}) is precisely the result conjectured in ref.~\cite{Vogt:2010cv}, including the full colour dependence. This result has also been obtained in section 3.2.3 of ref.~\cite{Beneke:2020ibj}, following an equivalent derivation, that leads from the expression given in eq.~\eqref{Tphiqresult3} to eqs.~(\ref{TZ3}, \ref{Pgqres}), exploiting eq.~\eqref{Tphiqsol}. \\

Using the all-order form of $P_{gq}$, and as in ref.~\cite{Beneke:2020ibj}, we can go further and derive the all-order LL form of the coefficient function $\widetilde{C}_{gq}$. Indeed, this is straightforward given the results of eqs.~(\ref{Cconstraint}, \ref{TZ1}). We have seen in eq.~(\ref{TZ2}) that the pole contributions from the first term in eq.~(\ref{Cconstraint}) have an explicit factor of
$\widetilde{C}_{\phi,g}$ in them. Thus, all the second term in eq.~(\ref{Cconstraint}) does is to remove the poles in the first term, and any other terms in which higher-order in $\epsilon$ contributions in $\widetilde{C}_{\phi,g}$ interact with the poles. We can thus find the coefficient function $\widetilde{C}_{\phi,q}$ by simply taking the ${\cal O}(\epsilon^0)$ piece of eq.~(\ref{TZ1}), ignoring the higher-order $\epsilon$ terms in $\widetilde{C}_{\phi,g}$ as we do so. The first term in eq.~(\ref{TZ1}) is finite, and taking $\epsilon\rightarrow 0$ gives
\begin{equation}
  -\frac{C_F}{C_F-C_A}\frac{1}{2N\ln N}\exp\left[4a_s C_F\ln^2 N\right].
  \label{Cphiq1}
\end{equation}
In the second term of eq.~(\ref{TZ1}), we can expand to get
\begin{align}
  \frac{C_F}{C_F-C_A}\frac{e^{2a_s C_A\ln^2 N}}{2N\ln N}\sum_{n=0}^\infty
  \frac{[-4a_s(C_F-C_A)\ln N]^n}{n!}\sum_{m=0}^\infty \epsilon^{m-n}
  \frac{B_m}{m!}(-\ln N)^m,
\label{Cphiq2}
\end{align}
such that the ${\cal O}(\epsilon^0)$ piece has $m=n$ in the second
sum, yielding
\begin{align}
  &\frac{C_F}{C_F-C_A}\frac{e^{2a_s C_A\ln^2 N}}{2N\ln N}
  \sum_{n=0}^\infty [-4a_s(C_F-C_A)\ln^2 N]^n \frac{B_n}{(n!)^2}\notag\\
  &=\frac{C_F}{C_F-C_A}\frac{e^{2a_s C_A\ln^2 N}}{2N\ln N}
    {\cal B}_0[4a_s(C_F-C_A)\ln^2 N]\,.
    \label{Cphiq3}
\end{align}
Putting things together, we get
\begin{equation}
  \widetilde{C}_{\phi,q}\Big|_{\rm LL}=
  \frac{1}{2\ln N}\frac{C_F}{C_F-C_A}\left[{\cal B}_0
    [4a_s(C_F-C_A)\ln^2 N]e^{2a_s C_A\ln^2 N}-e^{2a_s C_F\ln^2 N}\right],
  \label{Cphiqres}
\end{equation}
which matches the result in refs.~\cite{Vogt:2010cv,Beneke:2020ibj}. \\

Given the colour replacements in going from eq.~(\ref{Tphiqresult3}) to eq.~(\ref{T2gresult}), we can easily recycle our results to provide the counterparts of eqs.~\eqref{Pgqres}, \eqref{Cphiqres} in conventional DIS:
\begin{align}
P_{qg}\Big|_{\rm LL}&=\frac{2a_s n_f}{N}{\cal B}_0[4a_s(C_A-C_F)\ln^2 N];
\notag\\
  \widetilde{C}_{2,g}\Big|_{\rm LL}&=
  \frac{1}{2\ln N}\frac{n_f}{C_A-C_F}\left[{\cal B}_0
    [4a_s(C_A-C_F)\ln^2 N]e^{2a_s C_F\ln^2 N}-e^{2a_s C_A\ln^2 N}\right].
\label{Pqgres}
\end{align}

To summarise, in this section we have derived all-order LL forms for the kinematically subleading structure functions in (Higgs-induced) DIS. We used them to derive resummed results for the off-diagonal DGLAP splitting functions and the infrared-finite coefficient functions. Our results are in agreement with refs.~\cite{Vogt:2010cv,Beneke:2020ibj}, but our complementary approach -- constructing the structure functions using real emission contributions plus soft gluon unitarity -- means we do not have to make any assumptions about the exponentiation or otherwise of the virtual corrections. Furthermore, it seems clear that our arguments should generalise to other processes. Indeed they do, as we discuss in the following sections.

\section{Resummation of the $g\bar{q}$ channel in Drell-Yan production}
\label{sec:DY}

\begin{figure}
  \begin{center}
  	\mbox{	\captionsetup[subfigure]{oneside,margin={0.8cm,0cm},skip=0.5cm}
  		\subfloat[]{
  			\includegraphics[width=0.4\textwidth]{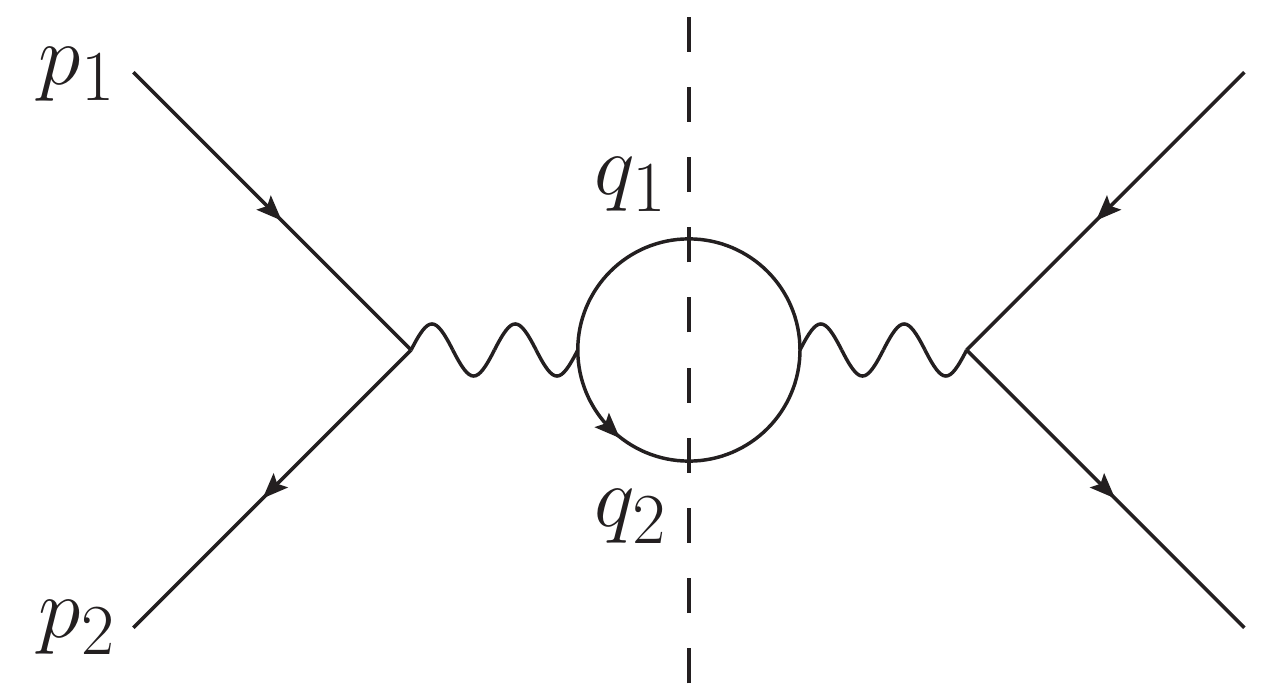}	\label{fig:DYLOa}
  		}
  	
  		\hspace{2.0cm}
  		\captionsetup[subfigure]{oneside,margin={0.8cm,0cm},skip=0.5cm}
  		\subfloat[]{
  			\includegraphics[width=0.4\textwidth]{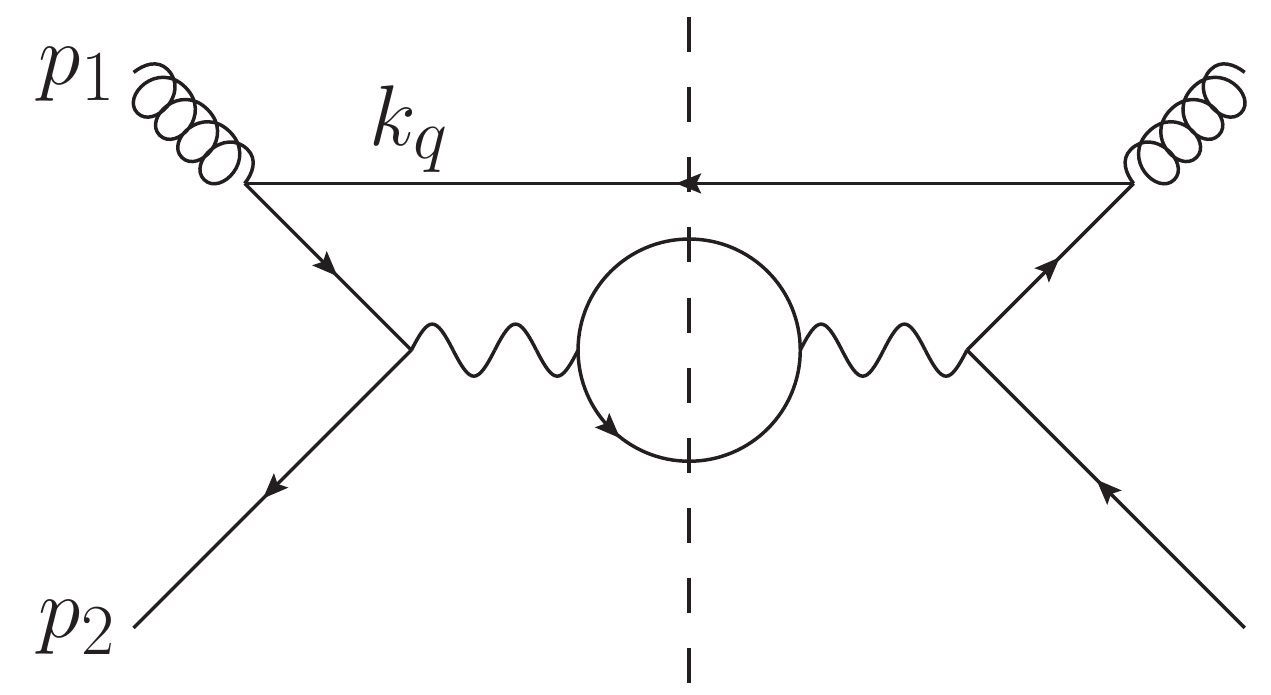}\label{fig:DYLOb}
  		}
  	}
    \caption{(a) LO squared amplitude for the DY production of an off-shell photon, with decay to a lepton pair; (b) the $g\bar{q}$ channel at NLO.}
    \label{fig:DYLO}
  \end{center}
\end{figure}

DY production of a vector boson is a canonical testbed for new resummation ideas, as well as being of phenomenological importance in its own right. Production of SM vector bosons is a key background to many new physics processes, and the production of new heavy bosons via an $s$-channel resonance is an important potential discovery mode of new physics that is actively being probed. For our purposes, we will consider the original DY process of production of an off-shell photon, where the latter decays to a lepton pair, which at LO corresponds to:
\begin{equation}
q(p_1)\,\bar{q}(p_2) \,\rightarrow \, \gamma^*(q)\rightarrow\,e^+(q_1)\, e^-(q_2)\,.
\label{DYproc}
\end{equation}
We quickly review the ingredients for the LO computation of this process, as their definitions will be needed in what follows. The squared amplitude for this process is shown in
figure~\ref{fig:DYLOa}.
The virtuality of the off-shell photon is conventionally written as
\begin{equation}
  Q^2=q^2\,,\quad q=q_1+q_2\,,
  \label{Q2def}
\end{equation}
and we also define the variable
\begin{equation}
  z=\frac{Q^2}{s}\,,
  \label{zdef}
\end{equation}
where $s=(p_1+p_2)^2$ is the partonic centre-of-mass energy. Thus, $z$ represents the fraction of the centre-of-mass energy that is carried by the vector boson, such that $z\rightarrow 1$ corresponds to threshold production. \\

The squared matrix element from figure~\ref{fig:DYLOa} summed (averaged) over final (initial) state colours and spins evaluates to
\begin{eqnarray}
\overline{|{\cal M}_{q\bar{q}\rightarrow \gamma^*}|^2} &=& \frac{1}{2C_A}
e^4 e_q^2 \left(\cos(2\theta)+3\right),
\label{MLODY}
\end{eqnarray}
where $e=\sqrt{4\pi \alpha_{\rm EM}}$ is the electromagnetic coupling, $e_q$ the charge of the quark in units of $e$, and $\theta$ the angle between the $z$-axis and the $e^+$ lepton. We may write the 2-body phase space  of the LO final state in $d=4$ dimensions as
\begin{align}
  \int {\rm d}\Phi^{(2)}(p_1+p_2; q_1, q_2)&=  \frac{1}{16\pi}\int_0^{\pi}{\rm d}\cos\theta\,,
	\label{eq:LOphasespace}
\end{align}
leading to the LO cross section:
\begin{eqnarray}
\sigma_{q\bar{q}\rightarrow \gamma^*} &=& 
\frac{1}{2s}\int {\rm d}\Phi^{(2)}(p_1 + p_2, q_1, q_2)	
\overline{|\mathcal{M}_{q\bar{q}\rightarrow \gamma^*}|^2}  \\
&=& \frac{1}{Q^2} \frac{4\pi \alpha^2_{\rm EM}e_q^2}{3C_A}, \nonumber  
\end{eqnarray}
where we have used the fact that $z=1$ at LO in the second line. In calculating higher-order corrections in what follows, we will normalise to this LO cross section. In particular, the following expression will be useful:
\begin{eqnarray}
	\int {\rm d}\Phi^{(2)}(p_1 + p_2, q_1, q_2) |\mathcal{M}_{q\bar{q}\rightarrow \gamma^*}|^2 = 
	8 Q^2 C_A\, \sigma_{q\bar{q}\rightarrow \gamma^*}\,,
	\label{eq:LOcontribution}
\end{eqnarray}
where the left-hand side contains the squared amplitude {\it before} spin and colour averaging. Note that this squared amplitude can be written as a contraction between a hadronic and leptonic tensor, defined as follows
\begin{equation}
|\mathcal{M}_{q\bar{q}\rightarrow \gamma^*}|^2 
= H_{\rm tree}^{\mu\nu} L_{\mu\nu},
\label{MNLO2}
\end{equation}
where 
\begin{equation}
H_{\rm tree}^{\mu\nu} = e^2 e_q^2
{\rm Tr}\left[\slsh{p}_2 \gamma^{\mu}\slsh{p}_1 \gamma^{\nu}\right],
\label{Hmunutree}
\end{equation}
and 
\begin{equation}
L_{\mu\nu} = \frac{e^2}{Q^4}
{\rm Tr}\left[\slsh{q}_1 \gamma_{\mu}\slsh{q}_2 \gamma_{\nu}\right].
\label{Lmunutree}
\end{equation}
We aim to show how LL logarithms can be resummed in the kinematically subleading $g\bar{q}$ channel for DY, which starts at NLP. We now proceed in the same way as before: we first compute the all-order phase space in the NLP approximation, then we examine the NLO $qg$ channel before moving to the all-order results. 

\subsection{All-order phase space for DY production}
\label{sec:DYphasespace}
In considering higher-order corrections to DY production, we will need to integrate over the multiparton phase space for a given number of real emissions. As we discussed for DIS in section~\ref{sec:phasespace}, we must then find suitable variables such that this phase space is tractable. The solution is again to use a Sudakov decomposition for the emitted parton momenta, and the analysis proceeds similarly to section~\ref{sec:phasespace}, albeit with minor changes due to having a different threshold variable. \\

Consider the emission of one additional soft quark with momentum $k_q$ and $n$ additional gluons with momenta $\{k_i\}$ dressing the LO process, which includes already the two leptons in the final state. As is well-known, the $(n+2)$-body phase space can then
be decomposed as follows:
\begin{eqnarray}
	\int {\rm d}\Phi^{(n+3)} = \frac{1}{2\pi}\int {\rm d}Q^2 \int {\rm d}\Phi^{(n+2)}(p_1+p_2; q, k_1 \dots k_n, k_q) \int {\rm d}\Phi^{(2)}(q; q_1, q_2)\,, 
\label{phasespacefac}
\end{eqnarray}
which has a straightforward physical interpretation. The second integral on the right-hand side is over the intermediate phase space of the off-shell photon (with fixed virtuality) and additional partons; the third integral corresponds to the decay of the photon into the lepton pair. Finally, one must integrate over all virtualities for the photon, and include an appropriate normalisation factor. Considering the second integral, we may evaluate this further as (suppressing the arguments for brevity)
\begin{eqnarray}
	\int  {\rm d}\Phi^{(n+2)} &=& (2\pi)^d \int \frac{ {\rm d} ^d q}{(2\pi)^{d-1}} \delta_+(q^2-Q^2) \left[\prod_{i=1}^n \int \frac{ {\rm d} ^d k_i}{(2\pi)^{d-1}} \delta_+(k_i^2)\right]\nonumber \\
	&& \hspace{1.5cm} \times \int \frac{ {\rm d} ^d k_q}{(2\pi)^{d-1}} \delta_+(k_q^2) \, \delta^{(d)}\left(p_1+p_2-q-k_q-\sum_{i=1}^n k_i\right).\,\,\,\,\,\,\,\,\,\,\,\,\,\,\,
\end{eqnarray}
We may then take $p_1$ and $p_2$ as the null vectors in our Sudakov decomposition, writing each parton momentum (including the quark momentum) as
\begin{equation}
  k_i=\bar{\alpha}_i p_1+\bar{\beta}_i p_2 +k_{i,\perp},\quad
  k_{i,\perp}\cdot p_1=k_{i,\perp}\cdot p_2=0\,.
  \label{ki}
\end{equation}
The $\delta^{(d)}$-function can be removed using the $ {\rm d} ^dq$ integral. The overall $\delta_+$ function then becomes
\begin{eqnarray}
\delta_+\left(\left(p_1+p_2-\sum_{i=1}^n k_i\right)^2-Q^2\right) = \frac{1}{s}\delta(1-z - \sum_{i=1}^n\bar{\alpha}_i- \sum_{i=1}^n\bar{\beta}_i)\,,
\end{eqnarray}
where, similarly to eq.~(\ref{deltaarg}), we have neglected terms that are quadratic in the soft parton momenta. Transforming from the usual momentum components to the Sudakov variables, one finds
\begin{eqnarray}
{\rm d}^d k_i = \frac{s}{4} {\rm d}\bar{\alpha}_i{\rm d}\bar{\beta}_i \left({\bf k}^2_{i,\perp}
\right)^{\frac{d-4}{2}} {\rm d} {\bf k}^2_{i,\perp} {\rm d}  \Omega^{(i)}_{d-2}\,.
\end{eqnarray}
As in the DIS analysis, all matrix elements we encounter will not depend on transverse angles, so that we can simply replace the differential solid angles with their integrated results of eq.~(\ref{omegad-2}). We find
\begin{eqnarray}
\int  {\rm d} \Phi^{(n+2)} &=& \frac{2\pi}{s} s^{n+1+(n+1)\frac{d-4}{2}} \frac{1}{(4\pi)^{(n+1)\frac{d}{2}}}\frac{1}{\Gamma^{n+1}\left(\frac{d-2}{2}\right)} \left[\prod_{i=1}^n \int {\rm d}\bar{\alpha}_i  {\rm d}\bar{\beta}_i \left(\bar{\alpha}_i\bar{\beta}_i\right)^{\frac{d-4}{2}}\right]  \nonumber \,\,\,\,\,\,\\
&& \hspace{2cm} \times \,\int {\rm d}\alpha_q  {\rm d}\beta_q \left(\alpha_q \beta_q\right)^{\frac{d-4}{2}} \delta^+(1-z  -\alpha_q -\beta_q - \sum_{i=1}^n(\bar{\alpha}_i+\bar{\beta}_i)).\,\,\,\,\,\,\,\,\,\,\label{eq:phasespaceandq}
\end{eqnarray}
This may be used in eq.~\eqref{phasespacefac} together with eq.~\eqref{eq:LOphasespace}.
  
\subsection{The $qg$ channel at NLO}
\label{sec:gqNLO}
To show how LL logarithms can be resummed in the $g\bar{q}$ channel for DY, we will use similar arguments to those used in the DIS analysis of section~\ref{sec:DIS}. There, we heavily made use of the results of refs.~\cite{Gribov:1972ri,Gribov:1972rt,Dokshitzer:1977sg} (reviewed in refs.~\cite{Dokshitzer:1978hw,Dokshitzer:1991wu}) to greatly streamline the effort involved in calculating all-order matrix elements. The key idea was to make a particular reference vector choice for the gluon polarisation sum, which in turn led to only pure ladder graphs being relevant for the real emission contributions at arbitrary order. As discussed in detail in refs.~\cite{Dokshitzer:1978hw,Dokshitzer:1991wu}, this idea readily generalises to DY production, and we will choose $p_2$ as our reference vector~\footnote{In fact,
refs.~\cite{Dokshitzer:1978hw,Dokshitzer:1991wu} advocate the use of a more general reference vector, involving a superposition of $p_1$ and $p_2$. Whilst this leads to a more physical interpretation of the resulting Feynman diagrams, it makes the phase space integrals more difficult, hence we will not adopt this here.}. With this choice, the squared matrix element of figure~\ref{fig:DYLOb} evaluates to
\begin{eqnarray}
	\overline{|\mathcal{M}_{g\bar{q}\rightarrow \gamma^*\bar{q}}|^2} &=& \frac{1}{2(d-2)} \frac{e^2 e_q^2 g_s^2 \mu^{4-d}C_F}{C_A^2-1} L_{\mu \nu} \frac{1}{(2p_1\cdot k_q)^2} \notag \\
	&& \times  {\rm Tr}\left[\slsh{p}_2\gamma^{\mu}\slsh{p}_1\gamma^{\sigma}\slsh{k}_q\gamma^{\rho}\slsh{p}_1\gamma^{\nu}\right] \left(-g_{\sigma \rho}+\frac{p_{1,\sigma}p_{2,\rho}+p_{2,\sigma}p_{1,\rho}}{p_1\cdot p_2}\right) \nonumber \\
	&=& \frac{T_R}{2C_A} \frac{g_s^2\mu^{4-d} }{2p_1 \cdot k_q}  |\mathcal{M}_{q\bar{q}\rightarrow \gamma^*}|^2\,.
        \label{MNLO}
\end{eqnarray}
Defining Sudakov variables for the emitted quark momentum via
\begin{equation}
  k_q=\alpha_q p_1+\beta_q p_2+k_{q,\perp}\,,
  \label{kqsuds}
\end{equation}
we obtain
\begin{eqnarray}
\overline{|\mathcal{M}_{g\bar{q}\rightarrow \gamma^*
		\bar{q}}|^2}  &=&  
\frac{4\pi \alpha_s\mu^{4-d} \, T_R}{2 C_A}\frac{1}{s} \frac{1}{\beta_q}|\mathcal{M}_{q\bar{q}\rightarrow \gamma^*}|^2   \nonumber .
\end{eqnarray}
We may integrate this over the phase space using
eq.~\eqref{phasespacefac} for the case $n=0$. Including also the flux factor and the integral representation of the $\delta$ function (eq.~\eqref{deltaid}), one finds a cross section
\begin{eqnarray}
\sigma_{g\bar{q}\rightarrow \gamma^*\bar{q}} 
&=& \frac{1}{2s} \int \frac{ {\rm d} Q^2}{s} \frac{1}{16\pi} \int   {\rm d} \cos\theta |\mathcal{M}_{q\bar{q}\rightarrow \gamma^*}|^2 s^{\frac{d-4}{2}} \frac{4\pi \alpha_s}{(4\pi)^{\frac{d}{2}}}\frac{\mu^{4-d}}{\Gamma\left(\frac{d-2}{2}\right)} 
\frac{T_R}{2 C_A }\nonumber \\
&&\hspace{2cm} \times \int_{-i\infty}^{+i\infty} \frac{ {\rm d} T}{2\pi i}{\rm e}^{T(1-z)}\int  {\rm d} \alpha_q \, {\rm e}^{-T\alpha_q}\, \left(\alpha_q\right)^{\frac{d-4}{2}} \int  {\rm d} \beta_q \, {\rm e}^{-T\beta_q}\,  \left(\beta_q\right)^{\frac{d-6}{2}}\,.\,\,\,\,\,\,\,\,\,\,\,\,\,\,\,\,\,
\end{eqnarray}
The required integrals are straightforward using the methods of section~\ref{sec:DIS}. After normalising to the LO cross section using eq.~(\ref{eq:LOcontribution}) one finds
\begin{eqnarray}
	\label{eq:NLOqgresult}
	\frac{1}{\sigma_{q\bar{q}\rightarrow \gamma^*}}\frac{ {\rm d} \sigma_{g\bar{q}\rightarrow \gamma^*\bar{q}}}{ {\rm d} z}&=&
\left(\frac{\alpha_s}{4\pi}\right)
\left[-\frac{2T_R}{\epsilon}(1-z)^{-2\epsilon} \right]+\ldots,
\end{eqnarray}
where we have kept only the LL dependence at NLP. Following ref.~\cite{Presti:2014lqa}, we will expand the cross section normalised to the LO result as in eq.~(\ref{Xdef}), writing
\begin{equation}
  \frac{1}{\sigma_{q\bar{q}\rightarrow \gamma^*}}\frac{{\rm d} \sigma_{{\rm DY}, g\bar{q}}}{{\rm d} z}
  \equiv {W}_{{\rm DY},g\bar{q}}=\sum_{n=1}^\infty a_s^n
  {W}^{(n)}_{{\rm DY},g\bar{q}}\,,
  \label{WDYdef}
\end{equation}
where we have shortened the notation $\sigma_{g\bar{q}\rightarrow \gamma^* \bar{q} g_1 \dots g_n}$ to $\sigma_{{\rm DY}, g\bar{q}}$. 
From eq.~(\ref{eq:NLOqgresult}), we then find the Mellin space result
\begin{equation}
  {W}^{(1)}_{{\rm DY},g\bar{q}}(N)=-\frac{2 T_R}{\epsilon}\frac{N^{2\epsilon}}
  {N}.
\label{WDY1}
\end{equation}

\subsection{All-order form for the $g\bar{q}$ cross section}
\label{sec:qgallorder}

\begin{figure}
  \begin{center}
  	
  	\mbox{	\captionsetup[subfigure]{oneside,margin={0.8cm,0cm},skip=0.5cm}
  		\subfloat[]{
  			\includegraphics[width=0.30\textwidth]{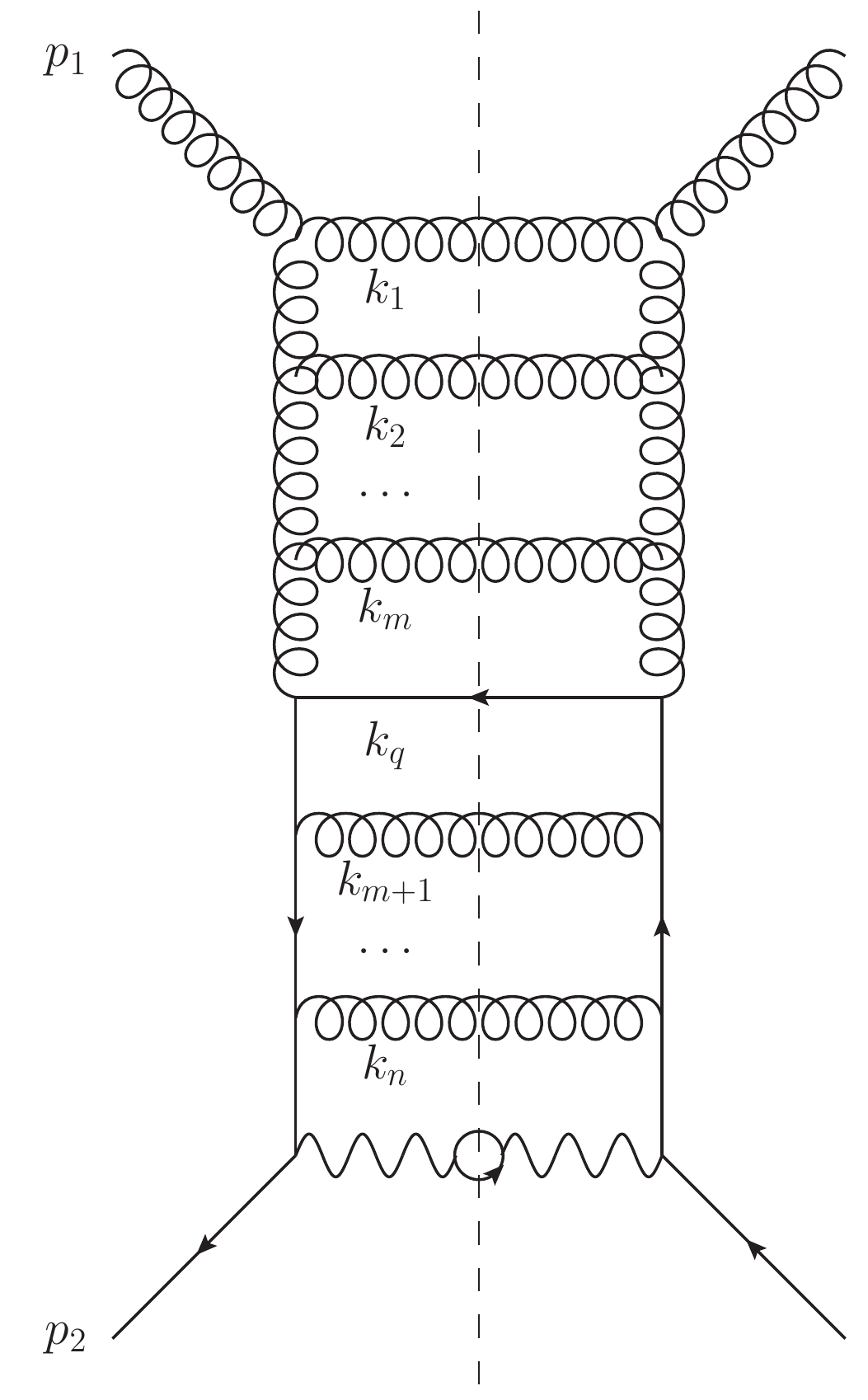}	\label{fig:DYladdera}
  		}
  	
  		\hspace{2.0cm}
  		\captionsetup[subfigure]{oneside,margin={0.8cm,0cm},skip=0.5cm}
  		\subfloat[]{
  			\includegraphics[width=0.30\textwidth]{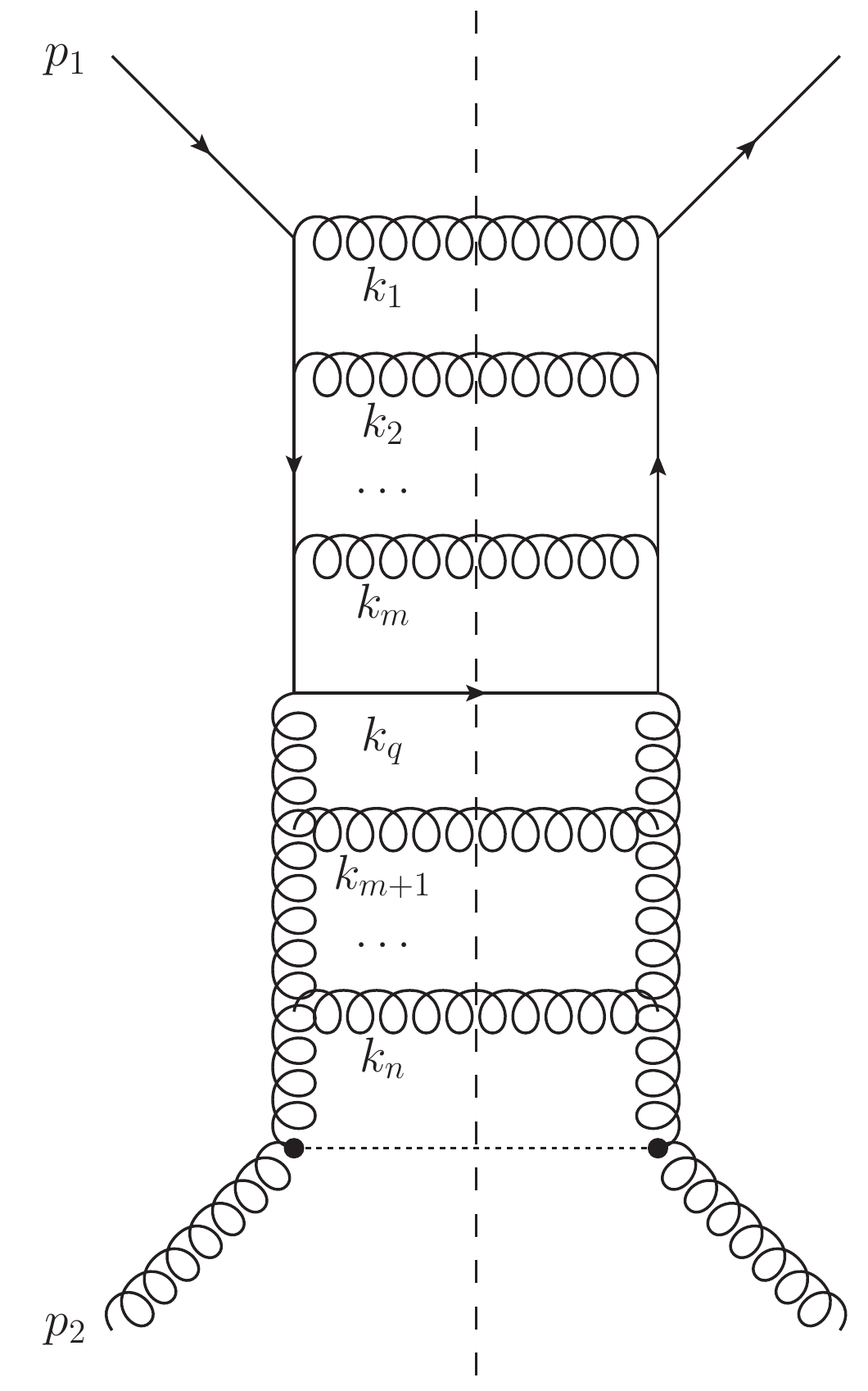}	\label{fig:DYladderb}
  		}
  	}
    \caption{(a) Ladder diagram contributing to the $g\bar{q}$ channel in DY production; (b) similar but for the $qg$ channel in Higgs production.}
    \label{fig:DYladder}
  \end{center}
\end{figure}

Having calculated the NLO result for the $g\bar{q}$ channel using the choice of $p_2$ as the reference vector, let us now generalise the calculation to higher orders. As for the DIS case, the emission of a soft quark at NLO has already placed us at next-to-leading power in the threshold variable, so that we need only include further emission of soft gluons at higher orders. As discussed above, only ladder graphs will be relevant. Furthermore, our choice of $p_2$ as a reference vector means that only gluons emitted from the upper half of the squared amplitude in figure~\ref{fig:DYLOb} will contribute~\footnote{It is for this reason that refs.~\cite{Dokshitzer:1978hw,Dokshitzer:1991wu} advocated using a more general reference vector, that interpolates between axial gauges in the upper and lower halves of the amplitudes, making each half of the DY amplitude look more DIS-like.}. Indeed, applying eikonal Feynman rules to emissions in the lower part of the diagram results in a vanishing factor
\begin{equation}
  p_2^\mu p_2^\nu \left(-\eta_{\mu\nu}+\frac{p_{2,\mu}k_\nu
  +p_{2,\nu}k_\nu}{p_2\cdot k}\right)=0\,.
\end{equation}
The most general ladder diagram we have to consider is shown in figure~\ref{fig:DYladdera}, 
and contributes to the summed-and-averaged matrix element
\begin{eqnarray}
	\overline{|\mathcal{M}_{{\rm DY},g\bar{q}}|^2} &=& e_q^2  \frac{C_A^m C_F^{n-m+1}}{C_A^2-1}\frac{g_s^{2(n+1)}\mu^{(4-d)(n+1)}}{2(d-2)} \notag\\
	&& \times \, \frac{L_{\mu\nu}}{\left(2 (p_1 \cdot (k_1+\dots k_m + k_q)\right)^2} {\rm Tr}\left[\slsh{p}_2\gamma^{\mu}\slsh{p}_1\gamma^{\rho}\slsh{k}_q\gamma^{\sigma}\slsh{p}_1\gamma^{\nu}\right] \nonumber \\
	&& \times \, \frac{p_1^{\mu_1}}{p_1 \cdot k_1} \dots \frac{p_1^{\mu_m}}{p_1 \cdot (k_1+\dots k_m)}\frac{p_1^{\mu_{m+1}}}{p_1 \cdot (k_1+\dots k_{m+1}+k_q)}   \dots \frac{p_1^{\mu_{n}}}{p_1 \cdot (k_1+\dots k_{n}+k_q)}  \nonumber  \\
	&& \times\,  \frac{p_1^{\nu_1}}{p_1 \cdot k_1} \dots \frac{p_1^{\nu_m}}{p_1 \cdot (k_1+\dots k_m)}\frac{p_1^{\nu_{m+1}}}{p_1 \cdot (k_1+\dots k_{m+1}+k_q)}   \dots \frac{p_1^{\nu_{n}}}{p_1 \cdot (k_1+\dots k_{n}+k_q)}  \nonumber \\
	&& \times\, \left(-g_{\rho \sigma} + \frac{p_{2,\rho}p_{1,\sigma}+p_{2,\sigma}p_{1,\rho}}{p_2 \cdot p_1}\right) \prod_{i=1}^n \left(-g_{\mu_i \nu_i} + \frac{p_{2,\mu_i}k_{i,\nu_i}+p_{2,\nu_i}k_{i,\mu_i}}{p_2 \cdot k_i}\right).\label{eq:qgallorder1}
\end{eqnarray}
This can be simplified enormously by realizing that in the the $-g_{\mu_i\nu_i}$ contributions vanish as $p_1^2 = 0$, and that the trace times the polarization sum for the incoming gluon evaluate to
\begin{align}
  &{\rm Tr}\left[\slsh{p}_2\gamma^{\mu}\slsh{p}_1\gamma^{\rho}\slsh{k}_q\gamma^{\sigma}\slsh{p}_1\gamma^{\nu}\right] \left(-g_{\rho \sigma} + \frac{p_{2,\rho}p_{1,\sigma}+p_{2,\sigma}p_{1,\rho}}{p_2 \cdot p_1}\right)= (d-2) 2p_1\cdot k_q \,\frac{H^{\mu\nu}}{e_q^2}\,.
\end{align}
Introducing again the Sudakov variables, we find
\begin{eqnarray}
		\overline{|\mathcal{M}_{{\rm DY},g\bar{q}}|^2}  &=& T_R C_A^{m-1} C_F^{n-m}\, g_s^{2(n+1)}\mu^{(4-d)(n+1)} \frac{2^{2(n+1)}}{8s^{n+1}}  |\mathcal{M}_{q\bar{q}\rightarrow \gamma^*}|^2\, \beta_q \left[\prod_{i=1}^n \frac{\bar{\beta}_i}{\bar{\alpha}_i}\right] \notag \\
		&& \times   \frac{1}{(\bar{\beta}_1)^2 \dots (\bar{\beta}_1 + \dots + \bar{\beta}_m)^2  (\bar{\beta}_1 + \dots + \bar{\beta}_m + \beta_q)^2 \dots (\bar{\beta}_1 + \dots + \bar{\beta}_n + \beta_q)^2 }\,.\,\,\,\,\,\,\,\,\,\,\label{eq:qgallorder2} 
\end{eqnarray}
To efficiently carry out the phase space integral, we can apply a similar trick to the DIS case of section~\ref{sec:DIS}, and symmetrise over crossed-ladder contributions, given that genuinely crossed ladders will be kinematically subleading~\footnote{This statement has been checked explicitly at NNLO. Moreover, if one computes the phase-space integrals without the symmetrisation over crossed-ladder contributions, one arrives at the same answer as demonstrated in appendix~\ref{app:DYphasespace}.}. The result is
\begin{align}
	\overline{|\mathcal{M}_{{\rm DY},g\bar{q}}|^2}  &= 
	T_R C_A^{m-1} C_F^{n-m}\, g_s^{2(n+1)} \mu^{(4-d)(n+1)} \frac{2^{2(n+1)}}{8s^{n+1}}
	\frac{1}{(n+1)!}
	|\mathcal{M}_{q\bar{q}\rightarrow \gamma^*}|^2
	\frac{1}{\beta_q}\prod_{i=1}^n\frac{1}{\bar{\alpha}_i\bar{\beta}_i}\,.
	\label{Mgqcalc}
\end{align}
We must combine this with the phase space of
eq.~\eqref{eq:phasespaceandq}. Upon rewriting the delta function according to eq.~(\ref{deltaid}), one obtains
\begin{align}
	\int{\rm d}\Phi^{(n+2)} \, \overline{|\mathcal{M}_{{\rm DY},g\bar{q}}|^2}  & \,= \,T_R C_A^{m-1} C_F^{n-m}\left(\frac{\alpha_s}{4\pi}\right)^{n+1}\, 2\pi\, 
	\frac{2^{2(n+1)}}{8s}
	\frac{1}{(n+1)!}\left(\frac{\mu^2}{s}\right)^{\frac{(4-d)(n+1)}{2}}
	|\mathcal{M}_{q\bar{q}\rightarrow \gamma^*}|^2\notag\\
	&\hspace{3cm} \times\, 
	\frac{\Gamma^{2n+1}\left(\frac{d-4}{2}\right)
		\Gamma\left(\frac{d-2}{2}\right)}{\Gamma\left(n(d-4)+d-3\right)}(1-z)^{-(n+1)(d-4)}\,.
	\label{Mgqcalc2}
\end{align}
Combining this with the remaining integrals in
eq.~(\ref{phasespacefac}) and using eq.~(\ref{eq:LOcontribution}), one finds that the normalised contribution to eq.~(\ref{WDYdef}) at ${\cal
  O}(\alpha_s^{n+1})$ is
\begin{equation}
  {W}_{{\rm DY},g\bar{q}}^{(n+1)}(z)=-\left(\sum_{m=0}^n C_A^m C_F^{n-m}\right)
  \frac{1}{(n+1)!}\frac{2
    T_R}{\epsilon^{2n+1}}4^n(1-z)^{-2\epsilon(n+1)}+\ldots,
  \label{WDYallorder1}
\end{equation}
where we have summed over all possible ladders, and kept LL terms only. In Mellin space, this result may be written as
\begin{equation}
   {W}_{{\rm DY},g\bar{q}}^{(n+1)}(N)=-\left(\sum_{m=0}^n C_A^m C_F^{n-m}\right)
  \frac{1}{(n+1)!}\,\frac{1}{N}\,\frac{2
    T_R N^{2\epsilon}}{\epsilon}\, \left(\frac{4N^{2\epsilon}}{\epsilon^2}
  \right)^{n}+\ldots,
  \label{WDYallorder2}
\end{equation}
which may be resummed via eq.~\eqref{colid} into the closed form
\begin{equation}
   {W}_{{\rm DY},g\bar{q}}\Big|_{\rm LL}=-\frac{2a_s T_R N^{2\epsilon}}{\epsilon}\frac{1}{N}
  \left(\frac{4a_s N^{2\epsilon}}{\epsilon^2}\right)^{-1}
  \frac{1}{C_F-C_A}\left\{\exp\left[\frac{4a_s C_F N^{2\epsilon}}
    {\epsilon^2}\right]-\exp\left[\frac{4a_s C_A N^{2\epsilon}}
    {\epsilon^2}\right] \right\}.
  \label{WDYgqLL}
\end{equation}
This is the pure real emission contribution, and must be complemented by virtual corrections. As in section~\ref{sec:DIS}, we may fix these using soft gluon unitarity, i.e.~the requirement that sufficiently
divergent infrared contributions must cancel when real and virtual corrections are combined, leaving only those collinear poles which can be absorbed into the parton distribution functions. The argument here
is necessarily more complicated than that of section~\ref{sec:DIS}, however, as each double pole in $\epsilon$ appearing in eq.~(\ref{WDYgqLL}) is accompanied by a factor of $N^{2\epsilon}$ rather than $N^\epsilon$ as a result of the different phase space in DY as opposed to DIS. We may modify each such pole according to
\begin{equation}
  \frac{N^{2\epsilon}}{\epsilon^2}\rightarrow
  \frac{N^{2\epsilon}+\lambda_i N^\epsilon
    +\lambda_j}{\epsilon^2}
  \label{doublepolemod}
\end{equation}
for some constants $\lambda_i$ and $\lambda_j$. Furthermore, we may again allow for an overall multiplicative factor involving both $C_A$ and $C_F$, which motivates the following ansatz:
\begin{align}
   {W}_{{\rm DY},g\bar{q}}\Big|_{\rm LL}&=-\frac{2a_s T_R N^{2\epsilon}}{\epsilon}\frac{1}{N}
  \exp\left[\frac{4a_s[C_F(\lambda_1 N^\epsilon+\lambda_2)
    +C_A(\lambda_3 N^\epsilon+\lambda_4)]}{\epsilon^2}\right] \notag \\
& \times 
  \left(\frac{4a_s (N^{2\epsilon}+\lambda_5 N^\epsilon+\lambda_6)}
       {\epsilon^2}\right)^{-1} \frac{1}{C_F-C_A} \notag\\
  &\times   \left\{\exp\left[\frac{4a_s C_F
           (N^{2\epsilon}+\lambda_7 N^\epsilon+\lambda_8)}
         {\epsilon^2}\right]-\exp\left[\frac{4a_s C_A
           (N^{2\epsilon}+\lambda_9 N^\epsilon+\lambda_{10})}
    {\epsilon^2}\right] \right\}.
  \label{WDYgqLL2}
\end{align}
One can directly simplify this expression by looking at the possible colour structures at each order. At the first order in $a_s$, there are no virtual corrections needed. At this order, the only possible colour structure (after factoring out the Born one) is proportional to $T_R$, which can only be created by eq.~\eqref{WDYgqLL2} by setting $\lambda_{9}+\lambda_{10} = \lambda_7 + \lambda_8$, whereas $\lambda_{9}+\lambda_{10} = \lambda_7 + \lambda_8 = \lambda_5+\lambda_6$ is needed to have the correct normalisation of the real emission contribution. Expanding to $\mathcal{O}(a_s^2)$ furthermore requires that $\lambda_7 = \lambda_9$ (and therefore that $\lambda_{10} = \lambda_8$), whereas at $\mathcal{O}(a_s^3)$ we find the constraint $\lambda_7 = \lambda_5$ (and therefore $\lambda_8 = \lambda_6$). We then arrive at the reduced ansatz
\begin{align}
	{W}_{{\rm DY},g\bar{q}}\Big|_{\rm LL}&=-\frac{2a_s T_R N^{2\epsilon}}{\epsilon}\frac{1}{N}
	\exp\left[\frac{4a_s[C_F(\lambda_1 N^\epsilon+\lambda_2)
		+C_A(\lambda_3 N^\epsilon+\lambda_4)]}{\epsilon^2}\right]\notag\\
	&\times 
	\left(\frac{4a_s (N^{2\epsilon}+\lambda_5 N^\epsilon+\lambda_6)}
	{\epsilon^2}\right)^{-1}\notag\\
	&\times     \frac{1}{C_F-C_A}\left\{\exp\left[\frac{4a_s C_F
		(N^{2\epsilon}+\lambda_5 N^\epsilon+\lambda_6)}
	{\epsilon^2}\right]-\exp\left[\frac{4a_s C_A
		(N^{2\epsilon}+\lambda_5 N^\epsilon+\lambda_{6})}
	{\epsilon^2}\right] \right\}.
	\label{WDYgqLL2b}
\end{align}
Contrary to the DIS case, here we find that expanding eq.~(\ref{WDYgqLL2b}) and requiring that higher-order poles vanish is not quite sufficient to fix all of the coefficients. However, there is more information that we can use. Mass factorisation implies that, at NLP order, the unfactorised cross section can be written as
\begin{equation}
  {W}_{{\rm DY},g\bar{q}}=\widetilde{C}_{{\rm DY},g\bar{q}}Z_{qq}Z_{gg}
  +\widetilde{C}_{{\rm DY},q\bar{q}} Z_{qg}Z_{qq}\,,
  \label{massfacDY}
\end{equation}
where we have introduced the relevant infrared finite coefficient functions $\widetilde{C}_{{\rm DY},ij}$, and the transition functions $Z_{ij}$, which are already known from the DIS analysis in section~\ref{sec:DIS}. Given that both $\widetilde{C}_{{\rm DY},g\bar{q}}$ and $Z_{qg}(=Z_{\bar{q}g})$ start at NLP, it is sufficient to know $\widetilde{C}_{{\rm DY},q\bar{q}}$ at LP, which is given by standard resummation as
\begin{equation}
  \widetilde{C}_{{\rm DY},q\bar{q}}=\exp\left[\frac{4a_s
      C_F(N^{2\epsilon}-1-2\epsilon\ln N)}{\epsilon^2}\right].
  \label{CDYqq}
\end{equation}
Rearranging eq.~(\ref{massfacDY}), we obtain the constraint
\begin{equation}
  \widetilde{C}_{{\rm DY},g\bar{q}}=\frac{{W}_{{\rm DY},g\bar{q}}}{Z_{qq}Z_{gg}}
  -\frac{\widetilde{C}_{{\rm DY},q\bar{q}}Z_{qg}}{Z_{gg}}\equiv {\cal O}(\epsilon^0)\,.
  \label{CDYgq}
\end{equation}
We have quoted the all-order LL form for $Z_{qq}(=Z_{\bar{q}\bar{q}})$ in eq.~(\ref{Zqq}), and
its counterpart for the gluon is
\begin{equation}
  Z_{gg}=\exp\left[\frac{4a_s C_A\ln N}{\epsilon}\right].
  \label{Zggform}
\end{equation}
The ratio $Z_{gq}/Z_{qq}$ may be obtained directly from eq.~(\ref{TZ3}): since the $\widetilde{C}_{\phi,q}$ coefficient is necessarily of $\mathcal{O}(\epsilon^0)$ and the transition functions capture all the poles in $\epsilon$, we find that 
\begin{equation}
  \frac{Z_{gq}}{Z_{qq}}=\sum_{n=1}^\infty
  \left.\frac{a_s^n}{2N\ln N}
  \frac{C_F}{C_F-C_A}\frac{[4(C_A-C_F)]^n\ln^n N}{n!\epsilon^n}
  f(-\epsilon\ln N)
  \right|_{\rm poles}.
  \label{Zgqoverqq}
\end{equation}
The explicit amplitude results of section~\ref{sec:DIS} imply that we can find a similar equation for the combination appearing in eq.~(\ref{CDYgq}) by relabelling $q\leftrightarrow g$, replacing $C_F$ by
$T_R$ in the numerator of the prefactor in eq.~(\ref{Zgqoverqq}), and replacing $C_A\leftrightarrow C_F$ elsewhere:
\begin{equation}
  \frac{Z_{qg}}{Z_{gg}}=\sum_{n=1}^\infty
  \left.\frac{a_s^n}{2N\ln N}
  \frac{T_R}{C_A-C_F}\frac{[4(C_F-C_A)]^n\ln^n N}{n!\epsilon^n}
  f(-\epsilon\ln N)
  \right|_{\rm poles}.
  \label{Zqgovergg}
\end{equation}
We may now substitute eqs.~(\ref{Zqq}, \ref{WDYgqLL2}, \ref{CDYqq}, \ref{Zggform}, \ref{Zqgovergg}) into eq.~(\ref{CDYgq}), and expand eq.~(\ref{CDYgq}) to fixed order to constrain the coefficients $\lambda_i$ by requiring that $\tilde{C}_{{\rm DY}, q\bar{q}}$ is of $\mathcal{O}(\epsilon^0)$. At $\mathcal{O}(a_s^2)$ we constrain $\lambda_1$ and $\lambda_2$ to be 
\begin{eqnarray}
\lambda_1 = -\frac{C_A(2\lambda_3 + \lambda_5 + 1)+C_F(\lambda_5-1)}{2C_F}\,,\quad \lambda_2 = -\frac{C_A(2\lambda_4 + \lambda_6)+C_F(\lambda_6+2)}{2C_F}\,.
\end{eqnarray}
At $\mathcal{O}(a_s^3)$ we find the additional constraints 
\begin{eqnarray}
\lambda_5 = \lambda_6 - 1\,, \quad \lambda_6 = 0\,.
\end{eqnarray}
After this order we do not find any new constraints, as the $\lambda_3$ and $\lambda_4$ coefficients are cancelled from $W_{{\rm DY}, g\bar{q}}$. We may therefore write the full LL form of the unfactorised $g\bar{q}$ cross section as
\begin{align}
  {W}_{{\rm DY},g\bar{q}}\Big|_{\rm LL}&=
  -\frac{T_R}{2(C_F-C_A)}\frac{1}{N} \frac{\epsilon(N^{\epsilon-1})}
  {N^\epsilon-1}
  \exp\left[\frac{4a_s C_F(N^\epsilon-1)}{\epsilon^2}\right]\notag\\
  &\hspace{3cm} \times
  \left\{\exp\left[\frac{4a_s C_F N^\epsilon(N^\epsilon-1)}
    {\epsilon^2}\right]
  -\exp\left[\frac{4a_s C_A N^\epsilon(N^\epsilon-1)}
    {\epsilon^2}\right]
  \right\}.
  \label{WDYLLcomplete}
\end{align}
We now come back to a point raised already in section~\ref{sec:virtual} concerning our implicit assumption on the exponentiation of the virtual contributions. Some readers may therefore be concerned that our ansatz of eq.~\eqref{WDYgqLL2} is not general enough. An alternative procedure would be to expand each order of the partonic cross section as follows:
\begin{equation}
	{W}_{{\rm DY},g\bar{q}}^{(n)}=\frac{1}{N\epsilon^{2n-1}}\sum_{l=2}^{2n}
	N^{l\epsilon} A^{(n,l)}_{{\rm DY},g\bar{q}},
\end{equation}
where the right-hand side includes all permissible powers of $N^\epsilon$. The coefficient $A^{(n,2n)}_{{\rm DY},g\bar q}$ is fixed from eq.~\eqref{WDYgqLL}, and reads
\begin{eqnarray}
	A^{(n,2n)}_{{\rm DY},g \bar q} = - \frac{1}{n!}\frac{T_R 4^{n}}{2}\sum_{m=0}^{n-1}C_A^m C_F^{n-m-1} = - \frac{2T_R}{n!}\sum_{m=0}^{n-1}(4 C_A)^m (4 C_F)^{n-m-1} \,.
\end{eqnarray}
The remaining coefficients at each order can be fixed using the constraint from eq.~\eqref{CDYgq}. Upon carrying out this exercise, we obtain precisely the coefficients $\{A_{{\rm DY},g\bar q}^{(n,l)}\}$ conjectured in ref.~\cite{Presti:2014lqa}, based on exact calculations of the fixed-order cross section. These coefficients agree with the terms we find from expanding eq.~\eqref{WDYLLcomplete}, which is obtained using the exponentiated ansatz for the virtual corrections. This shows that indeed we do not need to assume the exponentiated nature of the virtual corrections, although the procedure of obtaining $W_{{\rm DY}, g\bar{q}}$ is simplified if we do. Note a similar procedure may be employed for the DIS results of section~\ref{sec:DIS} as well. \\

We may now proceed to also find the NLP LL resummed form for the coefficient function $\widetilde{C}_{{\rm DY},g\bar{q}}$. From eqs.~(\ref{WDYLLcomplete}, \ref{Zqq}, \ref{Zggform}, \ref{CDYqq}), one finds for the first factor appearing on the right-hand side of eq.~\eqref{CDYgq}
\begin{align}
  \frac{{W}_{{\rm DY},g\bar{q}}}{Z_{qq}Z_{gg}}&=-\frac{T_R}{C_F-C_A}
  \frac{f(-\epsilon\ln N)}{2N\ln N}\exp\left[\frac{4a_s C_F(N^{2\epsilon}
      -1-2\epsilon\ln N)}{\epsilon^2}\right]\notag\\
&\hspace{0.5cm} \times  \left\{\exp\left[\frac{4a_s(C_F-C_A)\ln N}{\epsilon}
    \right]
  -\exp\left[\frac{4a_s (C_A-C_F)[N^\epsilon(N^\epsilon-1)-\epsilon\ln N]}
    {\epsilon^2}\right]
  \right\}.
  \label{WZZ}
\end{align}
Recognising that $\widetilde{C}_{{\rm DY},q\bar{q}}$ multiplies the entire result, we see that the only effect of the second term in eq.~(\ref{CDYgq}) is to remove the poles of the first term, and also any contributions from these poles hitting higher-order terms in $\widetilde{C}_{{\rm  DY},q\bar{q}}$. We can thus find the coefficient $\widetilde{C}_{{\rm DY},g\bar{q}}$ by taking the ${\cal O}(\epsilon^0)$ piece of eq.~(\ref{WZZ}), ignoring any higher order terms in $\widetilde{C}_{{\rm DY},q\bar{q}}$ as we do so (n.b.~this argument is similar to finding the coefficient $\widetilde{C}_{\phi,q}$ in eq.~(\ref{Cphiqres})). The result is
\begin{equation}
  \widetilde{C}_{{\rm DY},g\bar{q}}\Big|_{\rm LL}=\frac{T_R}{C_A-C_F}
  \frac{1}{2N \ln N}\left[e^{8C_Fa_s\ln^2 N}{\cal B}_0[4a_s(C_A-C_F)\ln^2 N]
    -e^{(2C_F+6C_A)a_s\ln^2N}\right],
  \label{CDYgqres}
\end{equation}
which precisely matches the conjecture in
ref.~\cite{Presti:2014lqa}. Given that the resummation of LL NLP terms in the kinematically leading channels in DY production has already been established previously, see refs.~\cite{Beneke:2018gvs,Bahjat-Abbas:2019fqa}, this analysis, for the first time, completes the resummation of the DY process at NLP LL order. We examine the closely related Higgs production process in the following section.

\section{Resummation of the $qg$ channel in Higgs boson production}
\label{sec:Higgs}

\begin{figure}
  \begin{center}
  	\mbox{	\captionsetup[subfigure]{oneside,margin={0.8cm,0cm},skip=0.5cm}
  		\subfloat[]{
  			\includegraphics[width=0.35\textwidth]{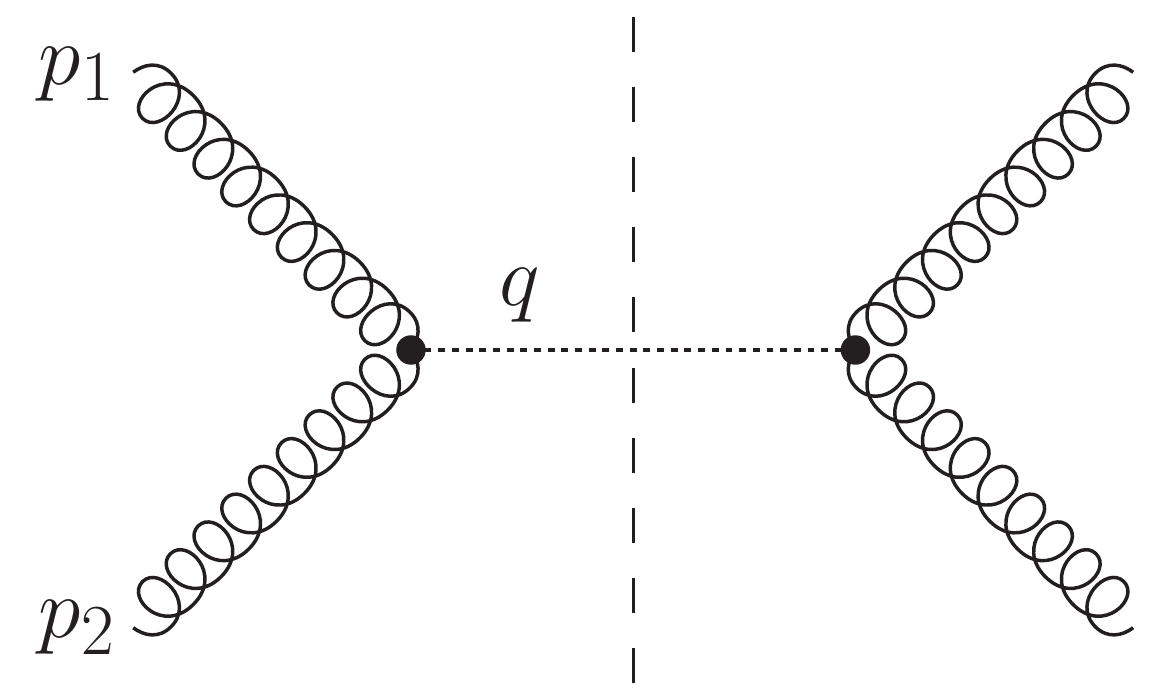}	\label{fig:HiggsprodLOa}
  		}
  	
  		\hspace{2.0cm}
  		\captionsetup[subfigure]{oneside,margin={0.8cm,0cm},skip=0.5cm}
  		\subfloat[]{
  			\includegraphics[width=0.35\textwidth]{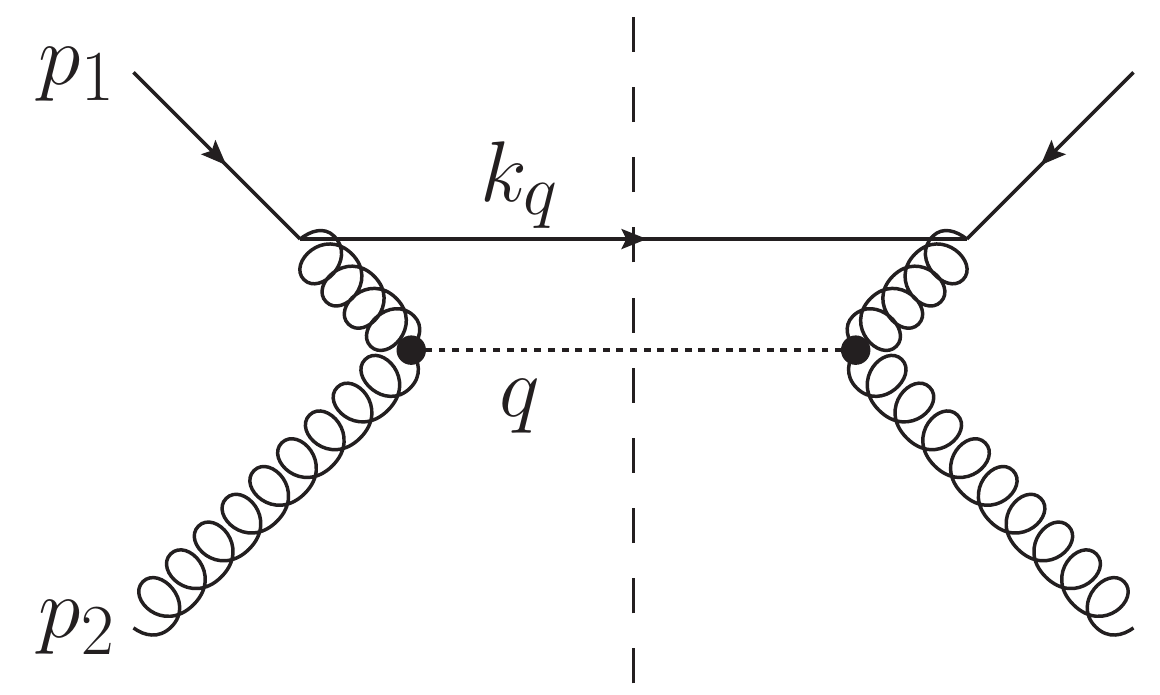}	\label{fig:HiggsprodLOb}
  		}
  	}
    \caption{(a) LO squared amplitude for Higgs boson production via gluon-gluon fusion; (b) the quark-gluon channel at NLO.}
    \label{fig:HiggsprodLO}
  \end{center}
\end{figure}

The final process we will consider in this paper is that of Higgs boson production via gluon-gluon fusion, which is closely related to the DY process from the resummation point of view. The LO squared amplitude is shown in figure~\ref{fig:HiggsprodLOa}, where we assume the same effective coupling as in figure~\ref{fig:DISLOb}.
The summed and averaged LO squared amplitude corresponding to figure~\ref{fig:HiggsprodLOa} is
\begin{equation}
\overline{|{\cal M}_{gg\rightarrow h}|^2}=\frac{|\lambda|^2}{(d-2)^2}\frac{1}{N_c^2-1}
\left(-\eta_{\mu\nu}+\frac{c_\mu p_{1,\nu}+c_{\nu}p_{1,\mu}}
{p_1\cdot c}\right)
\left(-\eta^{\mu\nu}+\frac{c^\mu p_{2}^{\nu}+c^{\nu}p_{2}^{\mu}}
{p_2\cdot c}\right),
\label{HiggsLOamp}
\end{equation}
where $c^\mu$ is the reference vector entering the gluon polarisation sum. In contrast to the DY case of the previous section, we now face the complication that we cannot simply choose $c=p_2$, as the second gluon polarisation sum is then ill-defined. However, it would be desirable to keep this choice for any final-state gluons, given that it means that we only have to consider ladder graphs associated with the upper part of the amplitude. A simple calculational fix is to choose {\it different} reference vectors for initial state and final state gluons, which amounts to defining polarisations differently for incoming or outgoing gluons. There is nothing to forbid this, although Bose symmetry demands that we treat all initial (or all final) gluons on an equal footing. For the initial-state gluons, let us leave $c_\mu$ general, but satisfying the conditions
\begin{equation}
c^2=0,\quad c\cdot p_1=c\cdot p_2=0\,.
\label{cconditions}
\end{equation}
For final state gluons, we will continue to use $p_2$ as the reference vector. \\

With the above general $c^\mu$, eq.~(\ref{HiggsLOamp}) can be simplified to
\begin{equation}
\overline{|{\cal M}_{gg\rightarrow h}|^2}=\frac{|\lambda|^2}{(d-2)}\frac{1}{N_c^2-1}\,,
\label{HiggsLOamp2}
\end{equation}
such that the LO cross section is found to be
\begin{align}
\sigma_{gg\rightarrow h}&\, =\, \frac{1}{2s}\int{\rm d}\Phi^{(1)}\, \overline{|{\cal M}_{gg\rightarrow h}|^2}
=\frac{\pi}{s^2}\frac{|\lambda|^2}{(d-2)}\frac{1}{N_c^2-1}\delta(1-z)\,,
\label{sigmaLOHiggs}
\end{align}
where we have defined 
\begin{equation}
z=\frac{q^2}{s}\equiv\frac{m_H^2}{s}\,.
\label{zdefHiggs}
\end{equation}
As for DY, the threshold variable is simply given by
$\xi=(1-z)$. Thus, we may reuse the same phase space for $n$ soft-gluon and $1$ soft-quark emissions, provided we ignore the additional two-body phase-space due to the decay of the virtual photon into the lepton pair. The relevant phase space is then given by eq.~\eqref{eq:phasespaceandq}.\\

The prefactor in eq.~(\ref{sigmaLOHiggs}) is needed to normalise higher-order contributions. Keeping with the theme of the paper, we will be concerned with the $qg$ channel that opens up for the first time at NLO, as shown in figure~\ref{fig:HiggsprodLOb}. How to resum NLP contributions at LL in the gluon-gluon channel has already been discussed in refs.~\cite{Bahjat-Abbas:2019fqa,Beneke:2019mua}. Analogous to the DY case, the LL contribution to the $qg$ channel at arbitrary order will originate from general ladder diagrams such as that shown in figure~\ref{fig:DYladderb}, which gives a summed and averaged squared matrix element
\begin{align}
\overline{|{\cal M}_{{\rm h}, qg}|^2}&=\frac{g_s^{2(n+1)}\mu^{(4-d)(n+1)}|\lambda|^2}{8(d-2)}
\frac{C_F^{m+1}C_A^{n-m}}{N_c^2-1}{\rm Tr}[\slsh{k}_q\gamma^\alpha
\slsh{p}_1\gamma^\beta]\left(-\eta_{\alpha\beta}
+\frac{c_\alpha p_{2,\beta}+c_\beta p_{2,\alpha}}{p_2\cdot c}\right)\notag\\
&\times \left(\prod_{i=1}^n\frac{2p_1\cdot p_2\,p_1\cdot k_i}
{p_2\cdot k_i}\right)\frac{1}
{[p\cdot k_1]^2\ldots[p\cdot(k_1+\ldots+k_n+k_q)]^2}\,.
\label{Hallorder1}
\end{align}
The second term in the gluon polarisation tensor will not contribute. To see this, note that contracting it with the trace and using the conditions of eq.~(\ref{cconditions}) yields
\begin{equation}
{\rm Tr}[\slsh{k}_q\slsh{c}\slsh{p}_1\slsh{p_2}]
+{\rm Tr}[\slsh{k}_q\slsh{p_2}\slsh{p}_1\slsh{c}]
=8 k_q\cdot c\,p_1\cdot p_2\,.
\label{term2}
\end{equation}
This projects out the transverse components of $k_q$ which, occurring linearly in the phase-space integral, will vanish. Simplifying the remainder of eq.~(\ref{Hallorder1}), we may convert to Sudakov variables and symmetrise over crossed-ladder contributions as in the DY case, finding
\begin{align}
\overline{|{\cal M}_{{\rm h}, qg}|^2}=\frac{g_s^{2(n+1)}\mu^{(4-d)(n+1)}|\lambda|^2C_F^{m+1}C_A^{n-m}}
{N_c^2-1}\frac{4^n s^{-n-1}}{(n+1)!}\frac{1}{\beta_q}\prod_{i=1}^n
\frac{1}{\bar{\alpha}_i\bar{\beta}_i}\,.
\label{Hallorder2}
\end{align}
We must combine this with the phase space of
eq.~\eqref{eq:phasespaceandq} and the flux factor to obtain the cross section. The integrals may all be carried out similarly to in section~\ref{sec:DY}, and after dividing by the LO normalisation in eq.~(\ref{sigmaLOHiggs}) we find the NLP LL contribution
\begin{equation}
{W}_{{\rm h},qg}^{(n+1)}(z)=-\frac{1}{(n+1)!}\frac{2C_F}{\epsilon}
\left(\frac{4}{\epsilon^2}\right)^n (1-z)^{-2\epsilon(n+1)}
\sum_{m=0}^n C_F^m C_A^{n-m}+\ldots,
\label{WHqg}
\end{equation}
which in Mellin space becomes
\begin{equation}
{W}_{{\rm h},qg}^{(n+1)}(N)=-\frac{1}{(n+1)!}\frac{2C_F N^{2\epsilon-1}}
{\epsilon}
\left(\frac{4N^{2\epsilon}}{\epsilon^2}\right)^n\sum_{m=0}^n C_F^m C_A^{n-m}
+\ldots.
\label{WHqgN}
\end{equation}
Perhaps unsurprisingly, this is exactly what one would obtain from the DY formula of eq.~(\ref{WDYallorder2}), upon making the colour replacements we have already seen, namely $T_R\rightarrow C_F$ in the prefactor and $C_F\rightarrow C_A$ elsewhere. The reason for these replacements should hopefully be clear: the replacement in the prefactor reflects the difference in colour structure (but similar kinematics) of the NLO processes. The second replacement follows from having interchanged the gluon and quark backbones of the ladders, and the fact that eikonal Feynman rules depend on the colour representation of the emitting particle, but not its spin. \\

The above replacements allow us to immediately recycle eqs.~(\ref{WDYLLcomplete}, \ref{CDYgqres}) to get the all-order LL form of the $qg$ cross section
\begin{align}
  {W}_{{\rm h},qg}\Big|_{\rm LL}&=
  -\frac{C_F}{2(C_A-C_F)}\frac{\epsilon(N^{\epsilon-1})}
  {N^\epsilon-1}
  \exp\left[\frac{4a_s C_A(N^\epsilon-1)}{\epsilon^2}\right]\notag\\
  &\times
  \left\{\exp\left[\frac{4a_s C_A N^\epsilon(N^\epsilon-1)}
    {\epsilon^2}\right]
  -\exp\left[\frac{4a_s C_F N^\epsilon(N^\epsilon-1)}
    {\epsilon^2}\right]
  \right\},
  \label{WHLLcomplete}
\end{align}
as well as the resummed coefficient function
\begin{equation}
  \widetilde{C}_{{\rm h},qg}\Big|_{\rm LL}=\frac{C_F}{C_F-C_A}
  \frac{1}{2N \ln N}\left[e^{8C_Aa_s\ln^2 N}{\cal B}_0[4a_s(C_F-C_A)\ln^2 N]
    -e^{(2C_A+6C_F)a_s\ln^2N}\right],
  \label{CHqgres}
\end{equation}
Again, these results match the conjectures of
ref.~\cite{Presti:2014lqa}.

\section{Discussion}
\label{sec:discuss}

In this paper, we have investigated the resummation of logarithmically enhanced contributions affecting the threshold production of heavy particles, specifically those contributions which are LL at NLP in the threshold variable. Previous studies have established that such terms can be resummed in those partonic channels which contribute already at LP in the threshold expansion. Here, we have completed the set of LL NLP terms by examining those partonic channels that start at NLP, due to the emission of a soft quark at NLO. In some sense, one expects these contributions to be straightforward to resum, in that they originate from an underlying NLO process which is dressed by purely soft gluon radiation. However, the combinatorics of the resummation is rather intricate, as has been previously discussed in
refs.~\cite{Vogt:2010cv,Presti:2014lqa,Beneke:2020ibj}.\\

Our resummation approach uses well-established diagrammatic arguments~\cite{Gribov:1972ri,Gribov:1972rt,Dokshitzer:1977sg} to efficiently calculate all-order forms for the purely real emission contributions to partonic structure functions and cross sections at LL order. We then fix the virtual corrections using a variant of the soft gluon unitarity procedure that has been applied at leading power~\cite{Catani:1989ne}, namely by requiring that virtual corrections cancel appropriate higher-order poles in the dimensional regularisation parameter $\epsilon$, leaving only those collinear singularities that can be absorbed into the parton distributions. Our all-order forms for the structure functions and cross sections lead straightforwardly to resummed splitting and coefficient functions, once mass factorisation is carried out.\\

As specific examples, we consider DIS induced by
both a virtual photon and a Higgs boson, and also DY and Higgs production. We confirm the conjectures for splitting and coefficient functions that were first given in refs.~\cite{Vogt:2010cv,Presti:2014lqa}. The DIS case was also obtained in ref.~\cite{Beneke:2020ibj}, under the assumption that the one-loop virtual corrections to the underlying NLO processes exponentiate, motivated by a re-factorisation and renormalisation-group evolution in SCET. Our approach is complementary to this, in that we focus primarily on an explicit calculation of the real emission contributions, with the virtual corrections uniquely determined by consistency requirements.\\

It would be very interesting to see if our methods can be applied to other scattering processes of immediate phenomenological interest, where the results could supplement existing LP resummation formulae at NLP LL. We hope that our study might be useful for those working on a variety of approaches to NLP resummation, including regarding how to relate these different methods.

\section*{Acknowledgments}
We are very grateful to Eric Laenen and Lorenzo Magnea for discussion and collaboration on related topics, and to Jort Sinninghe Damst\'{e}, some time ago, for conversations about subleading partonic channels. This work has been supported by the UK Science and Technology Facilities Council (STFC) Consolidated Grant ST/P000754/1 ``String theory, gauge theory and duality'', the STFC grant
452 number ST/T000864/1, and by the European Union
Horizon 2020 research and innovation programme under the Marie Sk\l{}odowska-Curie grant agreement No. 764850 ``SAGEX''. L.V. is supported by Fellini -- Fellowship for Innovation at INFN, funded by the 
European Union's Horizon 2020 research programme under the Marie Sk\l{}odowska-Curie Cofund Action, grant agreement no. 754496. 


\appendix

\section{Direct calculation of ladder integrals}
\label{app:integrals}
\subsection{DIS}
In this appendix, we show how to directly calculate the multiple integral appearing in eq.~(\ref{gluoncalc3}). One may start by transforming from the $\{b_i\}$ variables to the set $\{\xi_i\}$, given by
\begin{align}
\xi_1=b_1\, ,\quad \xi_2=b_1+b_2\, ,\quad \xi_n=b_1+b_2+\ldots b_n\, ,
\label{zidef}
\end{align}
or, in matrix form,
\begin{equation}
\left(\begin{array}{c}\xi_1\\\xi_2\\ \vdots \\ \xi_n\end{array}\right)
=\left(\begin{array}{ccccc}1 & 0 & 0 & \cdots & 0\\
1& 1 & 0 & \cdots & 0\\
\vdots & \vdots & \vdots & \vdots & \vdots \\
1 & 1 & 1 & \cdots & 1\end{array}\right)
\left(\begin{array}{c}b_1\\b_2\\ \vdots \\ b_n\end{array}\right).
\label{zimatrix}
\end{equation}
This allows us to quickly work out the Jacobian which, given the triangular form of the matrix, is simply equal to 1. The inverse transformation is
\begin{align}
b_1=\xi_1\,,\quad b_2=\xi_2-\xi_1\,,\quad \ldots\quad b_i=\xi_i-\xi_{i-1}\,,
\label{biinverse}
\end{align}
so that the $\{b_i\}$ integrals become
\begin{align}
\left(\prod_{i=1}^{n+1}\int {\rm d}\xi_i\right)\frac{\xi_1^{\frac{d-2}{2}}
(\xi_2-\xi_1)^{\frac{d-2}{2}}(\xi_3-\xi_2)^{\frac{d-2}{2}}\ldots
(\xi_{n+1}-\xi_{n})^{\frac{d-2}{2}}}{\xi_1^2\,\xi_2^2\,\ldots\,\xi_{n+1}^2}\,.
\label{ziint}
\end{align}
Given that the singularities are associated with all $b_i\rightarrow 0$, and thus $\xi_i\rightarrow 0$, we are free to set the upper limits of the $\{\xi_i\}$ integrals to arbitrary values. We may now evaluate the integrals in sequence, starting with the $\xi_1$ integral, which is
\begin{align}
\int {\rm d}\xi_1\,  \xi_1^{\frac{d-6}{2}}(\xi_2-\xi_1)^{\frac{d-2}{2}}&=
\xi_2^{d-3}\int {\rm d}\xi_1\,\xi_1^{\frac{d-6}{2}}(1-\xi_1)^{\frac{d-2}{2}}\notag\\
&=\xi_2^{d-3}\frac{\Gamma\left(\frac{d-4}{2}\right)\Gamma
\left(\frac{d}{2}\right)}{\Gamma(d-2)}\,,
\label{z1int}
\end{align}
where we have scaled $\xi_1\rightarrow \xi_1 \xi_2$, and then set the upper limit of the $\xi_1$ integral to 1. Next, we have the $\xi_2$ integral which, including the additional factor of $\xi_2$ from rescaling $\xi_1$, is
\begin{equation}
\int {\rm d}\xi_2\,\xi_2^{d-5}(\xi_3-\xi_2)^{\frac{d-2}{2}}=
\xi_3^{\frac{3d}{2}-5}\frac{\Gamma(d-4)\Gamma\left(\frac{d}{2}\right)}
{\Gamma\left(\frac{3d}{2}-4\right)}\,.
\label{z2int}
\end{equation}
If we carry on in this fashion then it is not too difficult to spot the pattern: the product of the first $n$ $\{\xi_i\}$ integrals gives
\begin{align}
\prod_{i=1}^n \int {\rm d}\xi_i\,\xi_i^{\frac{id}{2}-(2i+1)}(1-\xi_i)^{\frac{d}{2}-1}
&=\prod_{i=1}^n\frac{\Gamma[i(\frac{d}{2}-2)]\Gamma(\frac{d}{2})}
{\Gamma[\frac{(i+1)d}{2}-2i]}\notag\\
&=\frac{1}{n!}\left(-\frac{1}{\epsilon}\right)^n+\ldots\,.
\label{ziints}
\end{align}
There remains the final integral over $\xi_{n+1}$, which is 
\begin{equation}
\int {\rm d}\xi_{n+1}\,\xi_{n+1}^{\frac{(n+1)d}{2}-2(n+1)-1}e^{-\xi_{n+1}}
=\Gamma\left[(n+1)\left(\frac{d-4}{2}\right)\right]
=-\frac{1}{\epsilon(n+1)}+\ldots,
\label{zn+1int}
\end{equation}
so that the full $\{\xi_i\}$ integrals evaluate to
\begin{equation}
\frac{1}{(n+1)!}\left(-\frac{1}{\epsilon}\right)^{n+1}.
\label{fullzi}
\end{equation}
This is exactly the same result as found in eq.~(\ref{biints2}), as claimed.
\subsection{DY}
\label{app:DYphasespace}
Like in the DIS case, we may also prove eq.~\eqref{Mgqcalc} for DY without the symmetrisation of crossed-ladder contributions. We start with eq.~\eqref{eq:qgallorder2} and integrate over the phase space appearing in eq.~\eqref{eq:phasespaceandq}. Introducing the Laplace transformation of the $\delta$ function we find for the cross section and making the replacement of eq.~\eqref{bidef} we find
\begin{eqnarray}
	\sigma_{{\rm DY}, g\bar{q}} &=& \frac{1}{2s} \int \frac{{\rm d}Q^2}{s} \frac{1}{16\pi}\int {\rm d}\cos\theta |\mathcal{M}_{q\bar{q}\rightarrow \gamma^*}|^2 \, T_R C_A^{m} C_F^{n-m} g_s^{2(n+1)}\mu^{(4-d)(n+1)} \\ 
	&& \notag \times \, \int_{-i\infty}^{+i\infty}\frac{{\rm d}T}{2\pi i}{\rm e}^{T(1-z)}  s^{(n+1)\frac{d-4}{2}} \frac{2^{2(n+1)}}{8 (4\pi)^{(n+1)\frac{d}{2}}} \frac{1}{\Gamma^{n+1}\left(\frac{d-2}{2}\right)} 
	\\
	&& \notag 
	\times \, 	\left[\prod_{i=1}^n \int {\rm d}\bar{\alpha}_i  {\rm e}^{-T\bar{\alpha}_i} \left(\bar{\alpha}_i\right)^{\frac{d-6}{2}}\right]  \int {\rm d} \alpha_q  {\rm e}^{-T \alpha_q} \left(\alpha_q\right)^{\frac{d-4}{2}}  \,\,\,\,\,\,\\
	&& \notag  \times \,\left[\prod_{i=1}^{n+1} \int  {\rm d}b_i {\rm e}^{-T b_i} \left(b_i\right)^{\frac{d-2}{2}} \right] \frac{1}{(b_1)^2 \dots (b_1 + \dots + b_{n+1})^2 }.
\end{eqnarray}
Now we scale out the $T$ dependence from the Sudakov variables by making the replacement $\bar{\alpha}_i \rightarrow \frac{1}{T} \bar{\alpha}_i$ (and similary for $\alpha_q$, $b_i$). We find that the $T$-dependence becomes
\begin{eqnarray}
	\int_{-i\infty}^{+i\infty}\frac{{\rm d}T}{2\pi i}{\rm e}^{T(1-z)} T^{-n\frac{d-6}{2}}T^{-\frac{d-4}{2}}T^{-(n+1)\frac{d-2}{2}} = \frac{(1-z)^{(n+1)(d-4)}}{\Gamma\left(d(n+1)-4n-3\right)}\,.
\end{eqnarray}
The $\bar{\alpha}_i$ and $\alpha_q$ integrals may computed easily, whereas we transform the $b_i$ variables to $\xi_i$ as in eq.~\eqref{zidef}. As above, we may do the set of $\xi_i$ integrals in sequence to find the final result 
\begin{eqnarray}
	\frac{1}{\sigma_{q\bar{q}\rightarrow \gamma^*}}\frac{{\rm d}\sigma_{{\rm DY}, g\bar{q}}}{{\rm d}z } &=&  (1-z)^{-2\epsilon(n+1)} \, T_R C_A^{m} C_F^{n-m}\left(\frac{\alpha_s}{4\pi}\right)^{n+1}  \frac{2^{2n+1} }{(n+1)!} \left(-\frac{1}{\epsilon}\right)^{2n+1},\,\,\,\,\,\,\,\,\,\,\,
\end{eqnarray}
where we have neglected terms that come with a higher power of $\epsilon$. Summing over all possible ladder orderings and using \eqref{WDYdef}  to extract $W^{(n+1)}_{{\rm DY}, g\bar{q}}$ we indeed match the result of eq.~\eqref{WDYallorder1}. The fact that we may symmetrise over crossed-ladder contributions and find the same answer indeed proves that these crossed-ladder contributions are kinematically subleading. 

\section{LL form of the $Z_{gq}$ transition function}
\label{app:transition}

In this appendix, we prove eq.~(\ref{Zgqfinal}), which is needed to relate the transition function $Z_{gq}$ to the splitting function $P_{gq}$. Whilst this result may or may not be standard, we were unable to find an alternative proof in the literature, which is our reason for providing one here. We begin by noting that eqs.~(\ref{Peq}, \ref{gamdef}) imply
\begin{equation}
\frac{{\rm d}{\bf Z}}{{\rm d}\ln Q^2}=-{\bf \gamma}{\bf Z}\,.
\label{Zdiff}
\end{equation}
We may rewrite the left-hand side using the expression for the running coupling in $d=4-2\epsilon$ dimensions:
\begin{equation}
\frac{{\rm d} a_s}{{\rm d}\ln Q^2}=-\epsilon a_s+\beta(a_s)\,.
\label{running}
\end{equation}
Using standard arguments, we may ignore the $\beta$ function at LL level, since including it would introduce logarithms of the renormalisation scale, which replace $\ln(N)$ at a given order in $\alpha_s$, and thus result in subleading logarithmic contributions. Then
eq.~(\ref{Zdiff}) becomes
\begin{equation}
\frac{{\rm d}}{{\rm d}a_s}\left(\begin{array}{cc}
Z_{qq} & Z_{qg} \\ Z_{gq} & Z_{gg}
\end{array}\right)
=\frac{a_s}{\epsilon}\left(\begin{array}{cc} \gamma_{qq} & \gamma_{qg}\\
\gamma_{gq} & \gamma_{gg}\end{array}\right)
\left(\begin{array}{cc}
Z_{qq} & Z_{qg} \\ Z_{gq} & Z_{gg}
\end{array}\right),
\label{Zdiff2}
\end{equation}
so that $Z_{gq}$ satisfies
\begin{equation}
\frac{{\rm d} Z_{gq}}{{\rm d}a_s}-\frac{a_s}{\epsilon}\gamma_{gg}Z_{gq}
=\frac{a_s}{\epsilon}\gamma_{gq}Z_{qq}\, .
\label{Zgqeq}
\end{equation}
This is a first-order ordinary differential equation (ODE), and can be solved using the method of integrating factors. That is, given a first-order ODE in the form
\begin{equation}
\frac{{\rm d}y(x)}{{\rm d}x}+P(x)y(x)=Q(x)\,,
\label{intfac1}
\end{equation}
one may define the {\it integrating factor}
\begin{equation}
I(x)=\exp\left[\int^x dx' P(x')\right].
\label{Idef}
\end{equation}
Multiplying this factor with eq.~(\ref{intfac1}) yields
\begin{equation}
I(x)\frac{dy(x)}{dx}+P(x)I(x)y(x)=\frac{d}{dx}\left[I(x)y(x)\right]
=I(x)Q(x)\,,
\label{intfac2}
\end{equation}
which may be straightforwardly integrated to get
\begin{equation}
y(x)=I^{-1}(x)\int^x dx' I(x') Q(x')\,,
\label{intfac3}
\end{equation}
where the lower limit of integration on the right-hand side will be determined by the boundary conditions. Comparing with eq.~(\ref{Zgqeq}), we can immediately write down the solution:
\begin{equation}
Z_{gq}(a_s)=I^{-1}(a_s)\frac{1}{\epsilon}\int_0^{a_s}\frac{{\rm d}a'_s}{a'_s} \,
I(a'_s)\,\gamma_{gq}(a'_s)\,Z_{qq}(a'_s)\,,
\label{Zgqsol1}
\end{equation}
where we have implemented the boundary condition $Z_{gq}(0)=0$.  The integrating factor $I(a_s)$ is given by
\begin{equation}
I(a_s)=\exp\left[-\frac{1}{\epsilon}
\int^{a_s}_0\frac{{\rm d}a'_s}{a'_s}\gamma_{gg}(a'_s)\right].
\label{Igq}
\end{equation}
Note that the lower limit of integration is arbitrary, as it will cancel out on the right-hand side of eq.~(\ref{Zgqsol1}). We may thus choose it to be zero for convenience.  \\

Thus far, all of our statements have been exact. However, we can simplify things by restricting ourselves to LL order, in addition to noting (as above) that we may ignore NLP contributions in the diagonal anomalous dimensions and/or transition functions. Furthermore, the higher-order diagonal anomalous dimensions have the large $N$ behaviour~\footnote{Following ref.~\cite{Vogt:2010cv}, we adopt the $\msbar$ factorisation scheme throughout. The large $N$ behaviour of higher-order anomalous dimensions might be different in other factorisation schemes.}
\begin{equation}
\gamma_{qq,gg}^{(n-1)}(N)\sim \ln(N)+\ldots,
\label{PlargeN}
\end{equation} 
where the ellipsis denotes non-logarithmic or power-suppressed terms. That is, $\gamma_{qq}$ and $\gamma_{gg}$ only have a single logarithmic enhancement at all orders in $a_s$. Hence, as noted already in ref.~\cite{Vogt:2010cv}, leading logarithms in $Z_{gq}$ can only come from keeping the first-order terms in $\gamma_{gg}$ and $\gamma_{qq}$:
\begin{equation}
\gamma_{gg,qq}(a_s) \cong a_s \gamma^{(0)}_{gg,qq}(N)\,,\quad
\gamma_{qq}^{(0)}=4C_F \ln(N)\,,\quad \gamma_{gg}^{(0)}=4C_A\ln(N)\,.
\label{gamggform} 
\end{equation}
Using this result, the integrating factor of eq.~(\ref{Igq}) simplifies to
\begin{equation}
I(a_s)=\exp\left[-\frac{a_s}{\epsilon}\gamma_{gg}^{(0)}\right],
\label{Isimp}
\end{equation}
which is now a simple function of $a_s$, without an additional integration. Next, we can use the known all-order LL form for the diagonal transition function $Z_{qq}$ given in eq.~\eqref{Zqq}.  Substituting this equation along with eq.~(\ref{Isimp}) into eq.~(\ref{Zgqsol1}), the LL solution for $Z_{gq}$ takes the form
\begin{align}
Z_{gq}(a_s)&=\frac{1}{\epsilon}
\exp\left[\frac{a_s}{\epsilon}\gamma_{gg}^{(0)}\right] \int^{a_s}_0
\frac{{\rm d}a'_s}{a'_s}
\,\exp\left[-\frac{a'_s}{\epsilon}\gamma_{gg}^{(0)}\right]\,
\gamma_{gq}(a'_s)\,\exp\left[\frac{a'_s}{\epsilon}\gamma_{qq}^{(0)}\right]
\notag\\
&=\sum_{m=0}^\infty\frac{\gamma_{gq}^{(m)}(N)}{\epsilon}\exp\left[\frac{a_s}{\epsilon}\gamma_{gg}^{(0)}\right]
 \int_0^{a_s}{\rm d}a'_s \, (a'_s)^{m}\, 
\exp\left[-\frac{a'_s}{\epsilon}\left(\gamma^{(0)}_{gg}-\gamma^{(0)}_{qq}
\right)\right],
\label{Zgqsol2}
\end{align}
where in the second line we have substituted the perturbation series
\begin{equation}
\gamma_{gq}(a_s)=\sum_{m=0}^\infty a_s^{m+1}\gamma_{gq}^{(m)}(N).
\label{gamgqseries}
\end{equation}
Let us now focus on a given value of $m$ on the right-hand side. We
may first transform to the integration variable
\begin{equation}
x'=\frac{a'_s(\gamma_{gg}^{(0)}-\gamma_{qq}^{(0)})}{\epsilon}
\label{x'def}
\end{equation}
such that eq.~(\ref{Zgqsol2}) can be integrated to give
\begin{equation}
Z_{gq}\Big|_{\gamma_{gq}^{(m)}}=\frac{\epsilon^m \gamma_{gq}^{(m)}}{(\gamma_{gg}^{(0)}
-\gamma_{qq}^{(0)})^{m+1}}\exp\left[\frac{a_s}{\epsilon} \gamma_{gg}^{(0)}\right]\gamma\left(m+1,\frac{a_s(\gamma_{gg}^{(0)} -\gamma_{qq}^{(0)})}{\epsilon}\right).
\label{Zgqsol3}
\end{equation}
where we have introduced the {\it lower incomplete gamma function}
\begin{equation}
\gamma(\alpha,z)=\int_0^z {\rm d } t\, t^{\alpha-1}e^{-t}\,,
\label{lowgam}
\end{equation}
not to be confused with an anomalous dimension. It has the series expansion
\begin{equation}
\gamma(\alpha,z)=z^{\alpha} e^{-z}\alpha!\sum_{k=0}^{\infty}\frac{z^k}{(\alpha+k+1)!}\,.
\label{gammaexpand}
\end{equation}
Implementing this in eq.~(\ref{Zgqsol3}) leads to
\begin{align}
Z_{gq}\Big|_{\gamma_{gq}^{(m)}}&=\frac{a_s^{m+1} \gamma_{gq}^{(m)}}{\epsilon} \, m!\, \exp\left[
\frac{a_s\gamma_{qq}^{(0)}}{\epsilon}\right]\, \sum_{k=0}^\infty
\frac{a_s^k}{\epsilon^k}\frac{(\gamma_{gg}^{(0)}-\gamma_{qq}^{(0)})^k}
{(m+k+1)!}\notag\\
&=\gamma_{gq}^{(m)} m!\sum_{k=0}^\infty\sum_{l=0}^\infty
\frac{a_s^{k+l+m+1}}{\epsilon^{l+k+1}}\frac{(\gamma_{qq}^{(0)})^l
(\gamma_{gg}^{(0)}-\gamma_{qq}^{(0)})^k}{l!(m+k+1)!},
\label{Zgqsol4}
\end{align}
where in the second line we have also expanded the overall exponential factor. We are currently still working at all orders in $a_s$. But let us now isolate a particular coefficient, say of $a_s^n$, so that one may set $k=n-l-m-1$ on the right-hand side:
\begin{equation}
Z_{gq}^{(n)}\Big|_{\gamma_{gq}^{(m)}}= \frac{\gamma_{gq}^{(m)}m!}{\epsilon^{n-m}}
\sum_{l=0}^{n-m-1}
\frac{(\gamma_{qq}^{(0)})^l
(\gamma_{gg}^{(0)}-\gamma_{qq}^{(0)})^{n-l-m-1}}{l!(n-l)!}\,.
\label{Zgqsol5}
\end{equation}
To connect to eq.~\eqref{Zgqfinal}, we want to write the summand in terms of pure powers of $\gamma_{qq}^{(0)}$ and $\gamma_{gg}^{(0)}$. To do so, we may use the binomial theorem
\begin{equation}
(x+y)^n=\sum_{p=0}^n \left(\begin{array}{c}n\\p\end{array}\right)
x^p y^{n-p}\,,
\label{binomial}
\end{equation}
to arrive at 
\begin{equation}
Z_{gq}^{(n)}\Big|_{\gamma_{gq}^{(m)}}=\frac{\gamma_{gq}^{(m)}m!}{\epsilon^{n-m}}
\sum_{l=0}^{n-m-1}\sum_{k=0}^{n-l-m-1}\frac{(-1)^{n-l-m-k-1}}{l!(n-l)!}
\left(\begin{array}{c}n-l-m-1\\ k\end{array}\right)
\left(\gamma_{qq}^{(0)}\right)^{n-m-1-k}\left(\gamma_{gg}^{(0)}\right)^{k}\,.
\label{Zgqsol6}
\end{equation}
The last step consists of removing the double sum. First, we interchange the order of summation in $k$ and $l$, and then relabel $l\rightarrow n-m-1-k-l$, to rewrite eq.~(\ref{Zgqsol6}) as
\begin{equation}
Z_{gq}^{(n)}\Big|_{\gamma_{gq}^{(m)}}=\frac{\gamma_{gq}^{(m)}m!}{\epsilon^{n-m}}
\sum_{k=0}^{n-m-1}\sum_{l=0}^{k}\frac{(-1)^{k-l}}{l!(n-l)!}
\left(\begin{array}{c}n-l-m-1\\ n-k-m-1\end{array}\right)
\left(\gamma_{qq}^{(0)}\right)^{k}\left(\gamma_{gg}^{(0)}\right)^{n-m-1-k}.
\label{Zgqsol7}
\end{equation}
The sum over $l$ has the form
\begin{align}
\sum_{l=0}^k\frac{(-1)^{-l}}{l!(n-l)!}\left(\begin{array}{c}
n-l-m-1\\ n-k-m-1\end{array}\right)&=
\frac{1}{n!}\sum_{l=0}^k (-1)^{-l}\left(\begin{array}{c}n\\l\end{array}\right)
\left(\begin{array}{c}n-l-m-1\\n-k-m-1\end{array}\right)\notag\\
&=\frac{(-1)^k}{n!}\sum_{l=0}^k
\left(\begin{array}{c}n\\l\end{array}\right)
\left(\begin{array}{c}k+m-n\\k-l\end{array}\right),
\label{lsum1}
\end{align}
where in the second line we have used~\cite{Binomial1}
\begin{equation}
\left(\begin{array}{c}n\\k\end{array}\right)=(-1)^{n-k}
\left(\begin{array}{c}-k-1\\n-k\end{array}\right).
\label{binomid}
\end{equation}
We can then use the {\it Chu-Vandermonde identity}
\begin{equation}
\sum_{j=0}^k\left(\begin{array}{c}m \\ j\end{array}\right)
\left(\begin{array}{c}n-m\\k-j\end{array}\right)
=\left(\begin{array}{c}n\\k\end{array}\right)
\label{CV}
\end{equation}
to get 
\begin{equation}
\sum_{l=0}^k \left(\begin{array}{c}n \\ l\end{array}\right)
\left(\begin{array}{c}k+m-n \\ k-l\end{array}\right)
=\left(\begin{array}{c}k+m \\ k\end{array}\right).
\label{lsum2}
\end{equation}
Substituting our results back into eqs.~(\ref{Zgqsol7}), we find eq.~(\ref{Zgqfinal}) as required, upon reintroducing the sum over $m$. Recall that $\gamma^{(m)}_{gq}$ contributes at ${\cal O}(a_s^{m+1})$, so that the upper limit of the $m$ sum must be $n-1$ rather than $n$.


\bibliography{paper}
\end{document}